\documentclass[aps,prd,groupedaddress,showpacs,showkeys]{revtex4}
\usepackage{amssymb}
\usepackage{epsfig}

\newcommand{\eq}{\begin{eqnarray}}
\newcommand{\en}{\end{eqnarray}}
\newcommand{\bea}{\begin{eqnarray}}
\newcommand{\eea}{\end{eqnarray}}

\newcommand{\ra}{\rangle}
\newcommand{\la}{\langle}
\newcommand{\mnz}{\stackrel{\!\!\!\!\!\circ}{m_N}} 
\newcommand{\MPz}{\stackrel{\!\!\!\circ}{M_P}} 
\newcommand{\Dt}{D_\tau} 
 
\begin{document}

\title{\Large\bf Chiral Dynamics of Baryons in \\
       a Lorentz Covariant Quark Model}

\author{Amand \ Faessler, Th.\ Gutsche, V.\ E.\ Lyubovitskij, K.\ Pumsa-ard 
\vspace*{.8\baselineskip}}
\affiliation{Institut f\"ur Theoretische Physik, Universit\"at T\"ubingen, 
\\
Auf der Morgenstelle 14,D-72076 T\"ubingen, Germany\\}

\date{\today}

\begin{abstract} 
We develop a manifestly Lorentz covariant chiral quark model for the 
study of baryons as bound states of constituent quarks dressed
by a cloud of pseudoscalar mesons.
The approach is based on a non-linear
chirally symmetric Lagrangian, which involves effective degrees of
freedom - constituent quarks and the chiral (pseudoscalar meson) fields.
In a first step, this Lagrangian can be used to perform a dressing of
the constituent quarks by a cloud of light pseudoscalar mesons and other
heavy states using the calculational technique
of infrared dimensional regularization of loop diagrams.
We calculate the dressed transition operators with a proper
chiral expansion which are relevant for the interaction of quarks with
external fields in the presence of a virtual meson cloud.
In a second step, these dressed operators are used to calculate baryon 
matrix elements. Applications are worked out for the masses of the 
baryon octet, the meson-nucleon sigma terms, the magnetic moments of 
the baryon octet, the nucleon charge radii, 
the strong vector meson-nucleon couplings and the full momentum 
dependence of the electromagnetic form factors of the nucleon.
\end{abstract}


\maketitle

\newpage 

\section{Introduction} 
Chiral symmetry plays an important role in the low-energy 
(below 1 GeV) domain of Quantum Chromodynamics (QCD): it 
governs the strong interaction between hadrons. All known 
low-energy approaches (effective field theories, Lattice 
QCD, QCD sum rules, different types of quark models, etc.) 
in the study of the properties of light hadrons have to 
incorporate the concept of at least an approximate chiral 
symmetry to get reasonable agreement with data. 

The most convenient language for the treatment of light hadrons at 
small energies was elaborated in the context of Chiral 
Perturbation Theory (ChPT)~\cite{Weinberg:1978kz,Gasser:1983yg},  
the effective low-energy theory of the strong interaction. 
ChPT is based on a chiral expansion of the QCD Green functions, 
i.e. an expansion in powers of the external hadron momenta and 
quark masses. It was proved~\cite{Gasser:1983yg} that ChPT works 
perfectly in the meson sector (especially in the description of 
pion-pion interactions). A manifestly Lorentz invariant form of 
baryon ChPT was suggested in Ref.~\cite{Gasser:1987rb}. In the 
baryon sector a new scale parameter associated with the nucleon 
mass shows up and this leads to certain difficulties in the 
formulation of a consistent chiral expansion of matrix elements. 
In particular, the chiral expansion of the loop diagrams starts 
at the same order as the tree-level graphs. This leads to an 
inconsistency in the perturbation theory: the higher order graphs 
contribute to the low-order ones and the physical quantities (e.g., 
nucleon mass) require renormalization at every order of the 
expansion. Later, a method, referred to as Heavy Baryon Chiral 
Perturbation Theory (HBChPT)~\cite{Jenkins:1990jv}, was suggested, 
which is able to avoid the problems with chiral power counting. 
HBChPT keeps track of power counting at every step of the 
calculation. A disadvantage of HBChPT is the lack of manifest 
Lorentz covariance due to the nonrelativistic expansion of the 
nucleon propagator. As was pointed out in Ref.~\cite{Becher:1999he} 
this method also suffers from a further deficiency. Namely, the 
nonrelativistic expansion of the pion-nucleon scattering amplitude 
generates a convergence problem of the perturbative series in part 
of the low-energy region. 
In Refs.~\cite{Becher:1999he,Ellis:1997kc,Gegelia:1999gf,Schindler:2003xv} 
a new method for the study of baryons in ChPT was suggested. 
It is based on the infrared dimensional regularization of loop 
diagrams~\cite{Becher:1999he}, which exploits the advantages of 
the two frameworks formulated in Ref.~\cite{Gasser:1987rb} and 
Refs.~\cite{Jenkins:1990jv}, while avoiding their disadvantages. 
An equivalent formulation of Baryon ChPT based 
on the extended on-mass-shell renormalization was suggested in 
Refs.~\cite{Fuchs:2003qc,Fuchs:2003sh,Schindler:2005ke}. 
A  successful application of the improved versions of Baryon ChPT 
to nucleon properties has been performed in 
Refs.~\cite{Becher:1999he,Schweizer:2000di,Kubis:2000zd,Kubis:2000aa,%
Fuchs:2003sh,Schindler:2005ke}. In Ref.~\cite{Goity:2001ny} the method 
has been extended to the multi-nucleon sector. 
The consistent inclusion of vector mesons 
in Baryon ChPT has been done in Ref.~\cite{Fuchs:2003sh}, which 
helped to successfully improve the momentum behavior of the 
nucleon form factors up to approximately 
0.4 GeV$^2$~\cite{Schindler:2005ke}. 

Unfortunately, in the context of Baryon ChPT one is able to 
calculate the momentum dependence of hadronic matrix elements only 
in a sufficiently narrow region (e.g., an accurate description of 
nucleon electromagnetic form factors has been achieved up to about 
$Q^2 = 0.4$ GeV$^2$~\cite{Kubis:2000zd,Schindler:2005ke}).  
Also, chiral symmetry is 
not the only important feature of strong interactions in the low-energy 
domain. There are the problems of hadronization and confinement 
which are completely avoided in the effective field theories dealing 
with hadronic degrees of freedom. These additional effects certainly 
have a strong impact on the hadronic interactions at intermediate 
energies. As an illustration of the importance of hadronization and 
confinement phenomena related to hadronic properties we recommend 
the review~\cite{Efimov:1993ei}. In Ref.~\cite{Efimov:1993ei} 
the general principles of QCD-motivated relativistic quark models 
have been formulated with the explicit incorporation of the three 
low-energy key phenomena: hadronization, confinement and approximate 
chiral symmetry. Many properties of light and heavy flavored baryons 
(including such sophisticated characteristics as slope parameters 
and form factors) have been successfully described in this 
approach~\cite{Efimov:1993ei} and in a later developed similar 
model~\cite{Ivanov:1996pz,Ivanov:1996fj}. 

The main objective of the present work is to develop a Lorentz 
covariant chiral quark model~\cite{Faessler:2005fa} which is 
consistent with the latest developments in the baryon sector 
of ChPT and leaves space for the additional features of low-energy 
QCD - hadronization and confinement. The full 
approach~\cite{Faessler:2005fa} is an extension of the original 
chiral quark model suggested and developed in 
Refs.~\cite{Gutsche:1989vy,PCQM1,PCQM2,PCQM3}. We treat the 
constituent quarks as the intermediate degrees of freedom 
between the current quarks (building blocks of the QCD Lagrangian) 
and the hadrons (building blocks of ChPT). This concept dates back 
to the pioneering works of Refs. \cite{Altarelli:1973ff,Manohar:1983md}. 
Furthermore, our strategy in dressing the constituent quarks by a 
cloud of pseudoscalar mesons is motivated by the procedure pursued 
in Ref.~\cite{Manohar:1983md}. Recent analyses of experiments at 
Jefferson Lab (TJLAB)~\cite{Osipenko:2003bu}, Fermilab~\cite{Adams:1991cs}, 
BNL~\cite{Allgower:2002qi} and IHEP (Protvino)~\cite{Mochalov:2003qd} 
renewed the interest in the concept of constituent quarks. The obtained 
data can be interpreted in a picture, where the hadronic quasiparticle 
substructure is assumed to consist of constituent quarks with nontrivial 
form factors. These experiments also initiated new progress in the 
manifestation of constituent degrees of freedom in hadron phenomenology 
(see, e.g. Refs.~\cite{Petronzio:2003bw}).  

The broader concept of chiral quark models dates back to the work
of the early eighties~\cite{Theberge:1980ye,Oset:1984tv,Diakonov:1983hh},
where the nucleon is described as a bound system of valence quarks
with a surrounding pion cloud simulating the sea-quark contributions.
These models aim to include the two main features of low-energy hadron 
structure, confinement and chiral symmetry. With respect to the treatment 
of the pion cloud these approaches fall essentially into two categories. 
The first type of chiral quark models assumes that the valence quark 
content dominates the nucleon, thereby treating pion contributions 
perturbatively~\cite{Gutsche:1989vy,Theberge:1980ye,Oset:1984tv}.
Originally, this idea was formulated in the context of the cloudy bag
model~\cite{Theberge:1980ye}. By imposing chiral symmetry the MIT
bag model~\cite{Chodos:1974je} was extended to include the interaction
of the confined quarks with the pion fields on the bag surface. With the 
pion cloud treated as a perturbation on the basic features of the MIT bag, 
pionic effects generally improve the description of nucleon observables.
Later, similar perturbative chiral models~\cite{Gutsche:1989vy,Oset:1984tv} 
were developed where the rather unphysical sharp bag boundary is replaced 
by a finite surface thickness of the quark core. By introducing a static 
quark potential of general form, these quark models contain a set of free 
parameters characterizing the confinement (coupling strength) and/or the 
quark masses. The perturbative technique allows a fully quantized treatment 
of the pion field up to a given order in accuracy, usually evaluated in 
leading order. Although formulated on the quark level, where confinement 
is put in phenomenologically, perturbative chiral quark models are 
conceptually close to chiral perturbation theory on the hadron level.
Alternatively, when the pion cloud is assumed to dominate the nucleon 
structure this effect has to be treated non-perturbatively. 
The non-perturbative approaches are based for example on 
Refs.~\cite{Diakonov:1983hh}, where the chiral quark soliton model was 
derived. This model is based on the concept that the QCD instanton vacuum 
is responsible for the spontaneous breaking of chiral symmetry, which in 
turn leads to an effective chiral Lagrangian at low energy as "derived"
from QCD. On the phenomenological level the chiral quark soliton model 
tends to be advantageous in the description of the nucleon spin structure, 
that is for large momentum transfers, but is comparable to the original 
perturbative chiral quark models in the description of low-energy nucleon 
properties.

As a further development of chiral quark models with a perturbative
treatment of the pion cloud~\cite{Gutsche:1989vy,Theberge:1980ye,Oset:1984tv}, 
we extended the relativistic quark model suggested in~\cite{Gutsche:1989vy} 
for the study of the low-energy properties of the 
nucleon~\cite{PCQM1,PCQM2,PCQM3}. In the current manuscript we perform an 
extension of our previous approach in two directions: we formulate a chiral 
quark Lagrangian, which dynamically generates the dressing of the bare 
constituent quarks by meson degrees of freedom up to fourth order.
We also formulate a manifestly Lorentz-covariant version concerning
the structure of the bare constituent quarks, exemplified mainly
for the case of the electromagnetic form factors of the nucleon.
The resulting expectation values of dressed constituent quark operators
evaluated for nucleon states are matched and checked in their low-energy
behaviour with results of Baryon ChPT.

In the manuscript we proceed as follows. First, in Section II, we derive 
an effective Lagrangian, which is taken from Baryon 
ChPT~\cite{Gasser:1987rb,Becher:1999he,Fettes:1998ud,Kubis:1999xb}, and 
formulate it in terms of quark and mesonic degrees of freedom by also 
including external fields. Second, we use this Lagrangian to perform 
a dressing of the constituent quarks by a cloud of light pseudoscalar 
mesons and by other heavy states. In this vein we use the calculational 
technique developed by Becher and Leutwyler~\cite{Becher:1999he}.
We derive dressed transition operators with a proper chiral expansion, 
which in turn are relevant for the interaction of quarks with external 
fields in the presence of a virtual meson cloud. In Section III, we work 
out the implications for the chiral expansion of the nucleon mass and 
the meson-nucleon sigma terms, which, in consistency with Baryon ChPT, 
gives constraints on the parameters entering in the chiral quark 
Lagrangian. Applications to various quantities are worked out and 
presented in Section IV. We discuss the masses of the baryon octet and 
mainly the pion-nucleon sigma term. Using model-independent constraints
on the bare constituent quark distributions in the octet baryons,
we work out model predictions for the magnetic moments.
Using finally a full parameterization of the bare constituent quark
distributions in the nucleon, we give results for the full
momentum dependence of the electromagnetic form factors of the nucleon
and indicate the role of the meson cloud contributions.

\section{Approach} 

\subsection{Chiral Lagrangian} 

\noindent 
The chiral quark Lagrangian ${\cal L}_{qU}$ (up to order $p^4$), which 
dynamically generates the dressing of the constituent quarks by mesonic 
degrees of freedom, consists of the two main pieces ${\cal L}_{q}$ 
and~${\cal L}_{U}$: 
\eq\label{L_qU}
{\cal L}_{qU} \, = \, {\cal L}_{q} + {\cal L}_{U}\,, 
\hspace*{.5cm}
{\cal L}_q \, = \, {\cal L}^{(1)}_q + {\cal L}^{(2)}_q + 
           {\cal L}^{(3)}_q + {\cal L}^{(4)}_q + \ldots\,, 
\hspace*{.5cm}  
{\cal L}_{U} \, = \, {\cal L}_{U}^{(2)} + \ldots\,.   
\en
The superscript $(i)$ attached to ${\cal L}^{(i)}_{q(U)}$ 
denotes the low energy dimension of the Lagrangian: 
\eq\label{L_exp}
{\cal L}_{U}^{(2)} &=&\frac{F^2}{4} \la{u_\mu u^\mu + \chi_+}\ra\,,
\hspace*{.5cm} 
{\cal L}^{(1)}_q \, = \,  \bar q \left[ i \, \slash\!\!\!\! D - m
+ \frac{1}{2} \, g \, \slash\!\!\! u \, \gamma^5 \right] q\,, 
\nonumber\\[2mm] 
{\cal L}^{(2)}_q & = & c_1 \, \la{\chi_+}\ra  \, \bar q q 
\, - \, \frac{c_2}{4m^2} \, \la{u_\mu u_\nu}\ra \, 
(\bar q \, D^\mu \, D^\nu \, q \, + \, {\, \rm h.c. \,}) \, + \, 
\frac{c_3}{2} \, \la{u_\mu \, u^\mu}\ra \bar q q \nonumber\\
&+& \frac{c_4}{4}\, \bar q\, i\, \sigma^{\mu\nu}\, [u_\mu, u_\nu]\, q 
\, + \, \frac{c_6}{8m}\,\bar q\,\sigma^{\mu\nu}\, F_{\mu\nu}^+ \, q\,  
- \bar q {\cal M} q + c_5 \bar q \hat\chi_+  q\, + \, \ldots , \\[2mm]
{\cal L}^{(3)}_q &=& \frac{id_{10}}{2m} \,\bar q \, [D^\mu, F_{\mu\nu}^+] 
\, D^\nu \, q \, + \, {\, \rm h.c. \,} \, +  \ldots \,,\nonumber\\[2mm]
{\cal L}^{(4)}_q &=& - \frac{e_1}{16}\la \chi_+ \ra^{2} \, \bar q \, q \, 
+ \, \frac{e_2}{4}\la \chi_+ \ra \Box (\bar q \, q) 
- \frac{e_3}{16} \la \hat\chi_+^2 \ra  \bar q q 
- \frac{e_4}{16} \la \chi_+ \ra \bar q \hat\chi_+  q   
- \frac{e_5}{16} \bar q \hat\chi_+^2  q  \nonumber \\
&+&\frac{e_6}{2} \, \la \chi_+ \ra \, 
\bar q\,\sigma^{\mu\nu}\, F_{\mu\nu}^+ \, q\, 
+\, \frac{e_7}{4} \, 
\bar q\,\sigma^{\mu\nu}\, \{F_{\mu\nu}^+ \hat\chi_+\} \, q\,
+\, \frac{e_8}{2} \, 
\bar q\,\sigma^{\mu\nu}\, \la F_{\mu\nu}^+ \hat\chi_+ \ra \, q\, 
- \frac{e_{10}}{2}\,\bar q \, [D^\alpha, [D_\alpha, F_{\mu\nu}^+] ]   
\sigma^{\mu\nu} \, q \, + 
\ldots, \nonumber  
\en 
where $\hat\chi_+ = \chi_+ - \frac{1}{3} \la{\chi_+}\ra\,$, 
the symbols $\la \,\, \ra$, $[ \,\, ]$ and $\{ \,\, \}$ 
occurring in Eq.~(\ref{L_exp}) denotes the trace over flavor 
matrices, commutator and anticommutator, respectively.  

The couplings $m$ and $g$ denote the quark mass and axial charge 
in the chiral limit, $c_i$, $d_i$ and $e_i$ are 
the second-, third- and fourth-order low-energy 
coupling constants, respectively, which encode the contributions of 
heavy states. Note, that the inclusion of higher-dimensional terms 
in the chiral quark Lagrangian was originally suggested 
in Ref.~\cite{Manohar:1983md}. 
In particular, as a dimensional parameter in the higher-dimensional terms 
one can use the scale parameter of spontaneously broken chiral symmetry 
$\Lambda_{\chi} \simeq 4 \pi F \sim 1$ GeV
($F$ is the octet decay constant~\cite{Kubis:2000aa,Kubis:1999xb})
instead of the constituent quark mass.
This replacement is equivalent to a redefinition of the values of the 
low-energy constants, e.g. $c_2 \, \to \, c_2 \, (m/\Lambda_\chi)^2\,, 
\; d_{10} \, \to \, d_{10} \, (m/\Lambda_\chi)$, etc. 
Both scale parameters $m$ and $\Lambda_\chi$ are counted 
as the same order quantities in the chiral expansion, i.e 
as quantities of order $O(1)$.  
 
Additional terms in the Lagrangian denoted generically by dots do not 
contribute to the electromagnetic nucleon form factors and meson-nucleon 
sigma-terms, which are explicitly worked out later on in the applications.
Here $q$ is the quark field, the octet of pseudoscalar fields  
\eq 
\phi = \sum_{i=1}^{8} \phi_i\lambda_i = \sqrt{2} 
\left(
\begin{array}{ccc}
\pi^0/\sqrt{2} + \eta/\sqrt{6}\,\, & \,\, \pi^+ \,\, & \, K^+ \\
\pi^- \,\, & \,\, -\pi^0/\sqrt{2}+\eta/\sqrt{6}\,\, & \, K^0\\
K^-\,\, & \,\, \bar K^0 \,\, & \, -2\eta/\sqrt{6}\\
\end{array}
\right). 
\en 
is contained in the SU(3) matrix 
$U = u^2 = {\rm exp}(i\phi/F)$ where $F$ the octet decay 
constant~\cite{Kubis:2000aa,Kubis:1999xb}. We introduce the 
standard notations~\cite{Gasser:1987rb,Becher:1999he,Fettes:1998ud}
\eq 
& &D_\mu = \partial_\mu + \Gamma_\mu, \hspace*{.3cm} 
\Gamma_\mu = \frac{1}{2} [u^\dagger, \partial_\mu u] 
- \frac{i}{2} u^\dagger R_\mu u 
- \frac{i}{2} u L_\mu u^\dagger, \\[2mm] 
& &u_\mu = i u^\dagger \nabla_\mu U u^\dagger, \hspace*{.3cm} 
\chi_\pm = u^\dagger \chi u^\dagger \pm u \chi^\dagger u, \hspace*{.3cm} 
\chi = 2 B (s + ip), \,\,\, s = {\cal M} + \ldots \, .  \nonumber 
\en 
The fields $R_\mu$ and $L_\mu$ include sources 
\eq
R_\mu &=& v_\mu \, + \, a_\mu = e \, Q \, A_\mu \, - \, 
Q \, {\rm tan} \theta_W Z_\mu^0 + \ldots \,, \\ [2mm] 
L_\mu &=& v_\mu \, - \, a_\mu = e \, Q \, A_\mu \, + \, 
\biggl(\frac{e}{\sin^2\theta_W} \lambda_3 \, - \,  Q \biggr) \, 
{\rm tan}\theta_W \, Z_\mu^0 \, + \,  \frac{e}{\sin\theta_W\sqrt{2}} 
(W_\mu^+ T_+ \, + \, {\rm h.c.}) + \ldots \nonumber
\en 
where $s, p, v_\mu$ and $a_\mu$ denote the external scalar, 
pseudoscalar, vector and axial fields; \\ 
$A_\mu$, $W_\mu^\pm$ and $Z_\mu^0$ are the electromagnetic field, 
the weak charged and neutral boson fields.
The quark charge matrix is denoted by 
$Q = {\rm diag} \{ 2/3,-1/3,-1/3 \}$ and 
\eq
T_+ =
\left(
\begin{array}{ccc} 
0 & V_{ud} & V_{us} \\
0 & 0 & 0 \\
0 & 0 & 0 \\
\end{array}
\right) 
\en 
is the weak matrix containing the Cabibbo-Kobayashi-Maskawa quark-mixing 
matrix elements $V_{ij}$. The tensor $F_{\mu\nu}^+$ is defined as 
$F_{\mu\nu}^+ \, = \, u^\dagger F_{\mu\nu}  Q u +  
u F_{\mu\nu} Q u^\dagger$ where 
$F_{\mu\nu} = \partial_\mu A_\nu - \partial_\nu A_\mu$ is the 
conventional photon field strength tensor. 
Here ${\cal M} = {\rm diag}\{\hat m, \hat m, \hat m_s\}$ is the mass 
matrix of current quarks (we restrict to the isospin symmetry limit 
with $\hat m_{u}= \hat m_{d}=\hat{m}=7$ MeV and the mass of 
the strange quark $\hat m_s$ is related to the nonstrange one as  
$\hat m_s = 25 \, \hat m$).
The quark vacuum condensate parameter is denoted by
\eq 
B = - \frac{1}{F^2} \la 0|\bar u u|0 \ra  = 
      - \frac{1}{F^2} \la 0|\bar d d|0 \ra  \,. 
\en
To distinguish between 
constituent and current quark masses we attach the symbol 
$\ {\bf\hat{}}$ ("hat") when referring to the current quark masses. 
We rely on the standard picture of chiral symmetry 
breaking~($B \gg F$). In the leading order of the chiral expansion
the masses of pseudoscalar mesons are given by 
\eq\label{M_Masses}
M_{\pi}^2=2 \hat m B, \hspace*{.5cm} 
M_{K}^2=(\hat m + \hat m_s) B, \hspace*{.5cm} 
M_{\eta}^2= \frac{2}{3} (\hat m + 2 \hat m_s) B. 
\en 
In the numerical analysis we will use: $M_{\pi} = 139.57$ MeV, 
$M_K = 493.677$ MeV (the charged pion and kaon masses), 
$M_\eta = 574.75$ MeV and the canonical set of differentiated decay 
constants:  $F_\pi = 92.4$ MeV, $F_K/F_\pi = 1.22$ and 
$F_\eta/F_\pi = 1.3$~\cite{Gasser:1984gg}. 

The use of the physical masses and decay constants of pseudoscalar mesons 
incorporates only part of the corrections due to the breaking of 
unitary flavor symmetry. 
To generate another part of $SU(3)$ symmetry-breaking corrections 
we added a string of terms to the Lagrangian~(\ref{L_qU}):   
the current quark mass term $\bar q {\cal M} q$, terms containing LECs 
$c_5$, $e_4$, $e_5$, $e_7$ and $e_8$. 
The flavor-symmetry breaking terms containing term $\bar q {\cal M} q$ 
and LECS $c_5$, $e_4$ and $e_5$ allow to 
decouple the mass of the strange quark from the isospin-averaged 
mass $m$. The fourth-order couplings $e_7$ and $e_8$ incorporate 
explicit $SU(3)$ symmetry-breaking corrections in the magnetic moments 
of the constituent quarks and baryons.  
As was shown in Ref.~\cite{Kubis:2000aa} the inclusion of the  
$SU(3)$ symmetry-breaking terms is sufficient to obtain agreement 
with the experimental data for the magnetic moments of the baryon octet.  

\subsection{Inclusion of vector mesons}

Following Refs.~\cite{Gasser:1983yg,Kubis:2000zd,Ecker:1988te} 
we also include vector mesons in the chiral quark Lagrangian. 
In particular, we employ the tensor field representation of the 
spin-1 fields: vector mesons are written in terms of the antisymmetric 
tensor fields $W_{\mu\nu} = - W_{\nu\mu}$ where the three degrees of 
freedom $W_{ij} (i, j = 1, 2, 3)$ are frozen out. 
The latter representation is most convenient to construct the chirally 
invariant couplings of vector mesons to pions, photons and fermions 
(baryons, quarks): 
\eq 
{\cal L}_V = {\cal L}_V^0 + {\cal L}_{V}^{\rm int} 
\en 
where ${\cal L}_V^0$ is the free vector meson Lagrangian 
\eq 
{\cal L}_V^0 = - \frac{1}{2} \partial^\mu W^a_{\mu\nu} 
\partial_\rho W^{\rho\nu, a} + \frac{M_V^2}{4}  W^a_{\mu\nu} W^{\mu\nu, a} 
\en
and ${\cal L}_{V}^{\rm int} = {\cal L}_{V}^{\rm int, 1} \, + \,  
{\cal L}_{V}^{\rm int, 2}$ 
is the interaction Lagrangian of vector mesons with external vector and 
axial-vector sources $({\cal L}_{V}^{\rm int, 1})$ and with baryons
$({\cal L}_{V}^{\rm int, 2})$: 
\eq
{\cal L}_{V}^{\rm int, 1} &=& \frac{F_V}{2\sqrt{2}} 
\la \, W^{\mu\nu} \, F_{\mu\nu}^+ \, \ra  
\, + \, \frac{i G_V}{2\sqrt{2}} \la \, W^{\mu\nu} \, 
[u_\mu, u_\nu] \ra \,,\\[2mm] 
{\cal L}_{V}^{\rm int, 2} &=& 
\bar q \, \sigma^{\mu\nu} \, R_{\mu\nu} \, q \, + \, 
\bar q \, \gamma^{\mu}   \, S_\mu \, q \, + \, 
\bar q \sigma^{\alpha\beta} \, U_{\mu\beta} \, 
[ D_\alpha, [ D^\mu, q ] ] \,.\nonumber 
\en
Here we define
\eq
W_{\mu\nu} =
\left(
\begin{array}{ccc} 
(\rho^0 + \omega)/\sqrt{2} & \rho^+ & K^{\ast +} \\
\rho^-                    &(-\rho^0 + \omega)/\sqrt{2} & K^{\ast 0} \\
K^{\ast -} & \bar K^{\ast 0} & - \phi \\
\end{array}
\right)_{\mu\nu}\,\,,  
\en 
\eq 
R_{\mu\nu} = R_T W_{\mu\nu} + R_S \la  W_{\mu\nu}  \ra\,,\, 
S_\mu = S_T [D^\nu, W_{\mu\nu}] + S_S \la [D^\nu, W_{\mu\nu}]  
\ra\,,   
U_{\mu\beta} = U_T W_{\mu\beta} + U_S \la W_{\mu\beta} \ra\,, 
\nonumber
\en 
where $R_i$, $S_i$ and $U_i$ are the effective couplings related to 
the ones ($g_{Vqq}$ and $k_V$) used in the canonical interaction Lagrangian 
of vector mesons with quarks (for details on the nucleon-level 
Lagrangian see Ref.~\cite{Kubis:2000zd}): 
\eq\label{L_Vqq}  
{\cal L}_{Vqq} = \frac{g_{Vqq}}{\sqrt{2}} \, \bar q \biggl( 
\gamma^\mu \, V_\mu \, - \frac{k_V}{2 m} \, \sigma^{\mu\nu} \, 
\partial_\nu V_\mu \biggr) q \,. 
\en         
The standard nonet matrix of vector mesons is denoted by $V_\mu$ 
(its flavor content coincides with the one of the antisymmetric 
tensor $W_{\mu\nu}$). The matching conditions relating the couplings are: 
\eq 
R_S = 0\,, \hspace*{.3cm} 
R_T = - k_V \, g_{Vqq} \, \frac{M_V}{4 m \sqrt{2}} \,,  \hspace*{.3cm} 
U_S = \frac{2}{m} \, S_S \,, \hspace*{.3cm} U_T = \frac{2}{m} \, 
\biggl( \frac{g_{Vqq}}{M_V \, \sqrt{2}} \, +  \, S_T \biggr)\,. 
\en 
The couplings $F_V$ and $G_V$ are determined by the decays widths of
$\rho \to e^+ e^-$ and $\rho \to \pi \pi$. Using low-energy theorems, 
e.g. $\rho$-meson universality and 
the Kawarabayashi-Suzuki-Fayyazuddin-Riazuddin (KSFR) relation, one can 
express $F_V$ and $G_V$ by $g_{Vqq}$ and the vector meson mass 
$M_V$~\cite{Kubis:2000zd,Sakurai:1967jj}: $F_V = M_V/g_{Vqq}$ and 
$G_V = F_V/2$. Hence, in the vector-meson sector we only deal
with a single free parameter~$k_V$. 

\subsection{Power counting} 

The Lagrangian set up for constituent quarks, pseudoscalar and vector 
mesons is assumed to be valid 
in the region between the chiral symmetry breaking $\Lambda_{\chi}$ and 
confinement $\Lambda_{\rm QCD}$ scales. Such a Lagrangian 
is non-renormalizable due to the existence of an infinite tower of 
higher dimensional terms. A solution to this problem can be achieved if there 
is a dimensional parameter in the theory which suppresses corresponding 
non-renormalizable terms in the matrix elements. The problem of power 
counting in non-renormalizable effective chiral quark theories was 
discussed in detail in Ref.~\cite{Manohar:1983md}. It was shown in the 
framework of the so-called "naive dimensional analysis"  that higher-order 
corrections in the matrix elements can be suppressed by a dimensional 
parameter $\Lambda_{\chi} =  4 \pi F \sim 1$ GeV. 
Note, that a deficiency of quark models is the smallness of the constituent 
quark mass. Formally, we need to treat this quantity as ${\cal O}(q^0)$, 
while the pseudoscalar meson masses as ${\cal O}(q)$. 
Analytically the constituent quark mass survives in the chiral 
limit, while the pseudoscalar meson masses vanish.  
However, the numerical values of kaon and eta-meson masses are 
similar to the constituent quark mass. The contribution of the  
meson cloud is normally divided by the power of 
$\Lambda_\chi = 4 \pi F_P$ where $P = \pi, K, \eta$. The constituent 
quark mass can be finally removed from the expressions for the 
observables using its universal relation to the nucleon mass which is 
valid in the chiral limit and at one-loop (see details in Sec.~III): 
\eq\label{matching_mass} 
\frac{\mnz}{m}\, = \, \frac{m_N}{\bar m} \, = \,  
\biggl(\frac{g_A}{g}\biggr)^2 \,. 
\en 
In the last expression $m_N$ and $\mnz$ are the nucleon masses at one-loop 
(the physical mass) and in the chiral limit, respectively, 
$\bar m = m_u = m_d$ is the nonstrange constituent quark mass 
at one-loop (the dressed nonstrange constituent quark mass).  
Finally, all expressions for the baryonic observables do not contain 
the scale parameter - the constituent quark mass, but contain the 
dimensionless parameter - the axial charge of the constituent quark. 

\subsection{Dressing of quark operators} 

The total effective Lagrangian ${\cal L}_{\rm eff}$ includes the two 
terms ${\cal L}_{qU}$ and ${\cal L}_{V}$. The first term  $({\cal L}_{qU})$ 
is responsible for the dressing of quarks by a cloud of pseudoscalar mesons 
and heavy states, while the second one $({\cal L}_{V})$ generates the 
coupling to vector mesons. Any bare quark operator (both one- and two-body) 
can be dressed in a straightforward manner by use of the effective 
chirally-invariant Lagrangian ${\cal L}_{\rm eff}$. To illustrate the idea 
of such a dressing we consider the Fourier-transform of the electromagnetic 
quark operator: 
\eq\label{bare_V}
J_{\mu, \, {\rm em}}^{\rm bare}(q) = \int d^4x \, e^{-iqx} \, 
j_{\mu, \, {\rm em}}^{\rm bare}(x) \,, \quad \hspace*{.3cm}  
j_{\mu, {\rm em}}^{\rm bare}(x) = \bar q(x) \, \gamma_\mu \, Q \, q(x)\,. 
\en 
In Figs.1 and 2 we display the tree and loop diagrams which contribute 
to the dressed electromagnetic operator $J_{\mu, \, {\rm em}}^{\rm dress}$ 
up to fourth order. 

The dressed quark operator $j_{\mu, \, {\rm em}}^{\rm dress}(x)$ 
and its Fourier transform $J_{\mu, \, {\rm em}}^{\rm dress}(q)$ have the 
following forms 
\eq\label{Jmu_dress} 
j_{\mu, \, {\rm em}}^{\rm dress}(x) &=& \sum\limits_{q=u,d,s} \, 
\biggl\{ f_D^q(-\partial^2) \, [ \bar q(x) \gamma_\mu q(x) ] 
\, + \, \frac{f_P^q(-\partial^2)}{2m_q} \, \partial^\nu \, 
[ \bar q(x) \sigma_{\mu\nu} q(x) ] \biggr\} \,\\
J_{\mu, \, {\rm em}}^{\rm dress}(q) &=& \int d^4x \, e^{-iqx} \, 
j_{\mu, \, {\rm em}}^{\rm dress}(x) = \int d^4x \, e^{-iqx} \, 
\sum\limits_{q=u,d,s} 
\bar q(x) \, \biggl[ \, \gamma_\mu \, f_D^q(q^2) \, + \, \frac{i}{2m_q} \,  
\sigma_{\mu\nu} \, q^\nu \, f_P^q(q^2) \, \biggr] \, q(x)\,, \nonumber 
\en 
where $m_q$ is the dressed constituent quark mass (see details in 
Sec.~III);  $f_D^u(q^2)$, $f_D^d(q^2)$, $f_D^s(q^2)$ and 
$f_P^u(q^2)$, $f_P^d(q^2)$, $f_P^s(q^2)$ are 
the Dirac and Pauli form factors of $u$, $d$ and $s$ quarks. 
Here we use the appropriate sub- and superscripts with a definite 
normalization of the set of $f_D^q(0) \, \equiv \, e_q$ (quark charges) 
due to charge conservation. Note, that the dressed quark operator 
satisfies the current conservation: 
\eq\label{current_cons} 
\partial^\mu \, j_{\mu, \, {\rm em}}^{\rm dress}(x) =  
\sum\limits_{q=u,d,s} \, 
\biggl\{ f_D^q(-\partial^2) \partial^\mu [ \bar q(x) \gamma_\mu q(x) ]  
\, + \, \frac{f_P^q(-\partial^2)}{2m_q} \, \partial^\mu \partial^\nu 
[ \bar q(x) \sigma_{\mu\nu} q(x) ] \biggr\} = 0 \,. 
\en 
Evaluation of the diagrams in 
Figs.1 and 2 is based on the {\it infrared dimensional regularization} 
suggested in Ref.~\cite{Becher:1999he} to guarantee a straightforward 
connection between loop and chiral expansion in terms of quark masses and 
small external momenta. We relegate the discussion of the calculational 
technique~\cite{Kubis:2000zd} to the Appendices A (infrared regularization) 
and B (explicit form of the loop integrals). 

To calculate the electromagnetic form factors of the nucleon (or any baryon) 
we project the dressed quark operator between the nucleon (baryon) states. 
In the following we restrict to the case of the nucleon (the extension to 
any baryon is straightforward). The master formula is: 
\eq\label{master}
&&\la N(p^\prime) | \, J_{\mu, \, {\rm em}}^{\rm dress}(q) 
\, | N(p) \ra \, = \, (2\pi)^4 \, \delta^4(p^\prime - p - q) \, 
\bar u_N(p^\prime) \biggl\{ \gamma_\mu \, F_1^N(q^2) \, + \, 
\frac{i}{2 \, m_N} \, \sigma_{\mu\nu} q^\nu 
\, F_2^N(q^2) \biggr\} u_N(p) \, \nonumber\\
& = & (2\pi)^4 \, \delta^4(p^\prime - p - q) \sum\limits_{q = u,d}  
\biggl\{f_D^q(q^2) \, \la N(p^\prime)|\,j_{\mu, q}^{\rm bare}(0)\,|N(p) \ra
+  i \, \frac{q^\nu}{2 \, \bar m} \, f_P^q(q^2) \, 
\la N(p^\prime)| \, j_{\mu\nu, q}^{\rm bare}(0) \, |N(p) \ra 
\biggr\} \, ,
\en
where $N(p)$ and $u_N(p)$ are the nucleon state and spinor, 
respectively, normalized as 
\eq 
\la N(p^\prime) | N(p) \ra = 
2 E_N \, (2\pi)^3 \, \delta^3(\vec{p}-\vec{p}^{\,\prime}) 
\en 
and 
\eq 
\bar u(p) u(p) = 2 m_N 
\en 
with $E_N$ being the nucleon energy $E_N = \sqrt{m_N^2+\vec{p}^{\,2}}$. 
Here $F_1^N(q^2)$ and $F_2^N(q^2)$ are the Dirac and Pauli nucleon 
form factors. For convenience we present the expressions for the isoscalar 
$F_{1(2)}^S$ and isovector $F_{1(2)}^V$ nucleon form factors in Sec.~IVc.  
In Eq.~(\ref{master}) we express the matrix elements of the dressed quark 
operator by the matrix elements 
of the bare operators. In our case we deal with the bare quark operators of
the vector 
$j_{\mu, q}^{\rm bare}(0)$ and tensor $j_{\mu\nu, q}^{\rm bare}(0)$
structures defined as 
\eq\label{bare_operators} 
j_{\mu, q}^{\rm bare}(0) \, = \, \bar q(0) \, \gamma_\mu \, q(0)\,, 
\hspace*{1cm} j_{\mu\nu, q}^{\rm bare}(0) \, = \, \bar q(0) \, 
\sigma_{\mu\nu} \, q(0)\,. 
\en 
\noindent 
Eq.~(\ref{master}) contains our main result: we perform a model-independent 
factorization of the effects of hadronization and confinement contained in 
the matrix elements of the bare quark operators $j_{\mu, q}^{\rm bare}(0)$ 
and $j_{\mu\nu, q}^{\rm bare}(0)$ and the effects dictated by chiral 
symmetry (or chiral dynamics) which are encoded in the relativistic form 
factors $f_D^q(q^2)$ and $f_P^q(q^2)$. Due to this factorization 
the calculation of $f_D^q(q^2)$ and $f_P^q(q^2)$, on one side, 
and the matrix elements of $j_{\mu, q}^{\rm bare}(0)$ and 
$j_{\mu\nu, q}^{\rm bare}(0)$, on the other side, can be done 
independently. In particular, in a first step we derived 
a model-independent formalism based on the ChPT Lagrangian, 
which is formulated in terms of constituent quark degrees of freedom,
for the calculation of $f_D^q(q^2)$ 
and $f_P^q(q^2)$. The calculation of the matrix elements of the bare 
quark operators can then be relegated to quark models based on specific 
assumptions about hadronization and confinement. The explicit forms of 
$f_D^q(q^2)$ and $f_P^q(q^2)$ are given in Appendix C. 
Note that we also show in Fig.3 the diagrams contributing to the 
strong vector meson-nucleon interactions $\rho NN$ and $\omega NN$ 
at one loop which will be discussed in Sec.~IV(D). 

\subsection{Matching to ChPT} 

The matrix elements of the bare quark operators should be calculated 
using specific model assumptions about hadronization and confinement. 
However, the use of certain symmetry constraints, discussed
in the following, leads to a set
of relationships between the nucleon and the corresponding quark form factors
at their normalization point at zero momentum. 
In general, due to Lorentz and gauge invariance, the matrix elements 
in Eq.~(\ref{master}) can be written as 
\eq 
\la N(p^\prime) | \, j_{\mu, q}^{\rm bare}(0) \, | N(p) \ra &=& 
\bar u_N(p^\prime) \biggl\{ \gamma_\mu \, F_1^{Nq}(q^2) 
\, + \, \frac{i}{2 \, m_N} 
\, \sigma_{\mu\nu} \, q^\nu \,  F_2^{Nq}(q^2) \biggr\} u_N(p)\,, \\ 
i \, \frac{q^\nu}{2 \, m_q}  
\la N(p^\prime) | \, j_{\mu\nu, q}^{\rm bare}(0) \, | N(p) \ra &=& 
\bar u_N(p^\prime) \biggl\{ \gamma_\mu \, G_1^{Nq}(q^2) 
\, + \, \frac{i}{2 \, m_N}  
\, \sigma_{\mu\nu} \, q^\nu \,  G_2^{Nq}(q^2) \biggr\} u_N(p) \, ,
\nonumber 
\en 
where $F_{1(2)}^{Nq}(q^2)$ and $G_{1(2)}^{Nq}(q^2)$  are the Pauli and 
Dirac form factors describing the distribution of quarks of flavor 
$q=u, d$ in the nucleon. 

The first set of relations arise from charge conservation and isospin 
invariance: 
\eq 
& &F_1^{pu}(0) = F_1^{nd}(0) = 2\,, \hspace*{.5cm} 
F_1^{pd}(0) = F_1^{nu}(0) = 1\,, \hspace*{.5cm} 
G_1^{Nq}(0) = 0\,, \\
& &F_2^{pu}(0) = F_2^{nd}(0)\,, \hspace*{.5cm} 
F_2^{pd}(0) = F_2^{nu}(0)\,, \hspace*{.5cm} 
G_2^{pu}(0) = G_2^{nd}(0)\,, \hspace*{.5cm} 
G_2^{pd}(0) = G_2^{nu}(0)\,. \nonumber 
\en  
Note, that the quantities $G_2^{Nq}(0)$ are related to the bare 
nucleon tensor charges $\delta_{Nq}^{\rm bare}$: 
\eq 
& &G_2^{pu}(0) \, = \, G_2^{nd}(0) \, = \, 
\frac{m_N}{\bar m} \, \delta_{pu}^{\rm bare}  
\, = \, \frac{m_N}{\bar m} \, \delta_{nd}^{\rm bare}  \,, \\
& &G_2^{pd}(0) \, = \, G_2^{nu}(0) \, = \, 
\frac{m_N}{\bar m} \, \delta_{pd}^{\rm bare}  
\, = \, \frac{m_N}{\bar m} \, \delta_{nu}^{\rm bare}  
\,, \nonumber `
\en 
where $\delta_{Nq}^{\rm bare}$ are defined by~\cite{Jaffe:1991kp}:  
\eq
\la N(p) | \, j_{\mu\nu, q}^{\rm bare}(0) \, | N(p) \ra \, = \,  
\delta_{Nq}^{\rm bare} \, \bar u_N(p) \, \sigma_{\mu\nu} \, u_N(p) \,. 
\en 
The second set of constraints are the so-called {\it chiral symmetry 
constraints}. They are dictated 
by the infrared-singular structure of QCD in order to reproduce 
the leading nonanalytic (LNA) contributions to the magnetic moments and 
the charge and magnetic radii of nucleons~\cite{Kubis:2000zd,Beg:1973sc}: 
\eq\label{chiral_constr} 
\mu_p &=& - \frac{g_A^2}{8 \pi} \, \frac{M_\pi}{F_\pi^2} \, 
\mnz \, + \, \ldots \,, \nonumber\\ 
\la r^2 \ra^E_p &=& - \frac{1 + 5 g_A^2}{16 \, \pi^2 \, F_\pi^2} \,  
{\rm ln}\frac{M_\pi}{\mnz} \, + \, \ldots \,, \\ 
\la r^2 \ra^M_p &=& \frac{g_A^2}{16 \, \pi \, F_\pi^2 \, \mu_p} \, 
\frac{\mnz}{M_\pi} \, + \, \ldots \,,  \nonumber 
\en  
where $g_A$ and $\mnz$ are the axial charge and mass of the nucleon 
in the chiral limit.  
In particular, the LNA contribution to the magnetic 
moments is proportional to $M_\pi$.  
The nucleon radii are divergent in the chiral limit.   
The LNA contribution to the charge radii is proportional to the 
chiral logarithm ${\rm ln}(M_\pi/\mnz)$. The LNA contributions 
to the magnetic radii are represented by the same logarithm as in 
the case of the charge radii and by the singular term proportional to 
$1/M_\pi$. In order to fulfill the chiral symmetry 
constraints~(\ref{chiral_constr}) we derive the following identities 
involving the $F_2^{Nq}(0)$ and $G_2^{Nq}(0)$ form factors and
the low-energy constant (LEC) $d_{10}$:  
\eq\label{chiral_constr2}
& &1 + F_2^{pu}(0) - F_2^{pd}(0) = G_2^{pu}(0) - G_2^{pd}(0) = 
\biggl(\frac{g_A}{g}\biggr)^2\, \frac{m_N}{\bar m}\,, \\
& &1 + F_2^{nd}(0) - F_2^{nu}(0) = G_2^{nd}(0) - G_2^{nu}(0) = 
\biggl(\frac{g_A}{g}\biggr)^2 \, \frac{m_N}{\bar m}  
\nonumber  
\en 
and 
\eq\label{d_6} 
\bar d_6^{\rm ChPT} + \frac{1 + 5 g_A^2}{96 \pi^2 F_\pi^2} 
\, {\rm\ln}\frac{M_\pi}{\mnz} \equiv \bar d_{10} + 
\frac{1 + 5 g^2}{96 \pi^2 F_\pi^2}\,{\rm\ln}\frac{M_\pi}{\mnz}\,. 
\en 
Here $\bar d_6^{\rm ChPT}$ and $\bar d_{10}$ denote the renormalized LECs 
$d_6^{\rm ChPT}$ and $d_{10}$, respectively, at $\mu = \mnz$: 
\eq 
\bar d_6^{\rm ChPT} = d_6^{\rm ChPT} + 
\frac{1 + 5 g_A^2}{6 F_\pi^2} \bar\lambda \,, \quad\quad 
\bar d_{10} = d_{10} + \frac{1 + 5 g^2}{6 F_\pi^2} \bar\lambda \,  
\en 
where 
\eq 
\lambda(\mu) = \frac{\mu^{d-4}}{(4\pi)^2} \, 
\biggl\{ \frac{1}{d-4} \, - \, \frac{1}{2} ({\rm ln}4\pi \, + \, 
\Gamma^\prime(1) \, + \, 1 ) \biggr\}\,, \hspace*{.5cm} 
\bar\lambda = \lambda(\mnz)\,.  
\en 
Identity~(\ref{d_6}) represents the matching condition between the LEC 
of the ChPT Lagrangian and our quark-level Lagrangian. Analogous conditions 
involving other LECs can be derived when matching other physical 
amplitudes/quantities (see, e.g. Sec.~\ref{nucleon_mass}).   

Applying SU(6)-symmetry relations of the naive nonrelativistic quark 
model for the ratios of the magnetic moments and the tensor charges of 
the nucleon we can derive additional and well-known relations between 
$F_2^{Ni}(0)$ and $ G_2^{Ni}(0)$, respectively:
\eq\label{SU6_constr1} 
\frac{2+F_2^{pu}(0)}{1+F_2^{pd}(0)} \, = \, 
\frac{2+F_2^{nd}(0)}{1+F_2^{nu}(0)} \, = \,  
\frac{G_2^{pu}(0)}{G_2^{pd}(0)} \, = \, 
\frac{G_2^{nd}(0)}{G_2^{nu}(0)} \, = \, - \, 4 \,.
\en 
Substituting Eq.~(\ref{SU6_constr1}) into Eq.~(\ref{chiral_constr2}), 
we arrive at 
\eq\label{SU6_constr2} 
& & F_2^{pu}(0) = F_2^{nd}(0) \, = \,  
\frac{4}{5} 
\biggl(\frac{g_A}{g}\biggr)^2 \, \frac{m_N}{\bar m}  
\, - \, 2\,, \nonumber\\
& & F_2^{pd}(0) = F_2^{nu}(0) \, = \, 
- \frac{1}{5} 
\biggl(\frac{g_A}{g}\biggr)^2 \, \frac{m_N}{\bar m}  
\, - \, 1\,, \nonumber\\
& & G_2^{pu}(0) = G_2^{nd}(0) \, = \,  
\frac{4}{5} 
\biggl(\frac{g_A}{g}\biggr)^2 \, \frac{m_N}{\bar m}  
\,, \nonumber\\
& & G_2^{pd}(0) = G_2^{nu}(0) \, = \, - \ 
\frac{1}{5} 
\biggl(\frac{g_A}{g}\biggr)^2 \, \frac{m_N}{\bar m}  \, 
\,. 
\en 
Other interesting results are the predictions for the bare values 
of nucleon magnetic moments and tensor charges: 
\eq\label{SU6_constr3} 
\mu_p^{\rm bare} \, \equiv \, - \, \frac{3}{2} \, \mu_n^{\rm bare} 
\, = \, \sum\limits_{q = u, d} \, e_q \, [  \, F_1^{pq}(0) \, 
+  \, F_2^{pq}(0) \, ] \, = \, 
\frac{3}{5} \biggl(\frac{g_A}{g}\biggr)^2\, \frac{m_N}{\bar m}  
\en 
and 
\eq\label{SU6_constr3_tch} 
\delta_{pu}^{\rm bare}  \, \equiv \, - \, 4 \, \delta_{pd}^{\rm bare} 
\, = \, \frac{\bar m}{m_N} \, G_2^{pu}(0) \, = \, 
\frac{4}{5} \, \biggl(\frac{g_A}{g}\biggr)^2 \, , 
\en 
where $e_u = 2/3$ and $e_d= -1/3$ are the electric charges of $u$ and 
$d$ quarks. Below, in Sec.~\ref{nucleon_mass}, we also derive the 
constraint relating the axial charge of the nucleon $g_A$ and of the
constituent quark $g$.

\section{Nucleon mass and meson-nucleon $\sigma$-terms} 
\label{nucleon_mass}

In this chapter we consider two other important quantities
of low-energy nucleon 
physics: the nucleon mass and meson-nucleon $\sigma$-terms which are 
constrained by the Feynman-Hellmann theorem (FHT)~\cite{Feynman:1939pr}. 
In particular, the FHT relates the derivative of the nucleon mass with respect 
to the current quark masses to the pion-nucleon sigma-term 
$\sigma_{\pi N}$ and to the strange quark condensate in the nucleon: 
\eq\label{FHTh}
\sigma_{\pi N} \, \bar u_N(p) u_N(p) &\doteq&  
\hat m \la N(p)| \bar u(0) u(0) + \bar d(0) d(0) |N(p) \ra = 
\hat m \frac{\partial m_N}{\partial \hat m} \, \bar u_N(p) \, u_N(p) \,, \\ 
y_s \, \bar u_N(p) u_N(p) &\doteq& 
\la N(p)| \bar s(0) s(0) |N(p) \ra = 
\frac{\partial m_N}{\partial \hat m_s} \, \bar u_N(p) \, u_N(p)  \,. 
\nonumber 
\en 
In quantum field theory the nucleon mass $m_N$ is defined as the matrix 
element of the trace of the energy-momentum tensor $\Theta^{\mu\nu}(x)$:
\eq\label{QFT_mass} 
m_N \, \bar u_N(p) u_N(p) \doteq \la N(p) |\Theta_\mu^\mu(0) |N(p) \ra\,. 
\en 
When we restrict to one-body interactions between the 
quarks and neglect to the contribution of the heavy quarks 
in the constituent quark (CQ) approach the master 
formula~(\ref{QFT_mass}) spells as  
\eq\label{CQM_mass} 
m_N \, \bar u_N(p) u_N(p) \doteq \la N(p) | {\cal H}_{\rm mass}(0) 
|N(p) \ra\,. 
\en 
${\cal H}_{\rm mass}(x) = \bar q(x) \, m_q \, q(x)$ 
is the part of the Hamiltonian referred to as the quark mass term where 
$m_q = {\rm diag}\{m_u, m_d, m_s\}$ 
is the matrix of constituent quark masses with $m_u = m_d = \bar m$ due 
to isospin invariance. 
Note, Eq.(\ref{CQM_mass}) is valid in the one-body approximation. 
In general, one should include in the trace of the energy-momentum 
tensor also two- and three-body quark operators 
which encode quark-quark interaction contributions to the nucleon mass.  
In some CQ models the nontrivial dependence of $m_q$ (or $m_N$) on the 
current quark masses is missing. This leads to a contradiction with the 
low-energy behavior of the nucleon mass as a function of $\hat m (\hat m_s)$ 
and with the Feynman-Hellmann theorem. The use of the effective 
Lagrangian~(\ref{L_qU}) constrained by Baryon ChPT enables one to 
perform an accurate and consistent calculation of the nucleon mass and 
the corresponding sigma-terms. Again, as in the case of the electromagnetic 
form factors, the extension to other baryons is straightforward.  

In analogy with the electromagnetic form factors we define the nucleon 
mass and later on the sigma-terms as expectation values of the 
dressed operators. First, we write down the bare quark mass term:
\eq 
{\cal H}_{\rm mass}^{\rm bare}(x) =  m \, \bar q(x) \, q(x) 
\en 
i.e. the quark mass term at leading order of the chiral expansion 
(in the chiral limit). Here $m$ is the value of the constituent quark 
mass in the chiral limit introduced before in the Lagrangian~(\ref{L_qU}). 
The nucleon mass in the chiral limit $\mnz$ is defined by 
\eq 
\mnz \, \bar u_N(p) \, u_N(p) 
= \la N(p) | {\cal H}_{\rm mass}^{\rm bare}(0) |N(p) \ra 
=  m \la N(p) | \bar q(0) \, q(0) |N(p) \ra\,. 
\en 
The dressed quark mass term and the physical nucleon mass are given by  
\eq
{\cal H}_{\rm mass}^{\rm dress}(x) &=& \bar q(x) \, m_q \, q(x)\,, \\
m_N \,\bar u_N(p) \, u_N(p) &=& 
\la N(p) | {\cal H}_{\rm mass}^{\rm dress}(0) |N(p) \ra\, = 
\la N(p) | \bar q(0) \, m_q \, q(0) |N(p) \ra \nonumber 
\en 
where $m_q = {\rm diag}\{m_u, m_d, m_s\}$ is the matrix of the 
dressed (physical) constituent quark masses  (in our case the constituent 
quark mass at one loop with inclusion of chiral corrections) with 
$m_u = m_d$ due to isospin invariance.  
The constituent quark masses $m_q \doteq m_q(\hat m, \hat m_s)$ have 
a nontrivial dependence on the current quark masses $\hat m$ and $\hat m_s$ 
which can be accurately calculated with the use of the chiral 
Lagrangian~(\ref{L_qU}). For illustration we discuss the explicit 
expressions for the nonstrange and strange constituent quark masses 
at one loop and at $O(p^4)$ with  
\eq 
m_q = m + \Sigma_q(m) \,. 
\en 
The quark mass operator $\Sigma_q = {\rm diag}\{\Sigma_u, \Sigma_d, 
\Sigma_s\}$,  with  $\Sigma_u = \Sigma_d = \bar \Sigma$ 
due to isospin invariance, is evaluated on the mass-shell 
$\not\! p = m$, which is ultraviolet-finite 
by construction. All ultraviolet divergencies are removed via 
the renormalization of the fourth-order LECs $e_1$, $e_3$, $e_4$ and $e_5$ 
contributing to the $\Sigma_q$ operator. Here and in the following 
we identify the quark mass occurring in the loop integrals with its 
leading order value $m_q \to m$. The operator $\Sigma_q$ is described 
by the diagrams in Fig.4 and after expansion in powers of meson masses 
is given by: 
\eq\label{Sigma}  
\bar \Sigma &=& \hat m - \frac{3 g^2}{32 \pi} \, 
\biggl\{ \frac{M_\pi^3}{F_\pi^2} + \frac{2}{3} \frac{M_K^3}{F_K^2} 
+ \frac{M_\eta^3}{9 \, F_\eta^2} \biggr\} 
- \frac{3 g^2}{64 \pi^2 m} 
\biggl\{ \frac{M_\pi^4}{F_\pi^2} + \frac{2}{3} \frac{M_K^4}{F_K^2} 
+ \frac{M_\eta^4}{9 \, F_\eta^2} \biggr\} \\
&-& 4 c_1 M^2 + \frac{3c_2}{128 \pi^2}  
\biggl\{ \frac{M_\pi^4}{F_\pi^2} + \frac{4}{3} \frac{M_K^4}{F_K^2}  
+ \frac{M_\eta^4}{3 \, F_\eta^2} \biggr\}  
+ \frac{4}{3} c_5 (M_K^2 - M_\pi^2) 
\nonumber \\ 
&+& \bar e_1 M^4 + \frac{\bar e_3}{6} (M_K^2 - M_\pi^2)^2 
- \frac{\bar e_4}{3} M^2 (M_K^2 - M_\pi^2) 
+ \frac{\bar e_5}{36} (M_K^2 - M_\pi^2)^2 
\nonumber 
\en  
and 
\eq\label{Sigmas} 
\Sigma_s &=& \hat m_s - \frac{3 g^2}{32 \pi} \, 
\biggl\{ \frac{4}{3} \frac{M_K^3}{F_K^2} 
+ \frac{4}{9} \frac{M_\eta^3}{F_\eta^2} \biggr\} 
- \frac{3 g^2}{64 \pi^2 m} 
\biggl\{ \frac{4}{3} \frac{M_K^4}{F_K^2} 
+ \frac{4}{9} \frac{M_\eta^4}{F_\eta^2} \biggr\} \\
&-& 4 c_1 M^2 + \frac{3c_2}{128 \pi^2}  
\biggl\{ \frac{M_\pi^4}{F_\pi^2} + \frac{4}{3} \frac{M_K^4}{F_K^2}  
+ \frac{M_\eta^4}{3 \, F_\eta^2} \biggr\}  
- \frac{8}{3} c_5 (M_K^2 - M_\pi^2)\nonumber \\ 
&+& \bar e_1 M^4 + \frac{\bar e_3}{6} (M_K^2 - M_\pi^2)^2 
+ \frac{2 \bar e_4}{3} M^2 (M_K^2 - M_\pi^2)
+ \frac{\bar e_5}{9} (M_K^2 - M_\pi^2)^2 
\nonumber 
\en  
where 
\eq 
M^2 = M_\pi^2 -  M_K^2 + \frac{3}{2} M_\eta^2  
    = \frac{1}{2} M_\pi^2 + M_K^2    \,. 
\en 
Note, $\bar \Sigma$ and $\Sigma_s$ degenerate for $\hat m = \hat m_s$ and 
$M^2$ coincides with $M_\pi^2$ at $\hat m_s =0$. The contribution of the 
diagram in Fig.4(5) (the so-called mass-insertion term) is generated by 
the terms from second-order Lagrangian ${\cal L}^{(2)}_q$ and is absent 
in final results (see Eqs.(\ref{Sigma}) and (\ref{Sigmas})) after 
certain replacement the constituent quark mass in the free Lagrangian 
(see details in~\cite{Becher:1999he}). 
For simplicity the chiral logarithms are hidden in the renormalized 
LECs $\bar e_i$ with $i = 1,3,4,5$: 
\eq\label{LECs_e1e5} 
\bar e_i &=& e_i^r(\mu) - 
\frac{\beta_{e_i}}{32\,\pi^2\,F_\pi^2} {\rm\ln}{\frac{M_\pi^2}{\mu^2}}
\, = \, e_i - \frac{\beta_{e_i}}{F_\pi^2} \, \lambda_\pi 
\en 
where 
\eq 
\lambda_\pi = \frac{M_\pi^{d - 4}}{(4\pi)^2} \, 
\bigg\{ \frac{1}{d - 4} - \frac{1}{2} 
(\ln 4\pi + \Gamma^{\,\prime}(1) + 1 ) \biggr\}.
\en 
The LECs $e_i$ contain the poles at $d=4$ which are 
cancelled by the divergencies proportional to $\lambda_\pi$ in the r.h.s. of 
Eq.~(\ref{LECs_e1e5}). Therefore, the renormalized couplings 
$e_i^r(\mu)$ (or $\bar e_i$) are finite. 
The set of the $\beta_{e_i}$ coefficients is given by 
\eq\label{beta_LECs_e1e5}  
& &\beta_{e_1} = \frac{32g^2}{27m}  
- \frac{8}{9} (8c_1 - c_2 - 4 c_3) \,, \quad\quad 
\beta_{e_3} = \frac{52g^2}{9m} - \frac{10}{3} (8 c_1 - c_2 - 4 c_3)  
+ \frac{16}{9} c_5 \, \nonumber\\
& &\beta_{e_4} = \frac{20g^2}{9m} - \frac{32}{3} c_5 \,, 
\quad\quad \hspace*{1.95cm}
\beta_{e_5} = - \frac{4g^2}{m} - \frac{16}{3} c_5 \,. 
\en 
Note, that the $m_s - \bar m$ 
splitting is mainly generated by the difference in the values of the 
strange and the nonstrange current quark masses. 
The chiral symmetry constraints and the 
matching of the nucleon mass calculated within our approach to the 
model-independent derivation of~\cite{Becher:1999he} allow to
deduce certain 
relations between the set of our parameters and the ones in Baryon ChPT. 
In particular, the coefficient connected with $M_\pi^3/F_\pi^2$, 
referred to in 
the literature as the {\it leading nonanalytic coefficient} 
(LNAC)~\cite{Gasser:1983yg,Becher:1999he}, is model-independent and
constant due to dimensional arguments.
Precisely, this coefficient is equal to $- 3 g_A^2/(32\pi F_\pi^2)$. 
The cubic term in powers of the meson mass shows up due to 
the infrared singularity of the diagram in Fig.4(3). 
Using this requirement, 
we can relate the value of axial charge of the constituent 
quark in the chiral limit to the corresponding nucleon quantity: 
\eq\label{ga_matchning}  
g_A^2 \, \bar u_N(p) \, u_N(p) = g^2\la N(p)|\bar q(0) \, q(0)|N(p) \ra \,. 
\en 
The matching condition gives a constraint for the matrix element of the 
bare scalar-density operator in the nucleon. A rough estimate with 
$g_A = 1.25$~\cite{Gasser:1987rb}, $g \sim 1$ and by taking into account 
the normalization of the nucleon spinors gives a quite reasonable value 
for the scalar condensate: 
$\la N(p) | \bar q(0) \, q(0) |N(p) \ra \sim (25/8) \, m_N \,.$ 
Using the matching condition (\ref{ga_matchning}) we derive a final expression 
for the nucleon mass at one loop:
\eq\label{nucleon_mass_1loop_a}  
m_N \, = \, \mnz \, + \, \Sigma_N 
\en
where 
\eq\label{nucleon_mass_1loop_b}  
m_N \, = \, \biggl(\frac{g_A}{g}\biggr)^2 \, \bar m \,, 
\hspace*{1cm} 
\mnz \, = \, \biggl(\frac{g_A}{g}\biggr)^2 \, m 
\en 
and 
\eq\label{nucleon_mass_1loop_c}  
\Sigma_N &=& - \frac{3 g_A^2}{32 \pi} \, 
\biggl\{ \frac{M_\pi^3}{F_\pi^2} + \frac{2}{3} \frac{M_K^3}{F_K^2} 
+ \frac{M_\eta^3}{9 \, F_\eta^2} \biggr\} 
- \frac{3 g_A^2}{64 \pi^2 m} 
\biggl\{ \frac{M_\pi^4}{F_\pi^2} + \frac{2}{3} \frac{M_K^4}{F_K^2} 
+ \frac{M_\eta^4}{9 \, F_\eta^2} \biggr\} \\
&+& \biggl(\frac{g_A}{g}\biggr)^2 \, 
\biggl[ \hat{m} 
- 4 c_1 M^2 + \frac{3c_2}{128 \pi^2}  
\biggl\{ \frac{M_\pi^4}{F_\pi^2} + \frac{4}{3} \frac{M_K^4}{F_K^2}  
+ \frac{M_\eta^4}{3 \, F_\eta^2} \biggr\}  
+ \frac{4}{3} c_5 (M_K^2 - M_\pi^2) 
\nonumber \\ 
&+& \bar e_1 M^4 + \frac{\bar e_3}{6} (M_K^2 - M_\pi^2)^2 
- \frac{\bar e_4}{3} M^2 (M_K^2 - M_\pi^2) 
+ \frac{\bar e_5}{36} (M_K^2 - M_\pi^2)^2 \, \biggr]\,. 
\nonumber 
\en  
The constituent quark mass can be removed from the expressions for nucleon 
observables using the matching conditions~(\ref{nucleon_mass_1loop_b}). 
The same is true also for other baryonic observables and the constituent 
quark can be removed from the corresponding expressions.  
In the SU(2) picture (when neglecting the kaon and $\eta$-meson loops  
and putting $c_5$ and $e_5$ to be equal to zero) we reproduce 
the result of ChPT for the nucleon mass at one loop and at order 
$O(p^4)$~\cite{Becher:1999he}
\eq\label{mN_1loop} 
m_N &=& \mnz - 4 c_1^{\rm ChPT} - \frac{3 g_A^2}{32 \pi} \, 
\frac{M_\pi^3}{F_\pi^2} + k_1 M_\pi^4 
{\rm\ln}\frac{M_\pi}{\mnz} + k_2 M_\pi^4 + O(M_\pi^5)\,,\\
k_1 &=& - \frac{3}{32 \pi^2 F_\pi^2} \, \biggl(\frac{g_A^2}{\mnz} 
- 8 c_1^{\rm ChPT} + c_2^{\rm ChPT} + 4 c_3^{\rm ChPT} \biggr)\,, 
\nonumber\\
k_2 &=& \bar e_1^{\rm ChPT} - \frac{3}{128 \pi^2 F_\pi^2} \, 
\biggl(\frac{2 g_A^2}{\mnz} -  c_2^{\rm ChPT}  \biggr)\,, 
\nonumber\\ 
\bar e_1^{\rm ChPT} &=& e_1^{\rm ChPT} - 
\frac{3 \bar\lambda}{2 F_\pi^2} \biggl( 
\frac{g_A^2}{\mnz} 
- 8 c_1^{\rm ChPT} + c_2^{\rm ChPT} + 4 c_3^{\rm ChPT} \biggr) \, ,
\nonumber 
\en 
if we fulfill the following matching conditions between the LECs of 
the ChPT Lagrangian and our quark-level Lagrangian:
\eq\label{matching_LECs}
& &- 4 c_1^{\rm ChPT} M_\pi^2 = 
\biggl[ \hat m - 4 c_1 M_\pi^2 \biggr] 
\biggl(\frac{g_A}{g}\biggr)^2 \,,
\nonumber\\
& &8 c_1^{\rm ChPT}  - c_2^{\rm ChPT}  - 4 c_3^{\rm ChPT} 
- \frac{g_A^2}{\mnz} \equiv 
\biggl[ 8 c_1 - c_2 - 4 c_3 - \frac{g_A^2}{\mnz} \biggr] 
\biggl(\frac{g_A}{g}\biggr)^2 \,,\\ 
& &\bar e_1^{\rm ChPT} - \frac{3}{64 \, \pi^2 \, F_\pi^2} 
\biggl( \frac{2 g_A^2}{\mnz}  -  c_2^{\rm ChPT} \biggr) 
\equiv 
\biggl[ 
\bar e_1^\ast - \frac{3}{64 \, \pi^2 \, F_\pi^2} \, 
\biggl( \frac{2 g_A^2}{\mnz} -  c_2\biggr) \, \biggr]
\biggl(\frac{g_A}{g}\biggr)^2 \,,  \nonumber 
\en 
where $\bar e_1^\ast$ is a $SU(2)$ analogue of $\bar e_1$ 
derived in Eqs.~(\ref{LECs_e1e5}) and (\ref{beta_LECs_e1e5}) in 
the three-flavor picture: 
\eq 
\bar e_1^\ast 
\, = \, e_1^\ast - \frac{\beta_{e_1^\ast}}{F_\pi^2} \, \lambda_\pi\,, 
\quad\quad
\beta_{e_1^\ast} = \frac{3}{2} 
\biggl( \frac{g^2}{m} - 8 c_1 + c_2 + 4 c_3 \biggr) \,. 
\en 
Note, that there is an additional condition on $g$ and $g_A$ 
which shows up in the calculation of the axial nucleon charge:
\eq\label{axial_charge}
g_A \, \bar u_N(p) \, \gamma^\mu \, \gamma^5 \, \tau_3 \, u_N(p) 
= g \la N(p)|\bar q(0) \,  \gamma^\mu \, \gamma^5 \, \tau_3 \, 
q(0)|N(p) \ra \,. 
\en 
Eq.~(\ref{axial_charge}) gives a constraint on the matrix element of 
the isovector axial current. 

As an example for the sigma-terms we consider  
the pion-nucleon sigma-term  $\sigma_{\pi N}$. In QCD (see
Eq.~(\ref{FHTh})) this quantity is related to the expectation value 
of the scalar density operator. It is connected to the derivative of 
the part of the QCD Hamiltonian, explicitly breaking chiral symmetry, 
with respect 
to the current quark mass. This definition is consistent with the FH 
theorem~\cite{Feynman:1939pr}. In the context of CQ models, we should 
proceed with the dressed Hamiltonian 
${\cal H}_{\rm mass}^{\rm dress}(x)$ which already showed up in the 
calculation of the physical nucleon mass. In particular, the dressed scalar 
density operators $j_i^{\rm dress}(x)$ (where $i=u, d, s$ is the flavor 
index), relevant for the calculation of the meson-baryon sigma-terms within 
our approach, are defined as the  partial derivatives 
of ${\cal H}_{\rm mass}^{\rm dress}(x)$ with respect to the current quark 
mass $\hat m_i$ of $i$-th flavor: 
\eq\label{j_i}
j_i^{\rm dress}(x) \doteq 
\frac{\partial{\cal H}_{\rm mass}^{\rm dress}(x)}{\partial\hat m_i} 
\, = \, \bar q(x) \, \frac{\partial m_q}{\partial\hat m_i} \, q(x) 
\, = \, \bar q(x) \, \frac{\partial \Sigma_q}{\partial\hat m_i} \, q(x) \,. 
\en 
In the case of $\sigma_{\pi N}$ we have: 
\eq\label{sigma_piN} 
\sigma_{\pi N} \, \bar u_N(p) \, u_N(p) \doteq   
\hat m \, \la N(p)|  \, j_u^{\rm dress}(0) \, + \, 
j_d^{\rm dress}(0) \, |N(p) \ra \,. 
\en 
It should be clear that Eq.~(\ref{sigma_piN}) is consistent with the 
FH theorem: 
\eq 
\sigma_{\pi N} \, \bar u_N(p) \, u_N(p)  \, 
= \, \hat m \, \frac{\partial}{\partial\hat m} \,  
\underbrace{\la N(p)| \, {\cal H}_{\rm mass}^{\rm dress}(0) \, |N(p) 
\ra}_{= \, m_N \, \bar u_N(p) \, u_N(p)}    
 \, = \, \hat m \, \frac{\partial m_N}{\partial\hat m} \, \bar u_N(p) 
\, u_N(p)\,. 
\en 
Below we give the definitions of the strangeness content of the 
nucleon $y_N$, the kaon-nucleon $\sigma_{KN}$ and the eta-nucleon 
$\sigma_{\eta N}$ sigma-terms: 
\eq 
y_N  &=& 2 \, \frac{\partial m_N/\partial \hat m_s} 
{\partial m_N/\partial \hat m}\,, \nonumber\\
\bar u_N(p)  \, u_N(p) \, 
\sigma_{KN}^{u(d)} &=& \frac{\hat m + \hat m_s}{2} 
\la N(p)| \, j_{u(d)s; +}^{\rm dress}(0) \, | N(p) \ra\,, 
\nonumber\\[2mm]
\bar u_N(p)  \, u_N(p) \, 
\sigma_{KN}^{I=0} &=& \frac{\hat m + \hat m_s}{4} 
\la N(p)|  \, j_{us; +}^{\rm dress}(0) \, + 
\, j_{ds; +}^{\rm dress}(0) \, | N(p) \ra\,, 
\nonumber \\[2mm] 
\bar u_N(p)  \, u_N(p) \, 
\sigma_{KN}^{I=1} &=& \frac{\hat m + \hat m_s}{2} 
\la N(p)| \, j_{ud; -}^{\rm dress}(0) \, | N(p) \ra\,, 
\nonumber \\[2mm]
\bar u_N(p)  \, u_N(p) \, 
\sigma_{\eta N} &=& \frac{1}{3} 
\la N(p)| \, \hat m \, j_{ud; +}^{\rm dress}(0)  
\, + \, 4 \, \hat m_s \, j_s^{\rm dress}(0) \, | N(p) \ra\,, 
\en 
where $j_{qq^\prime; \pm}^{\rm dress} = 
j_{q}^{\rm dress} \pm  j_{q^\prime}^{\rm dress}$ and 
$|N(p)\ra$ denotes a proton state. 
All further details can be found in our previous papers~\cite{PCQM2} 
and in Appendix D. 

\section{Physical applications} 

In this chapter we consider the application of our approach to specific  
nucleon properties: magnetic moments, electromagnetic radii and form factors, 
nucleon mass, meson-nucleon sigma-terms and strong vector meson-nucleon 
form factors. 
We also calculate some canonical properties of hyperons: masses and magnetic 
moments. Note, that these properties have been studied within different 
approaches: 
QCD sum rules~\cite{Ioffe:1981kw}, ChPT and HBChPT~\cite{Gasser:1987rb,Jenkins:1990jv,Becher:1999he,Kubis:2000zd,Kubis:2000aa,Fuchs:2003qc,Gasser:1980sb,Borasoy:1996bx,Fuchs:2003ir,Frink:2005ru}, lattice QCD~\cite{Gusken:1988yi}, 
approaches based on vector dominance~\cite{Gari:1984rr} 
and on solutions of Schwinger-Dyson and 
Bethe-Salpeter equations~\cite{Hellstern:1997pg}, different types of chiral 
and quark models~\cite{Efimov:1993ei,Ivanov:1996pz,PCQM1,Theberge:1980ye,Oset:1984tv,Diakonov:1983hh,Chodos:1974je,Alberto:1988xj}, etc. 
Our main interest will be restricted to the region of small external momenta
where the contribution of the meson cloud is supposed to be largest.

\subsection{Masses of the baryon octet and the meson-nucleon sigma-terms} 

The nucleon mass and the meson-nucleon sigma-terms at one loop depend 
on the set 
of parameters: $g$, $m$, $c_1$, $c_2$, $c_5$, $\bar e_1$, $\bar e_3$,  
$\bar e_4$ and $\bar e_5$. 
First, we discuss a choice for $g$ and $m$. Note, that these parameters 
are constrained in our approach by the matching 
condition (\ref{nucleon_mass_1loop_b}): 
$\mnz \, = \, (g_A/g)^2 \, m$.  
In literature the value of the axial charge of the constituent quark varies
approximately from $0.9$ to $1$ (see detailed discussion in 
Refs.~\cite{Weinberg:1991gf}). 
The nucleon mass in the chiral limit was estimated in HBChPT: 
$\mnz = 770 \pm 110$ MeV~\cite{Borasoy:1996bx} and 
$\mnz = 890 \pm 180$ MeV~\cite{Frink:2005ru}. 
Inserting this range of values for $g$ 
and $\mnz$ and also the experimental value for 
$g_A = 1.267$~\cite{Eidelman:2004wy} into Eq.~(\ref{nucleon_mass_1loop_b}) 
gives the following limits for the constituent quark mass in the chiral limit:
\eq 
m  \, \simeq  \, 500 \, \pm \, 167 \  {\rm MeV}\,. 
\en 
Finally, we choose: 
\eq 
m = 420 \ {\rm MeV}\, \hspace*{.25cm} \text{and} \hspace*{.25cm} g = 0.9\,.  
\en 
Using the experimental value for the nucleon axial charge 
$g_A = 1.267$~\cite{Eidelman:2004wy} and Eq.~(\ref{nucleon_mass_1loop_b}) 
we get $\mnz = 832.4$ MeV which is rather close to our 
previous estimate done in the framework of the perturbative chiral quark
model (PCQM): 
$\mnz = 828.5$ MeV~\cite{PCQM2}. 
Here we do not pretend on a more accurate determination of 
$m$ and $g$. This will be done in a possible forthcoming paper where we intend 
to involve more constraints from data. However, we stress that 
we need a rather small value for $g$ in the interval $0.9 - 1$ 
to justify perturbation theory. In particular, the use of
$g=0.9$ gives the shift of the nonstrange constituent quark mass 
equal to $\Delta m = 53$ MeV. For $g=0.95$ and $g=1$ we get 
$\Delta m = 107$ MeV  and $\Delta m = 164$ MeV, respectively.   
On the other hand, the choice of $g$ is constrained by the bare magnetic 
moments of the nucleon (see Eq.~(\ref{SU6_constr3_2})). 
The use of $g=0.9$ gives a reasonable contribution of the valence quark to the 
nucleon magnetic moments (see discussion in the next section).
Finally, the couplings $c_1$, $c_2$, $c_5$ and $\bar e_1$ are fixed 
by four conditions: 
\eq\label{fit_c1_c2_c5_e1} 
m_N = 938.27 \ {\rm MeV}\,, \hspace*{.5cm} 
\sigma_{\pi N} = 45 \ {\rm MeV}\,, \hspace*{.5cm} 
y_N = 0.2\,, \hspace*{.5cm} 
m_s - \bar m \simeq 170 \ {\rm MeV}\,. 
\en 
The result of the fit is: 
\eq\label{e_1}
c_1 &=& (-0.317 - 0.004 \,\tilde e_3 
                 + 0.020 \,\tilde e_4 
                 + 0.001 \,\tilde e_5) \,\,\,{\rm GeV}^{-1}\,, 
\nonumber\\ 
c_2 &=& (1.093  - 1.354 \,\tilde e_3 
                 + 0.474 \,\tilde e_4 
                 - 0.338 \,\tilde e_5) \,\,\,{\rm GeV}^{-1}\,, 
\nonumber\\ 
c_5 &=& (-0.316 + 0.063 \,\tilde e_3 
                 + 0.005 \,\tilde e_5) \,\,\,{\rm GeV}^{-1}\,, 
\nonumber\\ 
\bar e_1 &=& (1.157 + 0.298 \,\tilde e_3
                    + 0.162 \,\tilde e_4 
                    + 0.088 \,\tilde e_5) \,\,\,{\rm GeV}^{-3}\,, 
\en   
where for convenience we introduce the dimensionless parameters 
$\tilde e_i = \bar e_i \times 1$ GeV$^3$. 
Note, the mass difference $m_s - m_{u(d)}$ is a crucial quantity to
roughly describe the
splittings between the octet baryon masses. 
Latter quantities are proportional 
to $m_s - \bar m$. Neglecting isospin-breaking and hyperfine-splitting 
effects we have in our approach:
\eq 
& &m_\Lambda - m_N = m_s - \bar m \simeq 170 \ 
{\rm MeV} \hspace*{1cm} {\rm (data: \ 177 \ MeV)}\,, \nonumber\\
& &m_\Sigma - m_N = m_s - \bar m \simeq 170 \ 
{\rm MeV} \hspace*{1cm} {\rm (data: \ 251 \ MeV)}\,, \\
& &m_\Xi - m_N = 2 [ m_s - \bar m ] \simeq 340 \ {\rm MeV} 
\hspace*{.6cm}  
{\rm (data: \ 383 \ MeV)}\,. \nonumber 
\en 
The meson cloud corrections are important to reproduce 
the full empirical value of $\sigma_{\pi N}$, 
contributing about 2/3 to the total value 
(for a detailed discussion see  Ref.~\cite{PCQM2}). 
In particular, our result for $\sigma_{\pi N}$  is compiled as: 
the total value is $\sigma_{\pi N}$ = 45 MeV, the contribution 
of the valence quarks is $\sigma_{\pi N}^{val}$ = 13.87 MeV 
(30\% of the total value), the contribution of the pion cloud 
is dominant 
\eq 
\sigma_{\pi N}^{\pi} = 26.69 \ {\rm MeV} \ 
+ 0.45 \ {\rm MeV} \, \tilde e_3  
- 2.58 \ {\rm MeV} \, \tilde e_4 
- 0.11 \ {\rm MeV} \, \tilde e_5 \,, 
\en 
the kaon and $\eta$-meson cloud generates
\eq 
\sigma_{\pi N}^{K + \eta} = 
  4.44 \ {\rm MeV} \ 
- 0.45 \ {\rm MeV} \, \tilde e_3  
+ 2.58 \ {\rm MeV} \, \tilde e_4 
+ 0.11 \ {\rm MeV} \, \tilde e_5 \,. 
\en 
The separate contributions 
$\sigma_{\pi N}^{\pi}$ and $\sigma_{\pi N}^{K + \eta}$ 
depend on the parameters $\tilde e_3$, $\tilde e_4$ and 
$\tilde e_5$, while the total 
contribution is independent on $\tilde e_i$ with $i=3,4,5$.   
Our predictions for the kaon-nucleon $\sigma_{KN}$ 
and eta-nucleon $\sigma_{\eta N}$ sigma-terms are: 
$\sigma_{KN}^{u} = 381.9$~MeV, $\sigma_{KN}^{d} = 351.8$~MeV, 
$\sigma_{KN}^{I=0} = 366.8$~MeV, $\sigma_{KN}^{I=1} = 15$~MeV,
and $\sigma_{\eta N} = 90$~MeV. 

\subsection{Magnetic moments of baryon octet} 

The magnetic moments of the nucleon can be written in terms of 
the Dirac and Pauli form factors as
\eq
\mu_N = F_1^N(0)+F_2^N(0),
\en
where in our formalism $F_1^N(0)$ and $F_2^N(0)$ are of the form
\eq
F_1^N(0) &=& \sum_{q=u,d} f_D^q(0) F_1^{Nq}(0) \,,\nonumber \\
F_2^N(0) &=& \sum_{q=u,d} [f_D^q(0) F_2^{Nq}(0)+f_P^q(0) G_2^{Nq}(0)]\,, 
\en 
where $f_D^q(0) \equiv e_q$ is the quark charge due to the 
charge conservation.  
It is convenient to separate the expressions for $F_1^N(0)$ and $F_2^N(0)$ 
into two contributions: the bare (valence-quark) and the meson cloud 
(sea-quark) contributions. 

The bare contribution is
\eq
F_1^{N \, {\rm bare}}(0) &=& \sum_{q=u,d} e_q \, F_1^{Nq}(0)\,, \nonumber \\
F_2^{N \, {\rm bare}}(0) &=& \sum_{q=u,d} e_q \, F_2^{Nq}(0)\,. 
\en 
The meson cloud gives rise to
\eq
F_1^{N \, {\rm cloud}}(0) &\equiv& 0\,, \nonumber \\
F_2^{N \, {\rm cloud}}(0) &=& \sum_{q=u,d} f_P^i(0) G_2^{Ni}(0)\,.
\en 
Note, that the contribution of the meson cloud to the Dirac form factor 
$F_{1}^N(0)$ is zero due to the charge conservation of the nucleon. 
Therefore, the nucleon magnetic moments are given in additive form with
\eq
\mu_N = \mu_N^{\rm bare}+\mu_N^{\rm cloud}.
\en
where 
\eq 
\mu_N^{\rm bare} = F_1^{N \, {\rm bare}}(0) + 
F_2^{N \, {\rm bare}}(0) 
\en
and 
\eq 
\mu_N^{\rm cloud} = F_1^{N \, {\rm cloud}}(0) + 
F_2^{N \, {\rm cloud}}(0) \,.
\en 
With the constraints laid out in Sec.II,  
we derived a relation for the bare contributions to 
the nucleon magnetic moments [see Eq.~(\ref{SU6_constr3})]: 
\eq\label{SU6_constr3_1} 
\mu_p^{\rm bare} \, \equiv \, - \, \frac{3}{2} \, \mu_n^{\rm bare} \, = \, 
\frac{3}{5} \biggl(\frac{g_A}{g}\biggr)^2 \, \frac{m_N}{\bar m} \,. 
\en 
Note that Eq.~(\ref{SU6_constr3_1}) can be further simplified using 
the constraints of (\ref{nucleon_mass_1loop_b}): 
\eq\label{SU6_constr3_2} 
\mu_p^{\rm bare} \, \equiv \, - \, \frac{3}{2} \, \mu_n^{\rm bare} \, = \, 
\frac{3}{5} \, \biggl(\frac{g_A}{g}\biggr)^4 \,. 
\en  
As already mentioned in the previous section, the choice of $g$ is constrained 
by the bare magnetic moments of the nucleon (see Eq.~(\ref{SU6_constr3_2})). 
The value of $g=0.9$ results in a reasonable contribution of the 
valence quarks to the nucleon magnetic moments with: 
\eq\label{SU6_constr3_3} 
\mu_p^{\rm bare} \, \equiv \, - \, \frac{3}{2} \, \mu_n^{\rm bare} \, = \, 
\frac{3}{5} \, \biggl(\frac{g_A}{g}\biggr)^4  \simeq 2.357  
\en  
which is about 84\% (for the proton) and about 82\% (for the neutron) 
of the experimental (total) values, where the remainder comes from the meson
cloud.

The main parameters contained in the meson cloud piece
that play an important role in fitting the magnetic moments
are the second-order coupling $c_6$ and the fourth-order 
flavor-breaking LECs $e_7$, and $e_8$. The coupling $e_6$ 
is absorbed in $c_6$ after the certain redefinition. 
To constrain these values we need besides the experimental values
for the nucleon (p and n) an additional value from the baryon octet.
Therefore, we will proceed with an extension of our formalism to
calculate the magnetic moments of the whole baryon octet.

In the following, for convenience, we introduce the bare Dirac 
$(F_{1}^{q}, G_{1}^{q})$ and Pauli $(F_{2}^{q}, G_{2}^{q})$ 
form factors of the quark of flavor $q$: 
\eq 
\la q(p^\prime) | \, j_{\mu, q}^{\rm bare}(0) \, | q(p) \ra &=& 
\bar u_q(p^\prime) \biggl\{ \gamma_\mu \, F_1^{q}(q^2) 
\, + \, \frac{i}{2 \, m_q} 
\, \sigma_{\mu\nu} \, q^\nu \,  F_2^{q}(q^2) \biggr\} u_q(p)\,, \\ 
i \, \frac{q^\nu}{2 \, m_q}  
\la q(p^\prime) | \, j_{\mu\nu, q}^{\rm bare}(0) \, | q(p) \ra &=& 
\bar u_q(p^\prime) \biggl\{ \gamma_\mu \, G_1^{q}(q^2) 
\, + \, \frac{i}{2 \, m_q}  
\, \sigma_{\mu\nu} \, q^\nu \,  G_2^{q}(q^2) \biggr\} u_q(p) \, .
\nonumber 
\en 
Note that the valence quark form factors (or valence quark contributions)
are constrained by certain symmetries including the infrared singularities
(see discussion in Section~II.E). Introducing the valence quark form factors
we fulfill matching conditions between our approach and ChPT and,
therefore, the structures due to chiral symmetry are not violated. 

The Sachs form factors of the quark of flavor $q$ are:   
\eq 
{\cal F}_q^E(t) = F_q^E(t) + G_q^E(t)\,, \quad\quad 
{\cal F}_q^M(t) = F_q^M(t) + G_q^M(t)\,, 
\en 
where 
\eq 
F\{G\}_q^E(t)= F\{G\}_1^{q}(t) -  \frac{t}{4m^2} F\{G\}_2^{q}(t)
\,, \quad\quad 
F\{G\}_q^M(t)= F\{G\}_1^{q}(t) +  F\{G\}_2^{q}(t)\,, \quad\quad 
t=-q^2 \,,  
\en 
are the contributions to the Sachs form factors 
associated with the expectation values of the vector and tensor currents, 
respectively. 

By using SU(6)-symmetry relations one can relate the Dirac and Pauli 
form factors 
describing the distribution of quarks of flavor $q=u, d, s$ in 
the baryon $"B"$, that is $F_{1(2)}^{Bq}(t)$ and $G_{1(2)}^{Bq}(t)$, 
to $F_q^{E(M)}(t)$ and $G_q^{E(M)}(t)$  by
\eq\label{FBi}
F_1^{Bq}(t) \, &=& \,  \frac{1}{1 + \tau_B} \,  
\biggl\{ \alpha_E^{Bq} F_q^E(t) +  \alpha_M^{Bq}\,\chi^{Bq}\, 
F_i^M(t) \tau_B \biggr\}\,, \nonumber\\
F_2^{Bq}(t) \, &=& \, \frac{1}{1 + \tau_B} \, 
\biggl\{ - \alpha_E^{Bq} F_q^E(t) +  \alpha_M^{Bq}\,\chi^{Bq}\, 
F_q^M(t) \biggr\}\,,\nonumber\\
G_1^{Bq}(t) \, &=& \, \frac{1}{1 + \tau_B} \, 
\biggl\{ \alpha_E^{Bq} G_q^E(t) +  \alpha_M^{Bq}\,\chi^{Bq}\, 
G_q^M(t) \tau_B \biggr\}\,, \nonumber\\
G_2^{Bq}(t) \, &=& \, \frac{1}{1 + \tau_B} \, 
\biggl\{ - \alpha_E^{Bq} G_q^E(t) +  \alpha_M^{Bq}\,\chi^{Bq}\, 
G_q^M(t) \biggr\}\,, 
\en
where $\tau_B = t/(4m_B^2)$ and $m_B$ is the baryon mass.  
In addition to the strict evaluation of SU(6) we have 
introduced the additional parameter $\chi^{Bq}$ for each quark of flavor $q$. 
The interpretation for adding these factors is such that to allow
the quark distributions 
for hyperons to be different from that for the nucleons.
In the case of the nucleons we
set $\chi^{Bq}=1$. The values for $\alpha_E^{Bq}$ and $\alpha_M^{Bq}$
for the baryon octet 
as derived from SU(6)-symmetry relations are given in Table~1. 

The quark Sachs form factors 
are modeled by the dipole characteristics with damping functions of 
an exponential form. This phenomenological form is required
to reproduce in particular the deviation
of the electromagnetic form factors of the nucleon from the dipole fit
as evident from recent experimental measurements.
We use the parameterization
\eq\label{quarkff}
& &F_q^E(t) = \frac{\rho_q^E(t)}{[1 + t/\Lambda_{qE}^2]^2}\,, 
\hspace*{2.5cm}
F_q^M(t) = \mu_q^F \, \frac{\rho_q^M(t)}{[1 + t/\Lambda_{qM}^2]^2}\,, 
\nonumber\\
& &G_q^E(t) = \gamma_q \, \rho_q^E(t) \, 
\frac{t/\Lambda_{qE}^2}{[1 + t/\Lambda_{qE}^2]^3}\,, 
\hspace*{1cm}
G_q^M(t) = \mu_q^G \, \frac{\rho_q^M(t)}{[1 + t/\Lambda_{qM}^2]^2}\,,
\en 
where $\rho_q^E(t) = \exp(-t/\lambda_{qE}^2)$ and 
$\rho_q^M(t) = \exp(-t/\lambda_{qM}^2)$. 
The parameters $\mu_q^F$ and $\mu_q^G$ are fixed by the symmetry 
constraints [see Eqs.~(\ref{SU6_constr2}) and (\ref{SU6_constr3_1})]: 
\eq
\mu_q^F \, = \, \mu_q^G \, = \, \mu_p^{\rm bare} \,.
\en
The remaining parameters $\gamma_q,\,\Lambda_{qE(M)}$ and
$\lambda_{qE(M)}$ are to be fixed later when we consider the full momentum
dependence of the nucleon electromagnetic form factors.  
Note, that in Ref.~\cite{Kelly:2002if} a similar 
parametrization of the nucleon form factors has been considered.
In Ref.~\cite{Friedrich:2003iz} the 
damping functions $\rho(t)$ have been parametrized with
constant values.

The magnetic moment of the octet baryon $"B"$ can be written in complete
analogy to the nucleon case as
\eq
\mu_B \, = \, \mu_B^{\rm bare} \, + \, \mu_B^{\rm cloud} 
\en
where
\eq
\mu_B^{\rm bare} &=& \sum_{q=u,d,s} e_q\, 
\left( F_1^{Bq}(0)+F_2^{Bq}(0) \right) \,, \nonumber\\
\mu_B^{\rm cloud} &=& \sum_{q=u,d,s} f_P^q(0) G_2^{Bq}(0) \,. 
\en
At $t=0$, the parameters $c_6$, $\bar e_6$, $\bar e_7$, $\bar e_8$, 
$c_2$, and $k_V$ contained in $f_P^q(0)$ will contribute, 
where the renormalized LECs $\bar e_6$, $\bar e_7$, $\bar e_8$ 
are given in Appendix~C.  
The parameter $\bar e_6$ can be absorbed in the coupling $c_6$ 
via 
\eq 
c_6 \to \tilde c_6 = c_6 - 16 m M^2 \bar e_6\,. 
\en 
Finally, form factors $f_P^q(0)$ depend on the parameters $\tilde c_6$, 
$\bar e_7$, $\bar e_8$, $c_2$ and $k_V$. 
Three of them ($\tilde c_6$, $\bar e_7$, $\bar e_8$) 
can now be fixed using the experimental values for the magnetic 
moments of the nucleons and of a hyperon in the baryon octet. 
We choose the $\Lambda^0$-hyperon with 
$\mu_{\Lambda^0}=-0.613 \pm 0.004$ (in units 
of the nuclear magneton). 
In our analysis of the magnetic moments of the octet baryons
we present two cases. First, we restrict to the SU(6) case by setting
all values $\chi^{Bi}=1$. Secondly, we allow the $\chi^{Bi}$ to be additional 
parameters. From now on we will name these two cases shortly 
as "Set I" and "Set II", respectively. 

A variation in the value of the axial charge
$g$ of the constituent quark
will also give rise to different contributions of the bare and the meson 
cloud parts to the total values of the magnetic moments. 
However, the total magnetic moments are the same in any case. 
As was discussed in the previous section, for the axial charge 
we choose $g=0.9$. 
In case of Set I we then obtain for $c_6$, $\bar e_7$, and $\bar e_8$:
\eq
\tilde c_6 &=& 0.859  - 0.106 \, \tilde c_2 - 1.254 \, k_V \,, \nonumber\\
\bar   e_7 &=& ( - 0.656 + 0.073\, \tilde c_2  + 0.607 \, k_V ) 
\,\,\,{\rm GeV}^{-3}\,, \\ 
\bar   e_8 &=& ( 0.033 - 0.008\, \tilde c_2  - 0.141 \, k_V ) 
\,\,\,{\rm GeV}^{-3}\,, \nonumber 
\en
where $\tilde c_2 = c_2 (1+k_V)$. 
The resulting values for the magnetic 
moments of the baryon octet for this case (Set I) are shown in Table 2, where 
reasonable agreement with data is obtained. Meson cloud contributions to the 
total values of the magnetic moments are about $5-30\%$ depending 
on the baryon.

For the case of Set II we start from the SU(6) result with the additional 
breaking parameter $\chi^{\Lambda s}$ for $\mu_{\Lambda^0}^{\rm bare}$ 
which can be written in terms of $\mu_{p}^{\rm bare}$ as 
$\mu_{\Lambda^0}^{\rm bare}=-\mu_{p}^{\rm bare} \chi^{\Lambda s}/3$. 
Then we have  
\eq
\chi^{\Lambda s}=-3\frac{\mu_{\Lambda^0}^{\rm bare}}{\mu_{p}^{\rm bare}} 
\approx -3\frac{\mu_{\Lambda^0}}{\mu_{p}}, 
\en
where we further assume that the contributions from the meson cloud part 
to the total magnetic moments both of $\Lambda^0$ and of $p$ 
are of the same order. With the experimental values 
for $\mu_{\Lambda^0}$ and $\mu_p$ we get $\chi^{\Lambda s}=0.66$. 
By fitting $\mu_p$, $\mu_n$, and $\mu_{\Lambda^0}$, 
with $\chi^{\Lambda s}=0.66$ as an additional input, we obtain
\eq
\tilde c_6 &=& 0.834  - 0.106 \, \tilde c_2 - 1.254 \, k_V \,, \nonumber\\
\bar   e_7 &=& ( - 0.832 + 0.073\, \tilde c_2  + 0.607 \, k_V )
\,\,\,{\rm GeV}^{-3}\,, \\ 
\bar   e_8 &=& ( 0.051- 0.008\, \tilde c_2  - 0.141 \, k_V ) 
\,\,\,{\rm GeV}^{-3}\,, \nonumber 
\en
where the constants contained in these parameters are slightly different from 
Set I. With appropriate values for the remaining $\chi^{Bq}$, the central 
values of the experimental results of the other magnetic moments in 
the baryon octet can be fully reproduced. For the set of $\chi^{Bq}$
we get 
\eq
\chi^{\Sigma u} &=& \chi^{\Sigma d}=0.963,\,\,\,\,\,\,\chi^{\Sigma s}=0.259, 
\nonumber \\
\chi^{\Xi u} &=& \chi^{\Xi d}=0.633,\,\,\,\,\,\,\chi^{\Xi s}=0.694, 
\nonumber \\
\chi^{\Sigma\Lambda u} &=& \chi^{\Sigma\Lambda d} = 0.988.
\en
relying on isospin symmetry. 
The isotriplet $\Sigma^+$, $\Sigma^0$, and $\Sigma^-$ shares the same set of 
$\chi^{\Sigma q}$ for the quark of flavor $q$, while
$\Xi^0$ and $\Xi^-$ contains the same set of $\chi^{\Xi q}$. 
The parameters $\chi^{\Sigma\Lambda u}$ and $\chi^{\Sigma\Lambda d}$ 
are directly related to the $\Sigma-\Lambda$
magnetic transition moment. For completeness,
our corresponding results for the magnetic moments
are also listed in Table~2. In the following we will use the last set
of values for $\tilde c_6$, $\bar e_7$, 
and $\bar e_8$, when considering the electromagnetic nucleon form factors.

\subsection{Nucleon electromagnetic form factors} 

Using the parameters from the analysis of the magnetic moments and
the masses of the baryon octet, further supplied by the ones from
the meson-nucleon sigma-terms, we now can consider
the full nucleon electromagnetic form 
factors. We recall that the Dirac and Pauli form factors for the nucleons are
\eq
F_1^N(t) &=& \sum_{q=u,d} [f_D^q(t) F_1^{Nq}(t)+f_P^q(t) G_1^{Nq}(t)]\,, 
\nonumber \\
F_2^N(t) &=& \sum_{q=u,d} [f_D^q(t) F_2^{Nq}(t)+f_P^q(t) G_2^{Nq}(t)]\,, 
\en
where $F_{1(2)}^{Nq}(t)$ and $G_{1(2)}^{Nq}(t)$ refer to the bare
constituent quark structure, while $f_D^q(t)$ and $f_P^q(t)$ contain
the chiral dynamics due to dressing of the quark operators. 
The isoscalar $F_i^S(t)$ and isovector $F_i^S(t)$  
nucleon form factors are written as 
\eq 
F_i^S(t) &=& F_i^p(t) + F_i^n(t) =   
\sum_{q=u,d} [ \, f_D^q(t) ( F_i^{pq}(t) + F_i^{nq}(t) ) 
+f_P^q(t) ( G_i^{pq}(t) + G_i^{nq}(t) ) \, ] \, \\  
F_i^V(t) &=& F_i^p(t) - F_i^n(t) =   
\sum_{q=u,d} [ \, f_D^q(t) ( F_i^{pq}(t) - F_i^{nq}(t) ) 
+f_P^q(t) ( G_i^{pq}(t) - G_i^{nq}(t) ) \, ] \, \nonumber 
\en 
where $i=1,2$. 
For clarity we recall the forms of $F_{1(2)}^{Nq}(t)$ and
$G_{1(2)}^{Nq}(t)$ of Eq.~(\ref{FBi}) with:
\eq\label{FNi}
& &F_1^{Nq}(t) \, = \,  \frac{1}{1 + \tau_N} \,  
\biggl\{ \alpha_E^{Nq} F_q^E +  \alpha_M^{Nq} F_i^M \tau_N \biggr\}\,, 
\hspace*{.3cm}
F_2^{Nq}(t) \, = \, \frac{1}{1 + \tau_N} \, 
\biggl\{ - \alpha_E^{Nq} F_q^E +  \alpha_M^{Nq} F_q^M \biggr\}\,, 
\nonumber\\
& &G_1^{Nq}(t) \, = \, \frac{1}{1 + \tau_N} \, 
\biggl\{ \alpha_E^{Nq} G_q^E +  \alpha_M^{Nq} G_q^M \tau_N \biggr\}\,, 
\hspace*{.3cm}
G_2^{Nq}(t) \, = \, \frac{1}{1 + \tau_N} \, 
\biggl\{ - \alpha_E^{Nq} G_q^E +  \alpha_M^{Nq} G_q^M \biggr\}\,, 
\en
where $\alpha_E^{Nq}$ and $\alpha_M^{Nq}$ are the SU(6) spin-flavor 
couplings ($\alpha_E^{pu} = \alpha_E^{nd} = 2$, 
$\alpha_E^{pd} = \alpha_E^{nu} = 1$, 
$\alpha_M^{pu} = \alpha_M^{nd} = 4/3$, 
$\alpha_M^{pd} = \alpha_M^{nu} = -1/3$); $t = - q^2$ is the Euclidean momentum 
squared and $\tau_N = t/(4m_N^2)$. The Dirac and Pauli form factors
$f_{D(P)}^q(t)$ of the quark of flavor $i$ are shown explicitly in Appendix C. 
Again, Eq.~(\ref{FNi}) can be separated into a bare part and 
a meson cloud part as
\eq
F_1^{N\,{\rm bare}}(t) &=& \sum_{q=u,d} e_q\,F_1^{Nq}(t)\,, \nonumber\\
F_2^{N\,{\rm bare}}(t) &=& \sum_{q=u,d} e_q\,F_2^{Nq}(t)\,,
\en
and the meson cloud contribution is 
\eq
F_1^{N\,{\rm cloud}}(t) &=& \sum_{q=u,d} 
[(f_D^q(t)-e_q) F_1^{Nq}(t)+f_P^q(t) G_1^{Nq}(t)], 
\nonumber \\
F_2^{N\,{\rm cloud}}(t) &=& \sum_{q=u,d} 
[(f_D^q(t)-e_q) F_2^{Nq}(t)+f_P^q(t) G_2^{Nq}(t)].
\en
where $e_q$ are the electric quark charges.

In the following we try to achieve a reasonable description
of the electromagnetic form factors of 
the nucleon by an appropriate set of parameters contained in $F_1^N(t)$ 
and $F_2^N(t)$. In general both the bare and the meson cloud parts
will contribute to the form factors. The bare part should well describe 
the form factors up to high $Q^2$ whereas the meson cloud part should 
play a role only for small $Q^2$. Our strategy is that we first
fit the bare part of all the form factors to the experimental data 
from $Q^2 \sim 0.5 \,\,\rm{GeV}^2$ to high $Q^2$. After that we 
reconsider the low $Q^2$ region by taking into account the meson 
cloud effect. At this point we do not pretend on an accurate 
analysis (or prediction) of the nucleon form factors at large Q$^2$. 
This is a task more appropriate for perturbative QCD (pQCD) which is the most 
convenient theoretical tool in this momentum region (for recent progress 
see e.g. Refs.~\cite{Brodsky:2003pw,Belitsky:2002kj}). 
In particular, in Ref.~\cite{Belitsky:2002kj} the following 
large Q$^2$-behavior for the ratio of the nucleon Pauli $F_2$ and 
Dirac $F_1$ form factors has been derived using pQCD: 
\eq 
\frac{F_2}{F_1} \propto \frac{{\rm ln}^2(Q^2/\Lambda^2)}{Q^2} \,, 
\en 
where $\Lambda$ is a soft scale related to the size of the nucleon. 
Our main idea is to use a physically reasonable parametrization of 
the nucleon form factors (valence quark contribution) which fits the data 
at intermediate and high $Q^2$ scale, while the derived constraints 
at zero recoil [see Sec.~II] are also satisfied. Then on top 
of the valence (or bare) form factors we place the meson cloud contribution 
which is relevant only at small $Q^2$ and calculated consistently 
using effective chiral Lagrangian~(\ref{L_qU}). 
Finally we might conclude on the role of the meson cloud 
in the infrared domain. 

Due to the finite size of the source of the meson fields, meson loops are 
expected to be strongly suppressed for large $Q^2$. To mimic the effect of 
additional regularisation of the integral, which leads to a restriction of 
the meson cloud contribution in the low $Q^2$ region,
we introduce the cutoff function $f_{cut}(t)$. The modification is
such that
\eq
F_{1(2)}^{N \, {\rm cloud}}(t) &\rightarrow& f_{cut}(t) \, 
F_{1(2)}^{N \, {\rm cloud}}(t) \,. 
\en
The specific form of $f_{cut}(t)$ should not modify
$F_{1(2)}^{N \, {\rm cloud}}(t)$ 
in the low $Q^2$ region, but should diminish its contribution
beyond a certain $Q^2$. This function $f_{cut}(t)$ should have zero
(or almost zero) 
slope at the point $Q^2=0\,\,\rm{GeV}^2$ to ensure that it will not 
artificially contribute to the slope of the form factors. 
First the step function seems to be an 
appropriate choice for $f_{cut}(t)$ but its sharp boundary affects 
the continuity of the form factors, and hence one must be careful. 
To avoid the problems associated with a sharp boundary we choose 
a smeared-out version for $f_{cut}(t)$ with 
\eq
f_{cut}(t)=\frac{1+\exp(-A/B)}{1+\exp[(t-A)/B]}\, ,
\en 
where the parameter $A$ characterizes the cut-off and
$B$ the smearing of the function.

In our fitting procedure we use the experimental data 
(Refs.~\cite{Simon:1980hu,Price:1971zk,Berger:1971kr,Hanson:1973vf,
Milbrath:1997de,Dieterich:2000mu,Jones:1999rz,Gayou:2001qd,Hohler:1976ax,
Janssens:1966,Bartel:1973rf,Walker:1993vj,Andivahis:1994rq,Litt:1969my,
Sill:1992qw,Lung:1992bu,Kubon:2001rj,Xu:2000xw,Anklin:1994ae,
Anklin:1998ae,Rock:1982gf,Markowitz:1993hx,Bruins:1995ns,Xu:2002xc,
Gao:1994ud,Herberg:1999ud,Passchier:1999cj,Eden:1994ji,Ostrick:1999xa,
Golak:2000nt,Madey:2003av,Zhu:2001md,Rohe:1999sh,Schiavilla:2001qe,Punjabi:2005wq}) 
on the ratios of the electromagnetic Sachs form 
factors of nucleons to the corresponding dipole form factors and on 
the neutron 
charge form factors. For the axial charge of the constituent quark $g=0.9$ 
we first fit the contribution of the bare part, where the parameters 
are contained in the ansatz of Eq.~(\ref{quarkff}). Our parameter 
results for the fit of the bare part are (in units of ${\rm GeV}$)
\eq
\lambda_{uE}=2.0043,\,\,\,\lambda_{dE}&=&0.9996,\,\,\,
\lambda_{uM}=7.3367,\,\,\,\lambda_{dM}=2.2954, \nonumber\\
\Lambda_{uE}=0.8616,\,\,\,\Lambda_{dE}&=&0.9234,\,\,\,
\Lambda_{uM}=0.9278,\,\,\,\Lambda_{dM}=1.0722.
\en
Note that the parameters $\gamma_u$ and $\gamma_d$ entering in 
Eq.~(\ref{quarkff}) cannot be fixed by considering only the bare part 
since they also occur in $F_1^{N\,{\rm cloud}}(t)$ and
$F_2^{N\,{\rm cloud}}(t)$. 
The complete fit to the full data on the electromagnetic form factors
of the nucleon fixes
the remaining low-energy constants in the effective Lagrangian as
\eq
\gamma_u &=& 1.081,\,\,\,\,\,\,\gamma_d \, = \, 2.596,\,\,\,\,\,\, 
c_2 \, = \, 2.502\,\,\,{\rm GeV}^{-1},\,\,\, 
c_4 \, = \, 1.693\,\,\,{\rm GeV}^{-1},\,\,\, 
\bar d_{10}\, = \, 1.110\,\,\,{\rm GeV}^{-2},\,\,\, \nonumber \\
\bar e_7 &=& - 0.650\,\,\,{\rm GeV}^{-3},\,\,\,\bar e_8 = 0.030
\,\,\,{\rm GeV}^{-3},\,\,\,\bar e_{10}=0.039\,\,\,{\rm GeV}^{-3}\,, 
\en 
where in the current manuscript we put $k_V = 0$ for simplicity. 
The couplings $\bar d_{10}$, $\bar e_{7}$, $\bar e_{8}$ and 
$\bar e_{10}$ are defined in Appendix C. 
With the choice $A=(0.4-0.5)\,\,\,{\rm GeV}^2$ and 
$B=0.025\,\,\,{\rm GeV}^2$ in 
$f_{cut}(t)$ the meson cloud contributions are suitably suppressed 
for values larger than $t \sim 0.5\,{\rm GeV}^2$. 
Note, that from the analysis of the 
nucleon mass and meson-nucleon sigma-terms we get the following constraint 
on the parameter $c_2 = (1.093 - 1.354 \,\tilde e_3 
              + 0.474 \,\tilde e_4 
              - 0.338 \,\tilde e_5) \,\,\,{\rm GeV}^{-1}$.  
The best description of the electromagnetic properties 
of the nucleon is obtained at 
the value of $c_2 = 2.502 \,\,\,{\rm GeV}^{-1}$ fixed from the 
data on the electromagnetic form factors. 
It means that we obtain the following 
constraint on the LECs $\tilde e_3$, $\tilde e_4$ and $\tilde e_5$: 
 $1.354 \,\tilde e_3 
 - 0.474 \,\tilde e_4 
 + 0.338 \,\tilde e_5  = - 1.409\,.$  

Results for the measured ratio of the electromagnetic form factors of the 
nucleon to the dipole form factor $G_D(t)= (1 + t/0.71$ GeV$^2)^{-2}$ 
are presented in Figs.~5-7. 
In Fig.~8 we present our result for the ratio of 
charge to magnetic form factor of the proton.
For completeness, the electromagnetic form factors of the nucleon are 
also shown in Fig.~9, where the meson cloud contributions are shown 
explicitly in comparison to the dipole form factor.
The role of $f_{cut}(t)$, which restricts the meson cloud contribution
to the low $Q^2$ region, is indicated in Fig.~10.  
Although the bare constituent quark contribution is fully
parameterized, a consistent explanation of the form factors
can only be achieved, when meson cloud corrections are included.
We stress that the bare constituent quark Sachs form factors
and hence the magnetic form factors at zero recoil ($t=0$) are
determined by the general constraints discussed in Section~II.E.
At this point the specific form of the quark form factors 
of Eq.~(\ref{quarkff}), leading to a satisfactory description  
in particular beyond $t \sim 0.5\,{\rm GeV}^2$, is not required. 
The results concerning the charge and magnetic radii of nucleons are
\eq
r_E^p &=& 0.881\,\,\,{\rm fm},\,\,\,\la r^2 \ra_E^n = -0.1177\,\,\,{\rm fm}^2,
\nonumber\\
r_M^p &=& 0.869\,\,\,{\rm fm},\,\,\,r_M^n = 0.847\,\,\,{\rm fm}.
\en
The experimental values reported in Ref.~\cite{Eidelman:2004wy}
for the charge radii of nucleons are
$r_E^p = 0.875 \pm 0.007 \,\, {\rm fm}$ and
$\la r^2 \ra_E^n = -0.1161 \pm 0.0022 \,\, {\rm fm}^2$. 
For the magnetic radii, 
the analysis of Refs.~\cite{Sick:2003gm,Kubon:2001rj} 
gives $r_M^p=0.855 \pm 0.035 \,\, {\rm fm}$ 
and $r_M^n=0.873 \pm 0.011 \,\, {\rm fm}$, respectively.

For further illustration we
follow Ref.~\cite{Kelly:2002if} to deduce the radial dependence
of the charge and magnetization densities of nucleons.
The charge and magnetization densities of nucleons in the rest frame are
\eq
\rho_E^N(r) = \frac{2}{\pi}\int_0^\infty dk\,k^2\,j_0(k\,r) 
\tilde \rho_E^N(k) \,, \quad\quad 
\rho_M^N(r) = \frac{2}{\pi}\int_0^\infty dk\,k^2\,j_0(k\,r) 
\tilde \rho_M^N(k),
\en
where $k^2=t/(1+\tau_N)$ and $j_0(kr)$ is the Bessel function. 
The intrinsic form factors $\tilde \rho_E^N(k)$ and 
$\tilde \rho_M^N(k)$ are 
\eq
\tilde\rho_E^N(k)=G_E^N(Q^2)(1+\tau_N)^{\lambda^E},\,\,\,\,\,\,
\tilde\rho_M^N(k)=\frac{G_M^N(Q^2)}{\mu_N}(1+\tau_N)^{\lambda^M}.
\en
Restricting to the discrete values
$\lambda^E=0,1,2$, where the full range is discussed in
Ref.~\cite{Kelly:2002if}, 
the results for the charge and magnetization densities of the nucleon 
are shown in Figs.~11 and 12. 
 
In Ref.~\cite{Friedrich:2003iz} the electromagnetic form factors of 
the nucleons are represented by the phenomenological ansatz
\eq
G_N(Q^2)=G_s(Q^2)+a_b\cdot Q^2\,G_b(Q^2),
\en
where the "smooth" part $G_s(Q^2)$ and the structured or "bump"
part $G_b(Q^2)$ are
parameterized as
\eq\label{FrWa}
G_s(Q^2) &=& 
\frac{a_{10}}{(1+Q^2/a_{11})^2}+\frac{a_{20}}{(1+Q^2/a_{21})^2}, 
\nonumber\\[2mm]
G_b(Q^2) &=& e^{-\frac{1}{2}\left(\frac{Q-Q_b}{\sigma_b}\right)^2}+
e^{-\frac{1}{2}\left(\frac{Q+Q_b}{\sigma_b}\right)^2}\,.
\en
The parameters $a_{10},\,a_{11},\,a_{20},\,a_{21},\,a_b,\,Q_b,$ and 
$\sigma_b$ are obtained by a fit to experimental data and 
the values are reported in Table~2 of Ref.~\cite{Friedrich:2003iz}.
The meson cloud part of our evaluation cannot be directly compared
or matched to $G_b (Q^2)$, since meson corrections contribute
to the magnetic form factors even at $Q^2=0$, which is not the case
in the treatment of Ref.~\cite{Friedrich:2003iz}. 
However, charge conservation 
restricts the meson cloud not to contribute to the charge form factors 
at zero recoil. Therefore, we can compare our results for the
charge form factors to 
the phenomenological ones of Ref.~\cite{Friedrich:2003iz}.  
Fig.~5 shows such 
a comparison for the case of the charge form factor of the proton. 
Based on the phenomenological ansatz 
of Ref.~\cite{Friedrich:2003iz} our result can be reproduced by readjusting 
the parameters of Eq. (\ref{FrWa}) as compiled in Table~3,
which are not so much different from 
the original analysis of Ref.~\cite{Friedrich:2003iz}.  

\subsection{Coupling of vector mesons to the nucleon} 
 
Finally, we calculate the strong vector meson-nucleon form 
factors $\rho NN$ and $\omega NN$ at one loop. 
We follow the strategy already developed for the
electromagnetic nucleon form factors. The corresponding bare operator 
is derived from the tree-level Lagrangian~(\ref{L_Vqq}):  
\eq\label{Jmu_Vqq_bare}  
J_{\mu, V}^{\rm bare}(q) \, = \, \int d^4x \, e^{-iqx} \, 
j_{\mu, V}^{\rm bare}(x)\,, \hspace*{.3cm}  
j_{\mu, V}^{\rm bare}(x) \, = \, g_{Vqq} \, \bar q \biggl( 
\gamma_\mu \, + \, \frac{k_V}{2 m_q} \, \sigma_{\mu\nu} \, 
\tensor{\partial^\nu} \biggr) \, \frac{\lambda_V}{2} \, q \,,  
\en 
where $\bar q \, \tensor{\partial^\nu} \, q \,  =  \,   
\bar q \, ( \loarrow{\partial^\nu}  +  \roarrow{\partial^\nu} ) \, q$ 
and $\lambda_V$ is the corresponding flavor matrix: 
$\lambda_\rho = {\rm diag}\{1,-1,0\}$ for the $\rho^0$ and 
$\lambda_\omega = {\rm diag}\{1,1,0\}$ for the $\omega$ meson.  
The diagrams contributing to these quantities 
are displayed in Fig.3: tree-level diagrams 
(Figs.3(1) and 3(2)) and one-loop 
diagrams due to the dressing by a cloud of pseudoscalar mesons 
(Figs.3(3) and 3(4)). 

The Fourier transform of the dressed vector-meson-quark transition 
operator has the following form 
\eq\label{Jmu_Vqq_dress} 
J_{\mu, V}^{\rm dress}(q) \, = \, \frac{1}{2} \, \int d^4x \, e^{-iqx} \, 
\bar q(x) \, \biggl[ \, \gamma_\mu \, f_{V_D}(q^2) \, + \, \frac{i}{2m_q} \,  
\sigma_{\mu\nu} \, q^\nu \, f_{V_P}(q^2) \, \biggr] \, q(x) 
\en 
where $f_{V_D}$ and $f_{V_P}$ are the matrices of the Dirac and Pauli 
form factors describing the coupling of $u$, $d$ and $s$ quarks to the
$\rho$ and $\omega$ vector mesons. These form factors
are given in terms of the Euclidean 
values of momentum squared $t = - q^2$ with: 
\eq 
f_{V_D}(t) &=& \sum\limits_{i = u, d, s} \, f_{V_D}^i(t) \,, 
\quad\quad 
f_{V_P}(t) \, = \, \sum\limits_{i = u, d, s} \, f_{V_P}^i(t) \,, \\
f_{\rho_D}^u(t) &=& - f_{\rho_D}^d(t) \, = \, 
g_{Vqq} \, \biggl\{1 - {\mathcal \epsilon}_5^\pi(t) 
+ \frac{1}{3} {\mathcal \epsilon}_5^\eta(t) \biggr\}\,, \nonumber\\ 
f_{\omega_D}^u(t) &=& f_{\omega_D}^d(t) \, = \, 
g_{Vqq} \, \biggl\{ 1 + 3 {\mathcal \epsilon}_5^\pi(t) 
+ \frac{1}{3} {\mathcal \epsilon}_5^\eta(t) \biggr\} \,, \nonumber\\  
f_{\rho_D}^s(t) &=& 0 \,,  \quad\quad\quad \;\, 
f_{\omega_D}^s(t) \, = \, g_{Vqq} \biggl\{ - 4 \,\, 
{\mathcal \epsilon}_5^K(t) \biggr\} \,, \nonumber \\ 
f_{\rho_P}^u(t) &=& - f_{\rho_P}^d(t) \, = \, 
k_V \, g_{Vqq} \, \biggl\{1 - m_{10}^\pi(t) 
+ \frac{1}{3} m_{10}^\eta(t) \biggr\}\,,\nonumber\\
f_{\omega_P}^u(t) &=& f_{\omega_P}^d(t) 
\, = \, k_V \, g_{Vqq} \, \biggl\{ 1 + 3 m_{10}^\pi(t) 
+ \frac{1}{3} m_{10}^\eta(t) \biggr\} \,, \nonumber  \\ 
f_{\rho_P}^s(t) &=& 0 \,,  \quad\quad\quad\quad \;\;   
f_{\omega_P}^s(t) \, = \, k_V \, g_{Vqq} \biggl\{ - 4 \,\, 
m_{10}^K(t) \biggr\} \,, \nonumber 
\en
where $\lambda_\phi = {\rm diag}\{0,0,1\}$.  
Here ${\mathcal \epsilon}_5^P(t)$ and $m_5^P(t)$ are the meson-cloud 
contributions given in Appendix C. 

To project the dressed quark operator (\ref{Jmu_Vqq_dress}) onto
the nucleon 
we proceed in analogy to the electromagnetic operator 
%(see Eq.~\ref{master}) as: 
\eq\label{master_V}
&\la&\!\!\! N(p^\prime) | \, J_{\mu, V}^{\rm dress}(q) 
\, | N(p) \ra \, = \, (2\pi)^4 \, \delta^4(p^\prime - p - q) \, 
\bar u_N(p^\prime) \, \frac{1}{2} \, \biggl\{ \gamma_\mu \, G_{VNN}(q^2) 
\, + \, \frac{i}{2 \, m_N} \, \sigma_{\mu\nu} q^\nu 
\, F_{VNN}(q^2) \biggr\} u_N(p) \, \nonumber\\ 
& = & (2\pi)^4 \, \delta^4(p^\prime - p - q) \, \frac{1}{2} \, \biggl\{ 
f_{V_D}^{ij}(q^2) \, \la N(p^\prime)|\,j_{\mu, ij}^{\rm bare}(0)\,|N(p) \ra
+  i \, \frac{q^\nu}{2 \, m_q} \, f_{V_P}^{ij}(q^2) \, 
\la N(p^\prime)| \, j_{\mu\nu, ij}^{\rm bare}(0) \, |N(p) \ra 
\biggr\} \,, 
\en 
where the bare matrix elements 
$\la N(p^\prime)| \, j_{\mu\nu, ij}^{\rm bare}(0) \, |N(p) \ra$  
are defined in Eq.~(\ref{bare_operators}).   
Here $G_{VNN}(q^2)$ and $F_{VNN}(q^2)$ are the vectorial 
and tensorial couplings of vector mesons to nucleons.  
We can express the strong $\rho NN$ and $\omega NN$ 
form factors through the bare electromagnetic nucleon form 
factors. After a simple algebra we arrive at: 
\eq\label{VNN_FF}
G_{\rho NN}(t) &=& g_{Vqq} \, [ \,  F_1^{p -}(t) \, f_{\rho_D}^u(t) 
\, + \, G_1^{p -}(t) \, f_{\rho_P}^u(t) \, ] \,, 
\nonumber\\ 
G_{\omega NN}(t) &=& g_{Vqq} \, [ \,  F_1^{p +}(t) \, f_{\omega_D}^u(t) 
\, + \, G_1^{p +}(t) \, f_{\omega_P}^u(t) \, ] \,, 
\nonumber\\
F_{\rho NN}(t) &=& g_{Vqq} \, [ \,  F_2^{p -}(t) \, f_{\rho_D}^u(t) 
\, + \, G_2^{p -}(t) \, f_{\rho_P}^u(t) \, ] \,, 
\nonumber\\ 
F_{\omega NN}(t) &=& g_{Vqq} \, [ \,  F_2^{p +}(t) \, f_{\omega_D}^u(t) 
\, + \, G_2^{p +}(t) \, f_{\omega_P}^u(t) \, ] \,, 
\en 
where $t=-q^2$ and 
$H_I^{p \pm} = H_I^{pu} \pm H_I^{pd}$ with $H=F$ or $G$ and $I=1$ or $2$. 

Finally, we present the expressions for the values of the 
vector-meson nucleon form factors at zero recoil or 
the coupling constants $G_{VNN}$ and $F_{VNN}$ which originate 
from the nucleon-level Lagrangian (for details on the nucleon-level 
Lagrangian see Ref.~\cite{Kubis:2000zd}): 
\eq\label{VNN}
{\cal L}_{VNN} \, = \, \frac{1}{2} \, \bar N \biggl\{  
\biggl( \gamma^\mu \, G_{\rho NN}  \, - \, \, 
\frac{F_{\rho NN}}{2 \, m_N} \, 
\sigma^{\mu\nu} \, \partial_\nu \biggr) \vec{\rho}_\mu \, \vec{\tau} 
\, + \, \biggl( \gamma^\mu \, G_{\omega NN}  \, - \, 
\frac{F_{\omega NN}}{2 \, m_N}  \, 
\sigma^{\mu\nu} \, \partial_\nu \biggr) \omega_\mu \, 
\biggr\} N \,. 
\en         
After a simple algebra we arrive at: 
\eq\label{VNN_1}  
& &G_{\rho NN} = g_{Vqq} \,, \quad\quad \frac{F_{\rho NN}}{G_{\rho NN}} 
\, = \, \mu_p^{\rm bare} \, - \, \mu_n^{\rm bare} \, - \, 1 \, + 
\, k_V \, \frac{m_N}{m_q}  \, ( \, \delta_{pu}^{\rm bare} \, 
- \, \delta_{pd}^{\rm bare} \,) \, [1 + \delta_\rho] \,, \nonumber\\
& &G_{\omega NN} \, = \, 3 \, g_{Vqq} \,, \quad\quad 
\frac{F_{\omega NN}}{G_{\omega NN}} \, = \, 
\mu_p^{\rm bare} \, + \, \mu_n^{\rm bare} \, - \, 1 
\, + \, k_V \, \frac{m_N}{m_q}  \, ( \, \delta_{pu}^{\rm bare} 
\, + \, \delta_{pd}^{\rm bare} \, ) \, [1 + \delta_\omega] \,,  
\en 
where 
\eq
\delta_\rho \, = \, - m_{10}^\pi(0) + \frac{1}{3} m_{10}^\eta(0) \,, 
\quad\quad
\delta_\omega \, = \, 3 m_{10}^\pi(0) + \frac{1}{3} m_{10}^\eta(0) \, 
\en 
are the corresponding one-loop corrections. 
For $k_V \equiv 0$ these equations reduce to the well-known 
SU(3) relations, relating the matrix elements of the vector current 
with different flavor content~\cite{deSwart:gc}. 
Using the actual expressions for the bare magnetic moments and
the tensor 
charges of the nucleons we finally obtain for the ratios of the tensor and 
vector couplings: 
\eq\label{VNN_2}  
\frac{F_{\rho NN}}{G_{\rho NN}} \, = \, 
\biggl(\frac{g_A}{g}\biggr)^4 \, 
\biggl( 1 \, + \, k_V \, [1 + \delta_\rho] \biggr) -1 \,, 
\quad\quad\quad \frac{F_{\omega NN}}{G_{\omega NN}} \, = \, \frac{1}{5} \, 
\biggl(\frac{g_A}{g}\biggr)^4 \, 
\biggl( 1 \, + \, 3 \, k_V \, [1 + \delta_\omega] \biggr) -1 \,. 
\en  
Note, that numerically the one-loop corrections to the 
tree-level results for the strong vector-meson nucleon form factors 
$F_{\rho NN}$ and $F_{\omega NN}$ are rather small: 
$\delta_\rho = 0.005$ and $\delta_\omega = 0.011$. 

\section{Summary} 

We developed a manifestly Lorentz covariant chiral quark model for the 
study of baryons as bound states of constituent quarks. 
The approach is based on the effective chiral Lagrangian involving 
constituent quarks and the chiral fields as effective degrees of freedom.  
This Lagrangian is used in the calculation of the dressed transition 
operators which are relevant for the interaction of quarks with external 
fields in the presence of a virtual meson cloud. Then the dressed 
operators are used in the calculation of baryon matrix elements. 

Our main result is as follows:
we perform a model-independent factorization of the 
effects of hadronization and confinement contained in the matrix elements 
of the bare quark operators and the effects dictated by chiral symmetry 
which are encoded in the corresponding relativistic form factors 
[see e.g. Eq.~(\ref{master})]. 
Due to this factorization the calculation of chiral effects and 
the effects of hadronization and confinement can be done independently. 
All low-energy theorems are reproduced in our approach due to the chiral 
invariance of the effective Lagrangian. 
The evaluated meson-cloud corrections are in agreement with the  
infrared-singular structure of the corresponding nucleon matrix 
elements~\cite{Kubis:2000zd,Beg:1973sc}. In particular, we reproduce the 
leading nonanalytic (LNA) contributions to the nucleon mass,  
to the pion-nucleon sigma-term, to the magnetic moments and to the charge
radii of the nucleons. 
The LNA contributions to the nucleon mass and magnetic moment are proportional 
to the $M_P^3$ and $M_P$, respectively, where $M_P$ is the pseudoscalar 
meson mass. The nucleon radii are divergent in the chiral limit.
Using model-independent constraints
on the bare constituent quark distributions in the octet baryons,
we work out model predictions for the magnetic moments.
Based on a full parameterization of the bare constituent quark
distributions in the nucleon, we give results for the full
momentum dependence of the electromagnetic form factors of the nucleon
and indicate the role of the meson cloud contributions.
Presently, the calculation of the matrix elements of the bare quark operators
is, besides the model independent constraints, based on parameterizations.
The direct calculation of these matrix elements should be performed
in quark models 
based on specific assumptions about hadronization 
and confinement.

\begin{acknowledgments}

The authors thank P.~Grabmayr, V.~Punjabi, H.~Gao, K.~de Jager and 
R.~G.~Milner for providing us the latest experimental data on 
electromagnetic nucleon form factors. This work was supported by the 
DFG under contracts FA67/25-3 and GRK683. 
This research is also part of the EU Integrated Infrastructure Initiative 
Hadronphysics project under contract number RII3-CT-2004-506078 and  
President grant of Russia "Scientific Schools"  No. 5103.2006.2. 
K.P. thanks the Development and Promotion of Science and Technology 
Talent Project (DPST), Thailand for financial support.  

\end{acknowledgments}

\newpage 

\appendix\section{Calculational technique of loop diagrams}

The calculational technique of the loop diagrams in Figs.1, 2, 3 and 4 
is referred to as  
{\it the infrared dimensional regularization} (IDR) and has been 
discussed in detail in Refs.~\cite{Becher:1999he,Kubis:2000zd}. Our case  
here differs by the use of the constituent quark degrees of freedom instead of 
the nucleon. We briefly recall the basic ideas of this technique. 
As we discussed before, in Baryon ChPT loop integrals are 
non-homogeneous functions of the mesonic momenta and the quark masses due 
to the presence of a new scale parameter, the nucleon mass. As a  result, 
the loop integral contains an infrared singular part containing 
the fractional powers of the meson masses and an infrared regular part 
involving the fractional powers of the nucleon mass. Due to the presence 
of the regular parts in the loop integrals, their chiral expansion starts 
at the same order as the tree graphs, it therefore spoils the power 
counting rules. The idea of the IDR method is to remove the infrared 
regular parts of the loop integrals from the consideration and to absorb 
them in the low-energy couplings of the underlying chiral Lagrangian. 
The IDR method is consistent with Lorentz and gauge invariance. 
Also, the chiral power countings are preserved and the Ward identities of 
chiral symmetry are fulfilled. 

The method can be explained by using the simple example of the self-energy 
graph shown in Fig.4(3) contributing to the quark mass 
operator~\cite{Becher:1999he,Kubis:2000zd}. 
We consider the scalar loop integral in $d$ dimensions: 
\eq 
H(p^2) = \int \frac{d^dk}{(2\pi)^d i} \, 
\frac{1}{[M^2 - k^2 - i\epsilon] \, [m^2 - (p-k)^2 - i\epsilon] }
\en
where $M$ and $m$ are the meson and constituent quark masses, 
respectively. Using the master formula in $d$-dimensions 
\eq 
\int \frac{d^dk}{(2\pi)^d} \, 
\frac{k^{2n}}{[M^2 - k^2]^m} = \frac{i^{2n + 1}}{(4\pi)^{d/2}}  \, 
\frac{\Gamma(n+d/2) \, \Gamma(m-n-d/2)}{\Gamma(d/2) \, \Gamma(m)} 
\, M^{2(n-m)+d} 
\en
we get at threshold $p^2 = (m+M)^2$: 
\eq 
H(p^2) = \underbrace{c_d \, \frac{M^{d-3}}{m+ M}}_{= I(p^2)} + 
\underbrace{c_d \, \frac{m^{d-3}}{m+ M}}_{= R(p^2)}\,, \hspace*{1cm} 
c_d = \frac{\Gamma(2-d/2)}{(4\pi)^{d/2} \, (d-3)} \, , 
\en 
Here $I(p^2)$ is the infrared singular piece which is characterized by 
fractional powers of $M$ and generated by the loop momenta of order 
of the meson mass. 
For $I(p^2)$ the usual power counting applies. Another piece 
in the decomposition of $H(p^2)$ defined by $R(p^2)$ is the infrared 
regular part which is generated by the loop momenta of the order of 
the constituent quark mass $m$ (in our counting $m$ is of the order of 
$\Lambda_{\chi SB} \sim 1$ GeV). As discussed before, we remove the 
regular part $R(p^2)$ by redefining the low-energy coupling constants 
in the chiral Lagrangian. In Ref.~\cite{Becher:1999he}  a recipe was 
suggested how to split the integral $H(p^2)$ into the singular and regular 
parts. We use the Feynman parametrization to combine 
two multipliers $a = M^2 - k^2 -i\epsilon$  and 
$b = m^2 - (p-k)^2 -i\epsilon$ in the denominator of $H(p^2)$: 
\eq 
\int \frac{d^dk}{(2\pi)^d} \, \frac{1}{a \, b} =  
\int \frac{d^dk}{(2\pi)^d} \,\int\limits_0^1 \, \frac{dx}{[a(1-x) + b x]^2} 
\en 
and then to write down the integral from 0 to 1 as the difference
of two integrals: 
\eq 
\int\limits_0^1 \, dx \, \ldots \, = \, 
\biggl[ \, \int\limits_0^\infty \, - \, 
\int\limits_1^\infty \, \biggr] \, dx \, \ldots
\en 
Then the integral from 0 to $\infty$ is exactly the infrared 
singular part and the integral from 1 to $\infty$ is the infrared 
regular one. This method can be applied to any general one-loop integral 
with adjustable number of meson and quark propagators. The calculational 
technique suggests that: one should separate numerator and denominator; 
simplify the numerator using the standard (invariant integration) 
methods. Finally, the result can be reduced to the master integral 
$I(p^2)$ and its derivatives (like in the conventional 
dimensional regularization). The integrals containing only the quark 
propagators do not contribute to the infrared singular parts, and 
therefore, vanish in the {\it infrared dimensional regularization}. 

The ultraviolet divergencies contained in the one-loop integrals are 
removed via the renormalization of the low-energy constants in the 
chiral Lagrangian. To perform the renormalization of the constituent 
quark at one loop and to guarantee charge conservation 
we need the $Z$-factor (the wave-function renormalization constant), 
which is determined by the derivative of the quark mass operator 
(generated by the diagrams in Fig.4) with respect to its momentum: 
\eq 
Z^{-1}_q = 1 - 
\frac{\partial\Sigma_q(\not\! p)}{\partial\not\! p}\bigg|_{\not p = m_q} 
\,, \hspace*{.5cm} q=u, d, s \,. 
\en 
The matrix $Z_q = {\rm diag}\{Z_u, Z_d, Z_s\}$ with 
$Z_u = Z_d = Z$ is given up to fourth order by 
\eq\label{Zq} 
Z_q \, = \, I \, + \, \sum\limits_{P = \pi, K, \eta} \, 
\frac{1}{F_P^2} \, \biggl[ \, - \, g^2 \, \alpha_P \, \Delta_P \, 
+ \, Q_P \, \biggr]   
\en 
where 
\eq 
& &\Delta_P  \, = \, 2 \, M_P^2 \, \biggl[ \lambda(\mu) \, + \, 
\frac{1}{16 \pi^2} \, {\rm ln}\biggl(\frac{M_P}{\mu}\biggr) \biggr]\,, 
\hspace*{.3cm} \lambda(\mu) = \frac{\mu^{d-4}}{(4\pi)^2} \, 
\biggl[ \frac{1}{d-4} \, - \, \frac{1}{2} ({\rm ln}4\pi \, + \, 
\Gamma^\prime(1) \, + \, 1 ) \biggr]\,,\nonumber\\[2mm] 
& &Q_P \, = \,  \frac{g^2 M_P^2}{24\pi^2} \, \alpha_P \, 
\biggl[ - 1 \, + \, \frac{3 \pi}{2} \, \frac{M_P}{m} 
\, + \, \frac{3}{2} \, \frac{M_P^2}{m^2}  \biggr] 
\, + \, \frac{3 c_2 M_P^4}{64 \pi^2 m} \, \beta_P \, I\,, \\[2mm]
& &\alpha_\pi = \frac{9}{2} Q + \frac{3}{2} I - \frac{9}{4} \lambda_3\,, 
\hspace*{.3cm}
\alpha_K = -3 Q + 2 I + \frac{3}{2} \lambda_3\,, 
\hspace*{.3cm}
\alpha_\eta = - \frac{3}{2} Q + \frac{I}{2}  + \frac{3}{4} \lambda_3\,, 
\nonumber\\[2mm] 
& &\beta_\pi = 1 \,, \hspace*{.3cm} 
   \beta_K = \frac{4}{3} \,, \hspace*{.3cm} 
   \beta_\eta = \frac{1}{3} \,. \hspace*{.3cm}
\nonumber 
\en 
For the evaluation of the form factor $f_D^q(q^2)$, the 
quark charge $Q$ [diagram in Fig.1(1)] has to be multiplied by $Z_q$. 
For $f_P^q(q^2)$, a second-order contribution to the quark 
anomalous magnetic moment proportional to the $c_6$ 
[diagrams in Figs.1(2) and 2(2$^\ast$)] has to be renormalized by the 
$Z_q$-factor. The term proportional to $c_2$ must be dropped, because 
the product $c_6 \, c_2$ is of higher-order when compared to 
the accuracy we are working in. 

\section{Loop integrals}

In this appendix, we present the loop integrals contributing 
to the electromagnetic transition operator between constituent 
quarks and evaluate them in the infrared regularization 
scheme~\cite{Becher:1999he}. Originally, these integrals have 
been introduced in chiral perturbation theory 
(ChPT)~\cite{Becher:1999he,Kubis:2000zd}.  

For the momenta of initial, final quark 
and photon field we introduce the notations: $p$, $p^\prime$ and 
$q = p^\prime - p$, respectively. We also introduce the 
notation $P = p^\prime + p$. Since the external quarks are on 
the mass shell $p^2 = p^{\prime \, 2} = m^2$, the structure integrals 
can be expanded through a set of scalar functions 
which exclusively depend on the transverse momentum squared 
$t = Q^2 = - q^2$ and  
the masses of meson $M$ and constituent quark $m$. 
As usual~\cite{Becher:1999he,Kubis:2000zd}, we identify the masses 
of particles occurring in the loop integrals with their leading order 
values, $M_P \to \MPz$ and $m_q \to m$. 
For universality, we calculate the integrals for adjustable values 
of the constituent quark mass inside the loop $(m^\ast)$ and for the 
external line $(m)$. Finally, in numerical calculations we will 
neglect the difference between the masses $m^\ast$ and $m$,
setting $m^\ast \equiv m$. 

\subsection{Infrared parts of loop integrals} 

We deal with the following loop integrals: 
\eq 
\int\limits_{I}&& \!\!\!\!\!\!\!
\frac{d^d k}{(2\pi)^d i} \, \frac{1}{M^2 - k^2} 
\, = \, \Delta_M\,,\\
&&\nonumber\\
\int\limits_{I}&&\!\!\!\!\!\!\! \frac{d^d k}{(2\pi)^d i} \, 
\frac{1}{[M^2 - k^2][M^2 - (k+q)^2]}  
\, = \, J^{(0)}(t,M^2)\,,\\
&&\nonumber\\
\int\limits_{I}&& \!\!\!\!\!\!\!\frac{d^d k}{(2\pi)^d i} \, 
\frac{k_\mu}{[M^2 - k^2][M^2 - (k+q)^2]}  
\, = \, - \, \frac{1}{2} \, q_\mu \, J^{(0)}(t,M^2)\,,\\
&&\nonumber\\
\int\limits_{I}&&\!\!\!\!\!\!\! \frac{d^d k}{(2\pi)^d i} \, 
\frac{k_\mu \, k_\nu}{[M^2 - k^2][M^2 - (k+q)^2]}  
\, = \, (q_\mu \, q_\nu \, - \, g_{\mu\nu} t) \, J^{(1)}(t,M^2) 
\, + \,  q_\mu \, q_\mu \, J^{(2)}(t,M^2)\,,\\
&&\nonumber\\
\int\limits_{I}&& \!\!\!\!\!\!\!\frac{d^d k}{(2\pi)^d i} \, 
\frac{1}{[M^2 - k^2][m^{\ast\, 2} - (p-k)^2]}  
\, = \, I^{(0)}(M^2,m^2,m^{\ast\,2})\,,\\
&&\nonumber\\
\int\limits_{I}&&\!\!\!\!\!\!\! \frac{d^d k}{(2\pi)^d i} \, 
\frac{k_\mu}{[M^2 - k^2][m^{\ast\, 2} - (p-k)^2]}  
\, = \, p_\mu \, I^{(1)}(M^2,m^2,m^{\ast\,2})\,,\\
&&\nonumber\\
\int\limits_{I}&&\!\!\!\!\!\!\! \frac{d^d k}{(2\pi)^d i} \, 
\frac{1}{[M^2 - k^2][m^{\ast\,2} - (p-k)^2]
[m^{\ast\,2} - (p^{\,\prime}-k)^2]}  \, = \, 
I_{12}^{(0)}(t,M^2,m^2,m^{\ast\,2})\,,\\
&&\nonumber\\
\int\limits_{I}&&\!\!\!\!\!\!\! \frac{d^d k}{(2\pi)^d i} \, 
\frac{k_\mu}{[M^2 - k^2][m^{\ast\, 2} - (p-k)^2]
[m^{\ast\, 2} - (p^{\,\prime}-k)^2]}  
\, = \, P_\mu \, I_{12}^{(1)}(t,M^2,m^2,m^{\ast\,2})\,,\\
&&\nonumber\\
\int\limits_{I} &&\!\!\!\!\!\!\!\frac{d^d k}{(2\pi)^d i} \,
\frac{k_\mu \, k_\nu}{[M^2 - k^2][m^{\ast\, 2} - (p-k)^2]
[m^{\ast\, 2} - (p^{\,\prime}-k)^2]} \, = \, 
g_{\mu\nu} \, I_{12}^{(2)}(t,M^2,m^2,m^{\ast\,2})\nonumber\\ 
&+&P_\mu \, P_\nu \, I_{12}^{(3)}(t,M^2,m^2,m^{\ast\,2}) 
\, + \, q_\mu \, q_\nu \, I_{12}^{(4)}(t,M^2,m^2,m^{\ast\,2})\,,\\
&&\nonumber\\
\int\limits_{I}&& \!\!\!\!\!\!\!\frac{d^d k}{(2\pi)^d i} \, 
\frac{1}{[M^2 - k^2][M^2 - (k+q)^2][m^{\ast\, 2} - (p-k)^2]}  
\, = \, I_{21}^{(0)}(t,M^2,m^2,m^{\ast\,2})\,,\\
&&\nonumber\\
\int\limits_{I}&& \!\!\!\!\!\!\!\frac{d^d k}{(2\pi)^d i} \, 
\frac{k_\mu}{[M^2 - k^2][M^2 - (k+q)^2][m^{\ast\, 2} - (p-k)^2]} 
\, = \, P_\mu \, I_{21}^{(1)}(t,M^2,m^2,m^{\ast\,2}) \nonumber\\ 
&-& \frac{1}{2} \, q_\mu \, I_{21}^{(0)}(t,M^2,m^2,m^{\ast\,2}) \,,\\
&&\nonumber\\
\int\limits_{I}&&\!\!\!\!\!\!\! \frac{d^d k}{(2\pi)^d i} \, 
\frac{k_\mu k_\nu}{[M^2 - k^2][M^2 - (k+q)^2][m^{\ast\, 2} - (p-k)^2]}  
\, = \,g_{\mu\nu} \, I_{21}^{(2)}(t,M^2,m^2,m^{\ast\,2}) \nonumber\\
&+& P_\mu \, P_\nu \, I_{21}^{(3)}(t,M^2,m^2,m^{\ast\,2}) \, + \, 
q_\mu \, q_\nu \, I_{21}^{(4)}(t,M^2,m^2,m^{\ast\,2}) \nonumber\\
&-&\frac{1}{2} \, (P_\mu \, q_\nu + P_\nu \, q_\mu) \, 
I_{21}^{(1)}(t,M^2,m^2,m^{\ast\,2}) \,, 
\en
where the symbol $\int\limits_{I}$ represents the loop integration 
according to the infrared dimensional regularization 
scheme~\cite{Becher:1999he,Kubis:2000zd}. 

\subsection{Reduction formulas for loop integrals} 

Higher-order tensorial integrals can be reduced to the 
basic scalar integrals using the invariant 
integration~\cite{Becher:1999he,Kubis:2000zd}:  
\eq 
J^{(1)}(t,M^2) &=& \frac{1}{4 (d - 1) t} \biggl[ (t + 4 M^2) 
J^{(0)}(t,M^2) \, - \, 2 \, \Delta_M \biggr]\,,\\
&&\nonumber\\
J^{(2)}(t,M^2) &=& \frac{1}{4} \, J^{(0)}(t,M^2) 
\, + \, \frac{1}{2 t} \, \Delta_M \,,\\
&&\nonumber\\
I^{(1)}(M^2,m^2,m^{\ast\,2}) &=& \frac{1}{2 m^2} \, 
\biggl[ M^{\ast\,2} \, I^{(0)}(M^2,m^2,m^{\ast\,2}) 
\, + \, \Delta_M \biggr]\,,\\
&&\nonumber\\
I_{12}^{(1)}(t,M^2,m^2,m^{\ast\,2}) &=& 
\frac{1}{4 m^2 + t} \biggl[ 
I^{(0)}(M^2,m^2,m^{\ast\,2}) \, + \, M^{\ast\,2} \, 
I_{12}^{(0)}(t,M^2,m^2,m^{\ast\,2}) \biggr]\,,\\
&&\nonumber\\
I_{12}^{(2)}(t,M^2,m^2,m^{\ast\,2}) &=& 
\frac{1}{d - 2} \biggl[ M^2 \, I_{12}^{(0)}(t,M^2,m^2,m^{\ast\,2}) 
\, - \, M^{\ast\,2} \, 
I_{12}^{(1)}(t,M^2,m^2,m^{\ast\,2}) \biggr]\,,\\
&&\nonumber\\
I_{12}^{(3)}(t,M^2,m^2,m^{\ast\,2}) &=& 
\frac{1}{(d - 2)\,(4 m^2 + t)} \biggl[  M^{\ast\,2}
\, (d - 1) \, I_{12}^{(1)}(t,M^2,m^2,m^{\ast\,2}) \nonumber\\
&-& M^2 \, I_{12}^{(0)}(t,M^2,m^2,m^{\ast\,2}) 
\, + \, \frac{d - 2}{2} \, I^{(1)}(M^2,m^2,m^{\ast\,2})\biggr]\,,\\
&&\nonumber\\
I_{12}^{(4)}(t,M^2,m^2,m^{\ast\,2}) &=& 
\frac{1}{(d - 2)\,t} \biggl[ - M^{\ast\,2}  
\, I_{12}^{(1)}(t,M^2,m^2,m^{\ast\,2}) \nonumber\\
&+& M^2 \, I_{12}^{(0)}(t,M^2,m^2,m^{\ast\,2}) 
\, + \, \frac{d - 2}{2} \, I^{(1)}(M^2,m^2,m^{\ast\,2})\biggr]\,,\\
&&\nonumber\\
I_{21}^{(1)}(t,M^2,m^2,m^{\ast\,2}) &=& 
\frac{1}{2 \, (4 m^2 + t)} \biggl[ 
(2 \,  M^{\ast\,2} \, + \, t)   
\, I_{21}^{(0)}(t,M^2,m^2,m^{\ast\,2}) \nonumber\\
&-& 2 \, I^{(0)}(M^2,m^2,m^{\ast\,2}) 
\, + \, 2 \, J^{(0)}(t,M^2)\biggr]\,,\\
&&\nonumber\\
I_{21}^{(2)}(t,M^2,m^2,m^{\ast\,2}) &=& 
\frac{1}{4 \, (d - 2)} \biggl[ 
\, (4 \, M^2 \, + \, t) \, I_{21}^{(0)}(t,M^2,m^2,m^{\ast\,2}) 
\nonumber\\
&-& 2 \, (2 \, M^{\ast\,2} \, + \, t)   
\, I_{21}^{(1)}(t,M^2,m^2,m^{\ast\,2}) 
\, - \, 2 \, I^{(0)}(M^2,m^2,m^{\ast\,2})\biggr]\,,\\
&&\nonumber\\
I_{21}^{(3)}(t,M^2,m^2,m^{\ast\,2}) &=& 
\frac{1}{4 \, (d - 2) \, (4 m^2 + t)} \biggl[ 
- \, (4 \, M^2 \, + \, t) \, I_{21}^{(0)}(t,M^2,m^2,m^{\ast\,2}) 
\nonumber\\
&+& 2 \, (d - 1) \, (2 \, M^{\ast\,2} \, + \, t)   
\, I_{21}^{(1)}(t,M^2,m^2,m^{\ast\,2}) \nonumber\\
&+& 2 \, I^{(0)}(M^2,m^2,m^{\ast\,2}) 
\, - \, 2 \, (d - 2) \, I^{(1)}(M^2,m^2,m^{\ast\,2})\biggr]\,,\\
&&\nonumber\\
I_{21}^{(4)}(t,M^2,m^2,m^{\ast\,2}) &=& 
\frac{1}{4 \, (d - 2)} \biggl[ 
- \, (4 \, M^2 \, + \, (d - 1) \, t) \, 
I_{21}^{(0)}(t,M^2,m^2,m^{\ast\,2}) \nonumber\\
&+& 2 \, (2 \, M^{\ast\,2} \, + \, t)   
\, I_{21}^{(1)}(t,M^2,m^2,m^{\ast\,2}) \nonumber\\
&-& 2 \, (d - 3) \, I^{(0)}(M^2,m^2,m^{\ast\,2}) 
\, + \, 2 \, (d - 2) \, I^{(1)}(M^2,m^2,m^{\ast\,2})\biggr]\,. 
\en
where $M^{\ast\,2} = M^2 + m^2 - m^{\ast\,2}$. 

\subsection{Scalar loop integrals}

The scalar loop integrals are given as~\cite{Becher:1999he,Kubis:2000zd}  
\eq
&&\Delta_M = 2 \, M^2 \lambda_M\,, \hspace*{1.5cm} 
\lambda_M = \frac{M^{d - 4}}{(4\pi)^2} \, 
\bigg\{ \frac{1}{d - 4} - \frac{1}{2} 
(\ln 4\pi + \Gamma^{\,\prime}(1) + 1) \biggr\}\,,\\
&&\nonumber\\ 
&&J^{(0)}(t,M^2) = - \, 2 \, \lambda_M \, - \, 
\frac{1}{16\pi^2} \, \biggl[1 \, + \,  
k\biggl(\frac{t}{M^2}\biggr) \biggr] \,, \\  
&&\nonumber\\ 
&&I^{(0)}(M^2,m^2,m^{\ast\,2}) = - \frac{M^{\ast\,2}}{m^2} \, 
\biggl[ \lambda_M \, - \, \frac{1}{32\,\pi^2} \biggr] 
- \frac{\mu^\ast}{8\pi^2} \, g(0,M,m,m^{\ast}) \, 
[ \, 1 \, - \, \Omega^2 \, ] \, r^{\ast\,3}\,, \\
&&\nonumber\\ 
&&I_{12}^{(1)}(t,M^2,m^2,m^{\ast\,2}) = 
-\frac{f\biggl(\displaystyle{\frac{t}{m^2}}\biggr)}{m^2} \, 
\biggl[ \lambda_M \, + \, \frac{1}{32\,\pi^2} \biggr] +   
\frac{M^{\ast\,2}}{32 \, \pi^2 \, M \, m^3} \, 
g\biggl(\frac{t}{m^2},M,m,m^{\ast}\biggr)\,, \\ 
&&\nonumber\\ 
&&I_{21}^{(1)}(t,M^2,m^2,m^{\ast\,2}) \, = \,  
\frac{f\biggl(\displaystyle{\frac{t}{m^2}}\biggr)}{m^2} \, 
\biggl[ \lambda_M \, + \, \frac{1}{32\,\pi^2} \biggr] 
+ \frac{1}{32 \, \pi^2 \, m^2} \, 
\biggl[ h_1\biggl(\frac{t}{M^2},M,m,m^{\ast}\biggr) \nonumber\\
&& \hspace*{3.7cm}+ \, 2 \,\biggl( \frac{1}{\mu^\ast} 
+ \Omega\biggr) \, h_2\biggl(\frac{t}{M^2},M,m,m^{\ast}\biggr) 
\biggr]\,, 
\en 
where 
$$\Omega = \frac{m^2 - m^{\ast\,2} - M^2}{2 m^{\ast} M}\,, 
\hspace*{.4cm}\mu = \frac{M}{m}\,, 
\hspace*{.4cm}\mu^\ast = \frac{M}{m^{\ast}}\,, 
\hspace*{.4cm}r^\ast = \frac{m^{\ast}}{m}\,.$$
The dimensionless functions $k$, $f$, 
$g$, $h_1$ and $h_2$ are given by the expressions 
\eq
& &k(s) = \int\limits_0^1 dx \, \ln (1 + x (1 - x) s) = 
\sqrt{\frac{4 + s}{s}} \, 
\ln \frac{\sqrt{4 + s } + \sqrt{s}}
{\sqrt{4 + s } - \sqrt{s}} - 2 \,, \\
&&\nonumber\\
& &f(s) = \int\limits_0^1 dx \, 
\frac{1}{1 + x (1 - x) s} = 
\frac{2}{\sqrt{ s (4 + s)}} \, 
\ln \frac{\sqrt{4 + s} + \sqrt{s}}
{\sqrt{4 + s} - \sqrt{s}} \,,\\
& &\nonumber\\
& & g(s,M,m,m^{\ast}) = \int\limits_0^1 dx \,  
\frac{\displaystyle{{\rm arccos}\biggl[ - r^\ast \, 
(\Omega + \mu^\ast)  
\frac{\sqrt{r^{\ast\,2} + x (1 - x) s}}{1 + x (1 -x) s}\biggr]}}  
{\displaystyle{[ 1 + x (1 - x) s) ] \,  
\sqrt{(1 - \Omega^2) r^{\ast\,2} + x (1 -x) s}}}\,,\\
& &\nonumber\\
& & h_1(s,M,m,m^{\ast}) = \int\limits_0^1 dx \, 
\frac{\displaystyle{{\rm ln}[ 1 + x (1 - x) s]}} 
{\displaystyle{1 \, + \, \mu^2 \, x (1 - x) \, s}}\,,\\
& &\nonumber\\
& & h_2(s,M,m,m^{\ast}) = \int\limits_0^1 dx \, 
{\rm arccos}\Biggl[ - \frac{\displaystyle{
[ \mu^\ast(1 + x (1 - x) s) + \Omega ] \, r^\ast}}
{\displaystyle{\sqrt{1 \, + \, \mu^2 \, x (1 - x) \, s} \, 
\sqrt{1 + s x (1 - x) }}} \Biggr]\nonumber\\
&&\hspace*{3cm}\times\frac{1}{[1 \, + \, \mu^2 \, x (1 - x) \, s] 
\sqrt{1 \, + \, \Omega^2 \, + \, x (1 - x) \, s}}\,.
\en

\section{Electromagnetic meson-cloud form factors} 

We explicitly show the contributions of the diagrams in Figs.~1 and 2 
to the Dirac and Pauli quark form factors. We use the following 
notations $f_D^{(i)}(t)$ and $f_P^{(i)}(t)$ where the superscript $i$ 
refers to the numbering of the
diagrams.  We expand 
the diagonal $3\times 3$ matrices $f_D^{(i)}(t)$ and $f_P^{(i)}(t)$ 
in the basis of the $3\times 3$ matrices $Q,\,I,$ and $\lambda_3$: 
\eq
Q = \left(\begin{array}{ccc}
2/3 & 0 & 0 \\
0 & -1/3 & 0 \\
0 & 0 & -1/3 
\end{array}\right),\,\,
I=\left(\begin{array}{ccc}
1 & 0 & 0 \\
0 & 1 & 0 \\
0 & 0 & 1 
\end{array}\right),\,\,
\lambda_3=\left(\begin{array}{ccc}
1 & 0 & 0 \\
0 & -1 & 0 \\
0 & 0 & 0 
\end{array}\right).
\en

\par
The contributions of the diagrams in Fig.1 to the 
Dirac form factors of the $u$, $d$ and $s$ quark are 
\eq
f_D^{(1)}(t) &=& Q \bar Z_q, \nonumber\\[3mm]
f_D^{(2)}(t) &=& f_D^{(4)}(t) \, = \, f_D^{(8)}(t) \, = \, f_D^{(12)}(t) 
\, = \,  0, \nonumber\\[3mm]
f_D^{(3)}(t) &=&  2 t \bar d_{10} Q \,, \nonumber\\[3mm]
f_D^{(i)}(t) &=& \sum\limits_{P = \pi, K, \eta} 
\lambda_i^P {\mathcal \epsilon}_i^P \,, \quad\quad 
i = 5,6,7,9,10,11 \,, 
\en
where $\bar Z_q$ is the finite part of the renomalization matrix 
$Z_q$ defined in Eq.~(\ref{Zq}),  
\eq
& & \lambda_5^\pi  = Q+\frac{1}{3}I-\lambda_3 \,, \,\,\, 
    \lambda_5^K    = -\frac{8}{3}Q-\frac{2}{9}I
                     +\frac{4}{3}\lambda_3 \,, \,\,\,
    \lambda_5^\eta = Q-\frac{1}{9}I-\frac{1}{3}\lambda_3 \,, 
\nonumber \\[3mm]
& & \lambda_6^\pi  = \lambda_7^\pi = \lambda_9^\pi = \lambda_3 \,, \,\,\, 
    \lambda_6^K    = \lambda_7^K = \lambda_9^K = 3Q - \lambda_3 \,, \,\,\, 
    \lambda_6^\eta = \lambda_7^\eta = \lambda_9^\eta = 0 \,, 
\nonumber\\[3mm]
& &\lambda_{10}^\pi  = c_6 \left(Q+\frac{1}{3}I-\lambda_3\right), \,\,\, 
\lambda_{10}^K    = c_6 \left(-\frac{8}{3}Q-\frac{2}{9}I 
                   + \frac{4}{3}\lambda_3\right), \,\,\, 
\lambda_{10}^\eta = c_6 \left(Q-\frac{1}{9}I-\frac{1}{3}\lambda_3\right)\,, 
\nonumber\\[3mm]
& &\lambda_{11}^\pi = Q\,, \quad\, \lambda_{11}^K = \frac{4}{3}Q\,, \quad 
\, \lambda_{11}^\eta = \frac{1}{3}Q\,. 
\en
\eq
{\mathcal \epsilon}_5^P &=& -\frac{g^2 m^2}{F_P^2}
\bigg\{ \bar I^{(1)}(M_P^2,m^2,m^2)+M_P^2 \bar I_{12}^{(0)}(t,M_P^2,m^2,m^2)
-2\bar I_{12}^{(2)}(t,M_P^2,m^2,m^2) \nonumber\\
&- & 8 m^2 \bar I_{12}^{(3)}(t,M_P^2,m^2,m^2) \bigg\}, \nonumber\\
{\mathcal \epsilon}_6^P &=& \frac{g^2}{F_P^2} \bigg\{ 
t \bar J^{(1)}(t,M_P^2)-4m^2 \bar I_{21}^{(2)}(t,M_P^2,m^2,m^2)
-16m^4 \bar I_{21}^{(3)}(t,M_P^2,m^2,m^2) \bigg\}, \nonumber\\
{\mathcal \epsilon}_7^P &=& -\frac{2 g^2 m^2}{F_P^2} 
\bar I^{(1)}(M_P^2,m^2,m^2), \nonumber\\
{\mathcal \epsilon}_9^P &=& - \frac{t}{F_P^2} \bar J^{(1)}(t,M_P^2,m^2,m^2), 
\nonumber\\
{\mathcal \epsilon}_{10}^P &=& - \frac{2 t g^2 m^2}{F_P^2} 
\bar I_{12}^{(3)}(t,M_P^2,m^2,m^2), \nonumber\\
{\mathcal \epsilon}_{11}^P &=& -\frac{3 c_2}{64 \pi^2 m F_P^2} M_P^4.
\en

The contributions of the diagrams in Fig.1 to the 
Pauli form factors of the $u$, $d$ and $s$ quark are 
\eq
f_P^{(1)}(t) &=& f_P^{(7)}(t) \, = \, f_P^{(8)}(t) \, = \, f_P^{(9)}(t) 
\, = \,  0, \nonumber\\
f_P^{(2)}(t) &=& \bar Z_q c_6 Q \,, \nonumber\\
f_P^{(3)}(t) &=& - 2 t \bar d_{10} Q\,, \nonumber\\
f_P^{(4)}(t) &=& -4 m t \bar e_{10}Q  
- 16m \biggl[ M^2 \bar e_{6} Q +  
\frac{1}{3} (M_K^2 - M_\pi^2) ( \bar e_{7} [ Q - \lambda_3 - I/3 ]
- \bar e_{8} I ) \biggr]\,, 
\nonumber\\
f_P^{(i)}(t) &=& \sum\limits_{P = \pi, K, \eta} 
\lambda_i^P m_i^P \,,  \quad\quad i = 5,6,10,11,12 \,, 
\en
where
\eq
& &\lambda_{12}^\pi= \lambda_3 \,, \quad  
\lambda_{12}^K = 3Q - \lambda_3  \,, 
\quad \lambda_{12}^\eta = 0 \,, \nonumber\\ 
\en
\eq 
m_5^P &=& -\frac{8 g^2 m^4}{F_P^2} \bar I_{12}^{(3)}(t,M_P^2,m^2,m^2), 
\nonumber\\
m_6^P &=& \frac{16 g^2 m^4}{F_P^2} \bar I_{21}^{(3)}(t,M_P^2,m^2,m^2), 
\nonumber\\
m_{10}^P &=& \frac{g^2 m^2}{F_P^2} \bigg\{ \bar I^{(1)}(M_P^2,m^2,m^2)
-M_P^2 \bar I_{12}(t,M_P^2,m^2,m^2)+4 \bar I_{12}^{(2)}(t,M_P^2,m^2,m^2) 
\nonumber\\
&+& 2 t \left(\bar I_{12}^{(3)}(t,M_P^2,m^2,m^2)
-\bar I_{12}^{(4)}(t,M_P^2,m^2,m^2) \right) \bigg\}, \nonumber\\
m_{11}^P &=& \frac{3 c_2}{64 \pi^2 m F_P^2} M_P^4, \nonumber\\
m_{12}^P &=& \frac{4 m t c_4}{F_P^2} \bar J^{(1)}(t,M_P^2,m^2,m^2).
\en
The contributions of the diagrams in Fig.2 (vector mesons contributions)  
to the Dirac and Pauli form factors of the $u$, $d$ and $s$ quark are 
\eq
f_D^{(2^\ast)}(t)&=& f_D^{(11^\ast)}(t) \, = \, 
f_D^{(12^\ast)}(t) \, = 0
 \,, \nonumber\\ 
f_D^{(3^\ast)}(t)&=& - \frac{t}{2}\, 
\biggl\{ \lambda_\rho \, D_\rho(t) \, + \, 
\frac{1}{3} \lambda_\omega \, D_\omega(t) \, + \, 
\frac{2}{3} \lambda_\phi \, D_\phi(t) \,\biggr\} \,, \nonumber\\ 
f_D^{(10^\ast)}(t)&=&\frac{k_V}{2} \,  \biggl\{
[\lambda_\omega D_\omega(t) - \lambda_\rho D_\rho(t)]
\epsilon_{10}^\pi 
- \frac{4}{3} \, [\lambda_\phi D_\omega(t) 
+ \lambda_\omega D_\phi(t)] \epsilon_{10}^K  \nonumber\\
&+& \frac{1}{9} [ 3 \lambda_\rho D_\rho(t) +
\lambda_\omega D_\omega(t) + 
8 \lambda_\phi D_\phi(t)] \epsilon_{10}^\eta \biggr\}\,, \nonumber\\
f_P^{(2^\ast)}(t) &=& \frac{k_V}{2} \, 
\biggl\{\bar Z \, \lambda_\rho \, D_\rho(t) + 
\frac{1}{3} \, \bar Z \, \lambda_\omega \, D_\omega(t) + 
\frac{2}{3} \, \bar Z_s \lambda_\phi D_\phi(t) \biggr\}\,, \nonumber\\ 
f_P^{(3^\ast)}(t)&=& f_P^{11^\ast}(t) \, = 0 \,, \nonumber\\ 
f_P^{(10^\ast)}(t) &=& \frac{k_V}{2} \, \biggl\{ 
[ \lambda_\omega D_\omega(t) - \lambda_\rho D_\rho(t)] m_{10}^\pi 
- \frac{4}{3} [ \lambda_\phi D_\omega(t) + \lambda_\omega D_\phi(t) ]  
m_{10}^K \nonumber\\
&+& \frac{1}{9} [ 3 \lambda_\rho \, D_\rho(t) \, + \, \lambda_\omega 
\, D_\omega(t) \, + \, 8 \lambda_\phi \, D_\phi(t)] m_{10}^\eta 
\biggr\}\,, \nonumber\\
f_P^{(12^\ast)}(t) &=&  - \, \frac{t}{2} \, k_V  
\biggl\{ \lambda_\rho D_\rho(t) 
\left[\frac{2}{F_\pi^2}\bar J^{(1)}(t,M_\pi^2)+
\frac{1}{F_K^2}\bar J^{(1)}(t,M_K^2)\right] \nonumber\\ 
&+& \frac{\lambda_\omega}{F_K^2} D_\omega(t)  \bar J^{(1)}(t,M_K^2)
+\frac{2\lambda_\phi}{F_K^2} D_\phi(t)\bar J^{(1)}(t,M_K^2)\biggr\} \,, 
\en 
where $\lambda_\rho = \lambda_3$\,, $\;$  
$\lambda_\omega = 2 Q + 2 I/3  - \lambda_3$\,, $\;$ 
$\lambda_\phi = 2 Q - I/3  - \lambda_3$  
and $\;$ $D_V(t) = 1/(M_V^2 + t)$.

\vspace*{.2cm} 

In the expressions above, the convergent parts of the structure integrals are
\eq
\bar J^{(1)}(t,M^2) &=& -\frac{1}{576\pi^2 t} 
\left[t \, + \, 3k(t)\left(t+4M^2 \right)\right]\,, \nonumber\\
\bar I^{(1)}(M^2,m^2,m^2) &=& 
\frac{\mu^3}{64\pi^2} \, [ \mu \, - \, g(0,M,m,m) ( 4-\mu^2 ) ]\,, 
\nonumber\\
\bar I_{12}(t,M^2,m^2,m^2) &=& \frac{1}{32\pi^2 m^2} 
[ g(t,M,m,m) \mu - f(t) ]\,, \nonumber\\
\bar I_{12}^{(2)}(t,M^2,m^2,m^2) &=& 
\frac{\mu^3}{256\pi^2}\biggl[ - 2 \mu \Dt  
\, + \, g(t,M,m,m) (4 - \mu^2  \Dt ) 
\, + \, g(0,M,m,m) (4 - \mu^2) \Dt \biggr], \nonumber\\
\bar I_{12}^{(3)}(t,M^2,m^2,m^2) &=& 
\frac{\mu^3 \, \Dt}{1024\pi^2 m^2} \biggl[ 
2 \mu \left( 1 + 2 \Dt \right)
\, - \, \mu f(t) \, - \, g(t,M,m,m) ( 4 - 3 \mu^2 \Dt) \nonumber \\
&+& g(0,M,m,m) ( 4 - \mu^2) ( 2 + 3 \Dt) \biggr]\,, 
\nonumber\\
\bar I_{12}^{(4)}(t,M^2,m^2,m^2) &=& \frac{\mu^3}{512\pi^2 m^2 \tau} 
\biggl[ \mu \tau \Dt \, - \, 2 g(t,M,m,m) (1 - \mu^2 \Dt) \nonumber\\ 
&-& g(0,M,m,m) (8 - 4 \mu^2 + (\mu^2 - 4) \Dt) \biggr] \,, \nonumber\\
\bar I_{21}^{(2)}(t,M^2,m^2,m^2) &=& 
\frac{1}{512\pi^2} \biggl[ 2 \Dt \mu ( \mu^2 - 2) 
( 2 \mu - g(0,M,m,m) (4 - \mu^2 )) \, + \, 
k(t) (2 + (\mu^2 - 2) \Dt)  \nonumber\\
&+& 2 H_{12}(t,M,m,m) \, ( 4 - (2- \mu^2)^2 \Dt ) \biggr]\,, \nonumber\\
\bar I_{21}^{(3)}(t,M_P^2,m^2,m^2) &=& \frac{\Dt}{1024\pi^2 m^2}
\biggl[ 2( 4 + 6 \mu^2 - \mu^4 + 2 (\mu^2 - 2) (\mu^2 +1) \Dt ) 
- 6 k(t) (2 + (\mu^2 - 2) \Dt)  \nonumber\\[1mm]
&+& 2 f(t) (2 + (\mu^2 -2) \Dt)^2  + g(0,M,m,m) \mu (4-\mu^2) 
[ 2 (\mu^2 + 2) + 3 (\mu^2 - 2) \Dt ] \nonumber\\[1mm] 
&-& H_{12}(t,M,m,m) \, ( 12 - 8 \mu^2 + 2 \tau - 3 (\mu^2 - 2)^2 D_\tau ) 
\biggr]\,.
\en
where 
\eq 
H_{12}(t,M,m,m) &=& h_1(t,M,m,m)+h_2(t,M,m,m) 
\left(\frac{2-\mu^2}{\mu}\right) \, \\
D_\tau &=& \frac{1}{1 + \tau/4} \,. \nonumber 
\en 

In the practical calculation, we keep the exact form of the function $f(t)$, 
while we expand the functions $k(t)$, $g(t,M,m,m)$, 
$h_1(t,M,m,m)$ and $h_2(t,M,m,m)$ in powers of $\theta$ and $\mu$ with  
$\theta=t/M^2$ and $\mu=M/m$. 

Finally, we keep our expressions for ${\mathcal \epsilon}_i^P$,    
and $m_i^P$ up to $O(\theta_P,\mu_P^4)$ which are explicitly
given as
\eq\label{expressions_e_m}
{\mathcal \epsilon}_5^P &=& \frac{g^2 m^2 \mu_P^2}{32\pi^2 F_P^2}
\bigg[f(t)\left(1-\frac{3}{4}\mu_P^2 D_\tau^2\right)+\pi\mu_P
\left(1-\frac{D_\tau}{2}-2D_\tau^2\right)
+\frac{\mu_P^2}{2}(1-D_\tau-\frac{3}{2}D_\tau^2)\bigg]\,, \\[2mm]
{\mathcal \epsilon}_6^P &=& \frac{g^2 m^2}{16 \pi^2 F_P^2}\bigg[\mu_P^2
\left(1-\frac{5\pi}{2}\mu_P-2\mu_P^2\right)+\tau\left(\frac{7}{2}
-\frac{35\pi}{24}\mu_P-\frac{13}{6}\mu_P^2+\frac{105\pi}{64}
\mu_P^3+\frac{107}{72}\mu_P^4\right)\bigg]\,, \nonumber\\[2mm]
{\mathcal \epsilon}_7^P &=& \frac{g^2 m^2 \mu_P^3}{16 \pi^2 F_P^2}
\left(\pi+\frac{\mu_p}{2}\right)\,, \nonumber\\[2mm]
{\mathcal \epsilon}_9^P &=& \frac{t}{192\pi^2 F_P^2}, \nonumber\\[2mm]
{\mathcal \epsilon}_{10}^P &=& \frac{g^2 m^2 \mu_P^3}{32\pi^2 F_P^2} 
(1-D_\tau)\bigg[\pi(1+2D_\tau)+\frac{\mu_P}{2}
\left(1+\frac{3}{2}D_\tau\right)+\frac{3}{4}\mu_P D_\tau f(t)\bigg]\,, 
\nonumber\\[2mm]
{\mathcal \epsilon}_{11}^P &=& -\frac{3 c_2 m^3 \mu_P^4}{64\pi^2 F_P^2}\,, 
\nonumber\\[2mm]
m_5^P &=& \frac{g^2 m^2 \mu_P^3}{16\pi^2 F_P^2}\bigg\{\frac{3}{8} \mu_P 
f(t) D_\tau^2 + D_\tau\bigg[\pi\left(\frac{1}{2}+D_\tau\right)+
\frac{3}{8}\left(\frac{2}{3}+D_\tau\right)\bigg]\bigg\}\,, \nonumber\\[2mm]
m_6^P &=& -\frac{g^2 m^2}{\pi F_P^2}\bigg[\frac{\mu_P}{8}
\left(\pi+\mu_P-\frac{15}{8}\pi\mu_P^2-\frac{4}{3}\mu_P^3\right)
\nonumber\\[2mm]
& & +\frac{\tau}{96\mu_P}\left(\pi+8\mu_P-\frac{105\pi}{8}\mu_P^2
-\frac{52}{3}\mu_P^3+\frac{1575\pi}{128}\mu_P^4+\frac{107}{10}
\mu_P^5\right)\bigg]\,, \nonumber\\[2mm]
m_{10}^P &=& \frac{g^2 m^2 \mu_P^2}{32\pi^2 F_P^2}\bigg\{f(t) 
\left[1-\frac{3}{4}D_\tau\mu_P^2(1-D_\tau)\right]-\pi\mu_P
\left(1+\frac{1}{2}D_\tau-2D_\tau^2\right) \nonumber\\[2mm]
& & -\frac{1}{2}\mu_P^2\left(1+\frac{1}{2}D_\tau
-\frac{3}{2}D_\tau^2\right)\bigg\}, \nonumber\\[2mm]
m_{11}^P &=& \frac{3 c_2 m^3 \mu_P^4}{64\pi^2 F_P^2}\,, \nonumber\\[2mm]
m_{12}^P &=& \frac{c_4 m t}{48\pi^2 F_P^2} \,. \nonumber 
\en
After absorbing divergences coming from 
vector meson-exchange diagrams the renormalized LECs 
are written as 
\eq
\bar d_{10}&=& d_{10}^r(\mu) - 
\frac{\beta_{d_{10}}}{32\pi^2 \, F_K^2} 
\, \ln\frac{M_K^2}{\mu^2} \,, \nonumber\\   
\bar e_{10} &=& e_{10}^r(\mu) - \frac{\beta_{e_{10}}}{32\pi^2 \, F_K^2} 
\, \ln\frac{M_K^2}{\mu^2} \,, \nonumber\\               
\bar e_6 &=& e_6^r(\mu) - \frac{\beta_{e_6}}{32\pi^2  \, F_\pi^2} 
\, \ln\frac{M_\pi^2}{\mu^2}\,, \nonumber\\
\bar e_7 &=& e_7^r(\mu) - \frac{\beta_{e_7}}{32\pi^2  \, F_\pi^2} 
\, \ln\frac{M_\pi^2}{\mu^2}\,, \nonumber\\
\bar e_8 &=& e_8^r(\mu) - \frac{\beta_{e_8}}{32\pi^2  \, F_\pi^2} 
\, \ln\frac{M_\pi^2}{\mu^2} 
\en
where the $\beta$-coefficients with taking into account of 
vector mesons are given by 
\eq
\beta_{d_{10}} &=& - \frac{1+5 g^2}{4} \,,\nonumber\\
\beta_{e_{10}} &=&  \frac{1}{8\, m} \left( 1 \, - \, 7 g^2 \, 
+ \, k_V \, + \, 4 m c_4 \right) \,,\nonumber\\
\beta_{e_6} &=&  - \frac{35}{144m} ( c_6 g^2 + k_V) 
- \frac{1}{16m} \, (c_6 - 4 c_4m) - \frac{5g^2}{9m}  \,, \nonumber\\
\beta_{e_7} &=&  - \frac{11}{48m} ( c_6 + k_V) \, g^2 
- \frac{3}{16m} \, (c_6 - 4 c_4m) - \frac{7g^2}{6m}  \,, \nonumber\\
\beta_{e_8} &=&  - \frac{1}{16m} ( c_6 + k_V) g^2 
+ \frac{1}{16m} \, (c_6 - 4 c_4m) + \frac{g^2}{4m} 
\en 
and
\eq 
M^2 = M_\pi^2 -  M_K^2 + \frac{3}{2} M_\eta^2 \,. 
\en 
\section{Meson-nucleon sigma-terms} 

Two important identities are involved in the evaluation of the meson-nucleon 
sigma-terms. In particular, the derivatives with respect to the 
current quark masses are equivalent to the ones   
with respect to the meson masses~\cite{Gasser:1980sb}: 
\begin{eqnarray}
\hat{m}\frac{\partial}{\partial\hat{m}} \Sigma_q &=& 
M_\pi^2 \, \biggl(\frac{\partial}{\partial M_\pi^2} + 
\frac{1}{2}\frac{\partial}{\partial M_K^2} + 
\frac{1}{3}\frac{\partial}{\partial M_\eta^2} \biggr) \Sigma_q \,, \\[2mm]
\hat{m}_s\frac{\partial}{\partial\hat{m}} \Sigma_q &=& M_{K\eta}^2 \, 
\biggl(\frac{\partial}{\partial M_\pi^2} + 
\frac{1}{2}\frac{\partial}{\partial M_K^2} + 
\frac{1}{3}\frac{\partial}{\partial M_\eta^2} \biggr) \Sigma_q \,, \\[2mm]
\hat{m}\frac{\partial}{\partial\hat{m}_s} \Sigma_q &=& M_\pi^2 \, 
\biggl(\frac{1}{2} \frac{\partial}{\partial M_K^2} + 
\frac{2}{3} \frac{\partial}{\partial M_\eta^2} \biggr) \Sigma_q \,, \\[2mm]
\hat{m}_s\frac{\partial}{\partial\hat{m}_s} \Sigma_q &=& M_{K\eta}^2 \, 
\biggl(\frac{\partial}{\partial M_K^2} + 
\frac{4}{3} \frac{\partial}{\partial M_\eta^2} \biggr) \Sigma_q \, ,
\end{eqnarray} 
with $M_{K\eta}^2 = - M_K^2 + 3M_\eta^2/2\,.$

Below we give the exact expression for the typical type of derivatives: 
\eq 
\hat{m}\frac{\partial}{\partial\hat{m}} \bar\Sigma &=&  \hat{m} -  
\frac{9 g^2 M_\pi^2}{64 \pi}
\biggl\{ \frac{M_\pi}{F_\pi^2} + \frac{M_K}{3 \, F_K^2} 
+ \frac{M_\eta}{27 \, F_\eta^2} \biggr\} 
- \frac{3 g^2 M_\pi^2}{32 \pi^2 m} 
\biggl\{ \frac{M_\pi^2}{F_\pi^2} + \frac{M_K^2}{3 \, F_K^2} 
+ \frac{M_\eta^2}{27 \, F_\eta^2} \biggr\} \nonumber\\[2mm]
&-& 4 c_1 M_\pi^2 + \frac{3c_2}{64 \pi^2} 
\biggl\{ \frac{M_\pi^4}{F_\pi^2} + \frac{2}{3} \frac{M_K^4}{F_K^2}  
+ \frac{M_\eta^4}{9 \, F_\eta^2}  \biggr\} 
- \frac{2}{3} c_5 M_\pi^2 \nonumber \\[2mm] 
&+& 2 \bar e_1 M_\pi^2 M^2 
- \frac{\bar e_3}{6} M_\pi^2 (M_K^2 - M_\pi^2) 
+ \frac{5}{12} \bar e_4  M_\pi^4 
- \frac{\bar e_5}{36} M_\pi^2 (M_K^2 - M_\pi^2) \,. 
\en 
One should note that there are additional useful relations between different 
sigma-terms~\cite{PCQM2}. In particular, with the definitions of 
$y_N$ and $\sigma_{KN}^{I=1}$ we can relate $KN$ and $\eta N$ sigma-terms 
to the $\pi N$ sigma-term as 
\eq 
\sigma_{KN}^u=\sigma_{\pi N}(1+y_N)\frac{\hat m+m_s}{4\hat m}
+\sigma_{KN}^{I=1}\,, \hspace*{1cm} 
\sigma_{KN}^d=\sigma_{K N}^u - 2 \sigma_{KN}^{I=1}\,, \hspace*{1cm}  
\sigma_{\eta N}=\sigma_{\pi N} \frac{\hat m + 2 y_N m_s}{3\hat m}\,. 
\nonumber
\en

\newpage
\noindent 
\begin{center}
{\bf Table 1.} SU(6) couplings $\alpha_E^{Bi}$ and $\alpha_M^{Bi}$. \\

\vspace*{.3cm}
\def\arraystretch{1.5}
\begin{tabular}{|c|c|c|c|c|c|c|}
\hline
 & $\alpha_E^{Bu}$\,\,\, & $\alpha_E^{Bd}$\,\,\, & $\alpha_E^{Bd}$\,\,\, 
& $\alpha_M^{Bu}$\,\,\, & $\alpha_M^{Bd}$\,\,\, & $\alpha_M^{Bs}$\,\,\, \\
\hline
$p$ & 2 & 1 & 0 & $\frac{4}{3}$ & -$\frac{1}{3}$ & 0 \\
$n$ & 1 & 2 & 0 & -$\frac{1}{3}$ & $\frac{4}{3}$ & 0 \\
$\Lambda^0$ & 1 & 1 & 1 & 0 & 0 & 1 \\
$\Sigma^+$ & 2 & 0 & 1 & $\frac{4}{3}$ & 0 & -$\frac{1}{3}$ \\
$\Sigma^-$ & 0 & 2 & 1 & 0 & $\frac{4}{3}$ & -$\frac{1}{3}$ \\
$\Xi^-$ & 0 & 1 & 2 & 0 & -$\frac{1}{3}$ & $\frac{4}{3}$ \\
$\Xi^0$ & 1 & 0 & 2 & -$\frac{1}{3}$ & 0 & $\frac{4}{3}$ \\
$\Sigma^0 \Lambda^0$ & 0 & 0 & 0 & $\frac{1}{\sqrt{3}}$ 
& -$\frac{1}{\sqrt{3}}$ & 0 \\
\hline
\end{tabular}
\end{center}

\vspace*{2cm}
\noindent 
\begin{center} 
\def\arraystretch{1.5}
{\bf Table 2.} Magnetic moments of the baryon octet 
(in units of the nucleon magneton $\mu_N$) 

\vspace*{.3cm}
\begin{tabular}{|c||c|c|c||c|c|c||c|}
\hline
& \multicolumn{3}{c||}{Set I} & \multicolumn{3}{c||}{Set II} & Exp. \\
\cline{2-7}
& 3q & Meson Cloud & Total & 3q & Meson Cloud & Total & \\
\hline\hline
$\mu_p$ &\, 2.357 \,&\, 0.436 \,&\, 2.793 \,&\, 2.357 \,&\, 0.436 \,
&\, 2.793 \,&\, 2.793 \\
$\mu_n$ & -1.571 & -0.342 & -1.913 & -1.571 & -0.342 & -1.913 & -1.913 \\
$\mu_{\Lambda^0}$ & -0.786 & 0.173 & -0.613 & -0.518 & -0.095 & -0.613 
&-0.613 $\pm$ 0.004 \\
$\mu_{\Sigma^+}$ & 2.357 & 0.317 & 2.674 & 2.085 & 0.373 & 2.458 
& 2.458 $\pm$ 0.010\\
$\mu_{\Sigma^0}$ & 0.786 & 0.005 & 0.791 & 0.570 & 0.073 & 0.643 & - \\
$\mu_{\Sigma^-}$ & -0.786 & -0.306 & -1.092 & -0.935 & -0.225 & -1.160 
& -1.160 $\pm$ 0.025\\
$\mu_{\Xi^0}$ & -1.571 & 0.136 & -1.435 & -1.058 & -0.192 & -1.250 
&-1.250 $\pm$ 0.014 \\
$\mu_{\Xi^-}$ & -0.7855 & 0.2921 & -0.4934 & -0.5580 & -0.0927 
& -0.6507 & -0.6507 $\pm$ 0.003 \\
$|\mu_{\Sigma^0 \Lambda^0}|$ & 1.36 & 0.27 & 1.63 & 1.34 & 0.27 
& 1.61 & 1.61 $\pm$ 0.08 \\
\hline
\end{tabular}
\end{center}

\vspace*{1cm}
\noindent 
\begin{center} 
\def\arraystretch{1.5}
{\bf Table 3.} Comparison of the parameters of the original
phenomenological ansatz in 
Ref.~\cite{Friedrich:2003iz} to an equivalent set, which is
obtained by fitting to our results for the
charge form factor of the proton.

\vspace*{.3cm}
\begin{tabular}{|c|c|c|}
\hline
  &\,\,\,\,\,\, Ref.~\cite{Friedrich:2003iz} \,\,\,\,\,\,& Present result \\
\hline
$a_{10}$ & 1.041 & 1.053 \\
$a_{11}$ & 0.765 & 0.768 \\
$a_{20}$ & -0.041 & -0.053 \\
$a_{21}$ & 6.2 & 4.7 \\
$a_b$ & -0.23 & -0.38 \\
$Q_b$ & 0.07 & 0.14 \\
$\sigma_b$ & 0.21 & 0.20 \\
\hline
\end{tabular}
\end{center}

\newpage
\begin{center}
\epsfig{figure=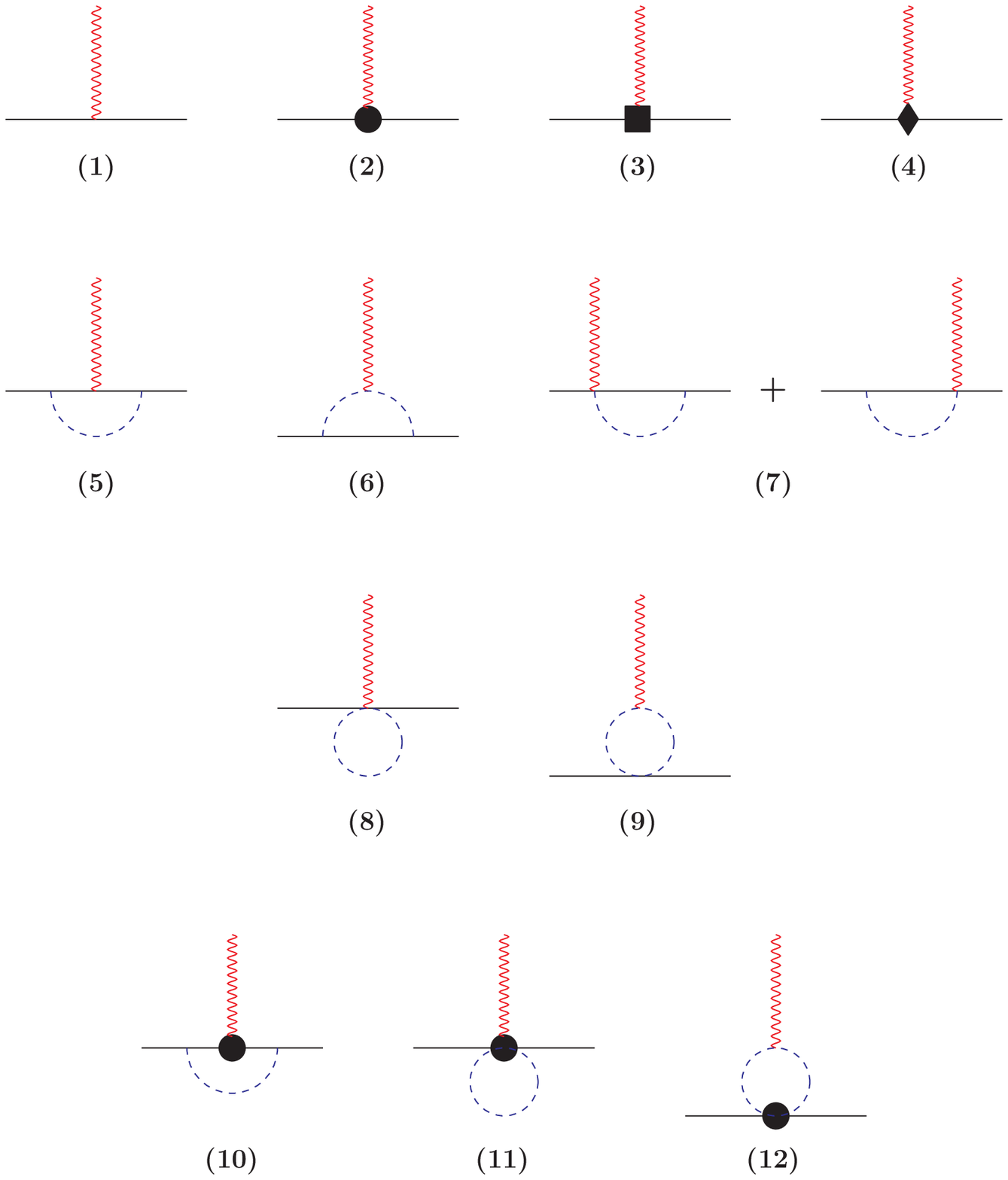,height=17cm}
\end{center}
\vspace*{0.5cm}
\noindent 
{\bf Fig. 1.} {\em Diagrams including pseudoscalar meson contributions  
to the EM quark transition operator up to fourth order.
Solid, dashed and wiggly lines refer to quarks, pseudoscalar mesons 
and the electromagnetic field, respectively. Vertices denoted by a black 
filled circle, box and diamond correspond to insertions from the second, 
third and fourth order chiral Lagrangian.   
\label{fig1}}

\newpage
\begin{center} 
\epsfig{figure=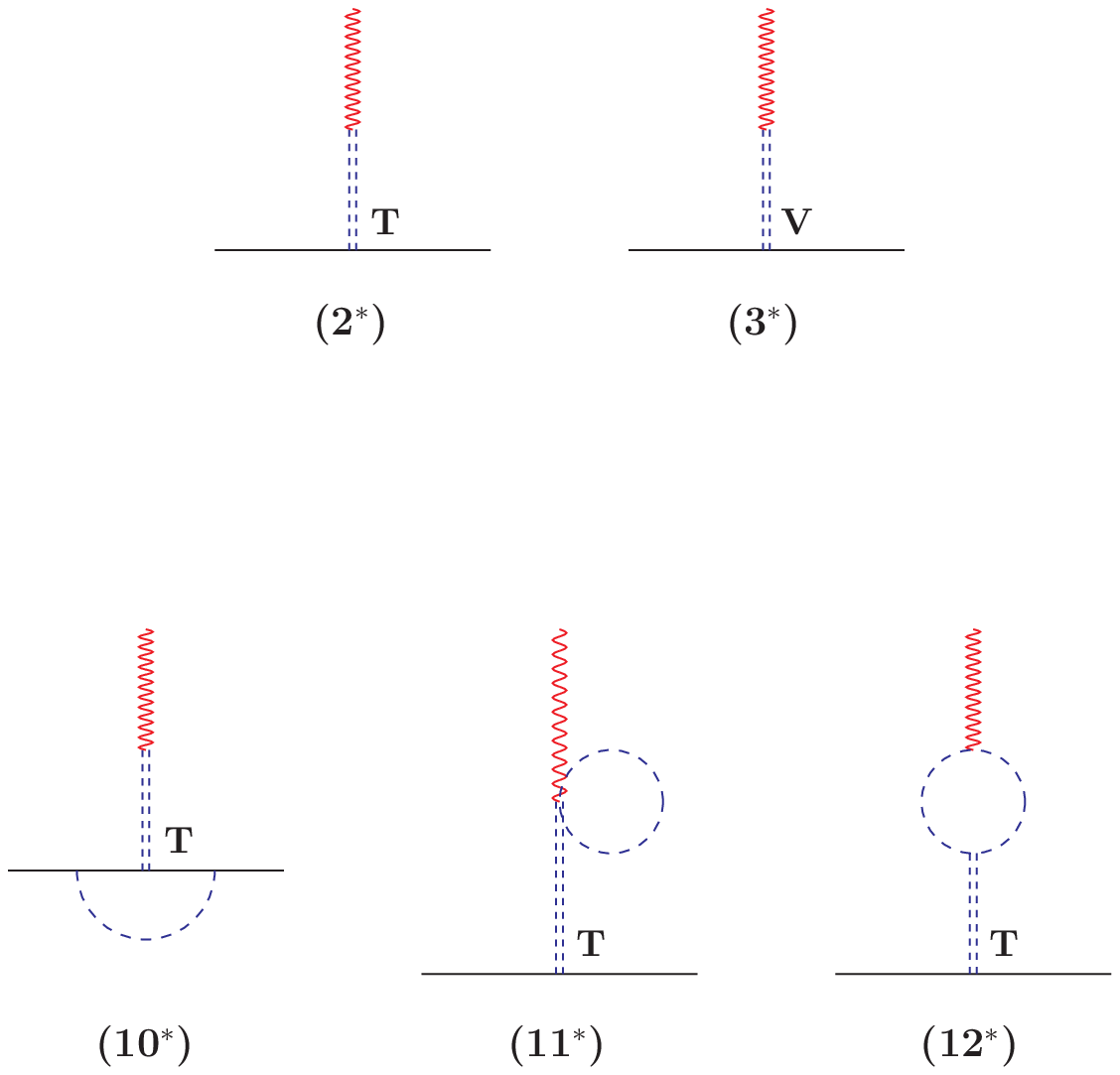,height=10cm} 
\end{center} 
\vspace*{0.5cm}
\noindent {\bf Fig. 2.} {\em Diagrams including vector-meson contributions 
to the EM quark transition operator. 
Double-dashed lines correspond to vector mesons. 
The symbols {\bf V} and {\bf T} refer to the vectorial 
and tensorial couplings of vector mesons to quarks. 
\label{fig2}}
\vspace*{0.5cm}
\begin{center}
\label{diagram_Vqq}
\epsfig{figure=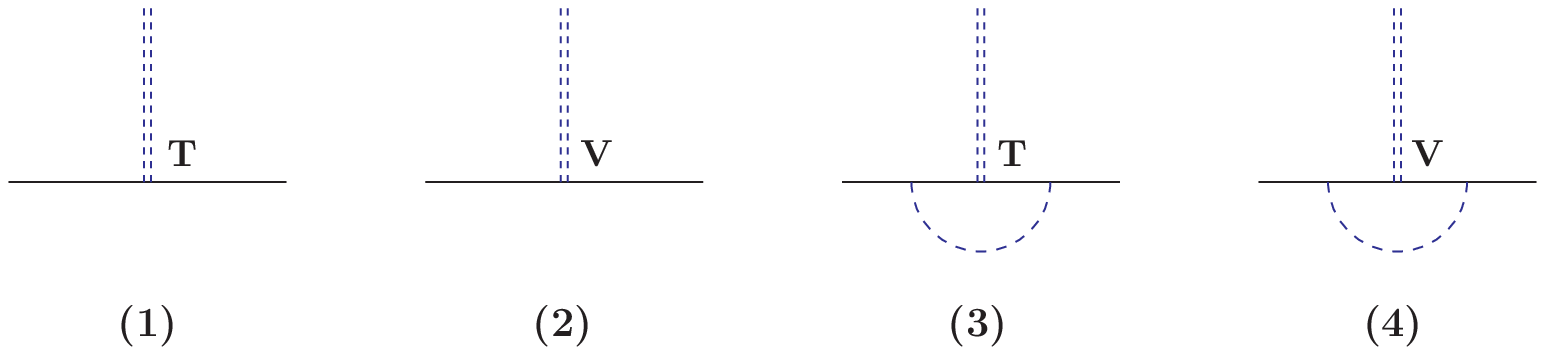,height=3.2cm}
\end{center} 
\vspace*{0.5cm}
\noindent {\bf Fig. 3.} 
{\em Diagrams contributing to the quark operator describing 
stong interaction of vector mesons to quarks. 
The symbols {\bf V} and {\bf T} refer to the vectorial 
and tensorial couplings of vector mesons to quarks. 
\label{fig3}}
\vspace*{0.5cm}
\begin{center} 
\epsfig{figure=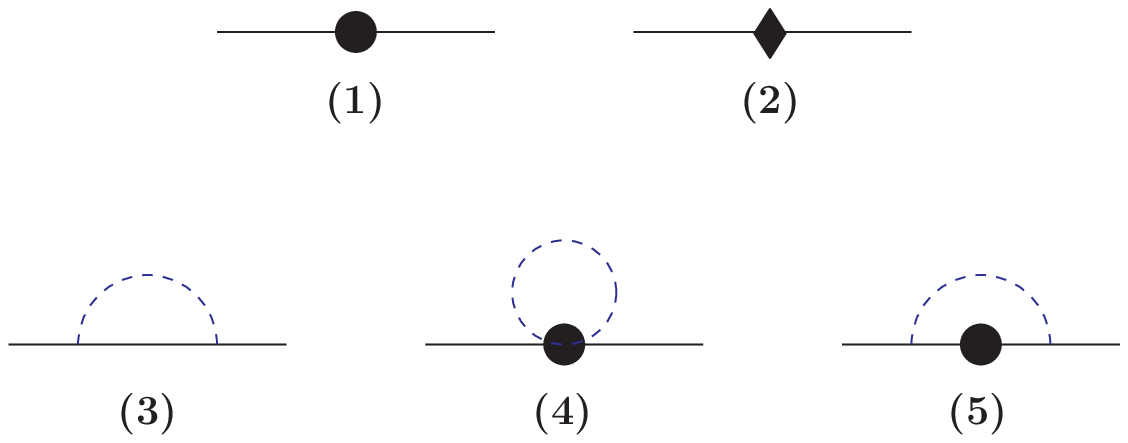,height=4.cm} 
\end{center} 
\vspace*{0.5cm}
\noindent {\bf Fig. 4.} {\em Diagrams contributing to the 
mass operator of the quark at one loop. Vertices denoted by a black 
filled circle and diamond correspond to insertions from the second 
and fourth order chiral Lagrangian.   
\label{fig4}}

\newpage
\vspace*{-1cm}
\begin{center}
\epsfig{figure=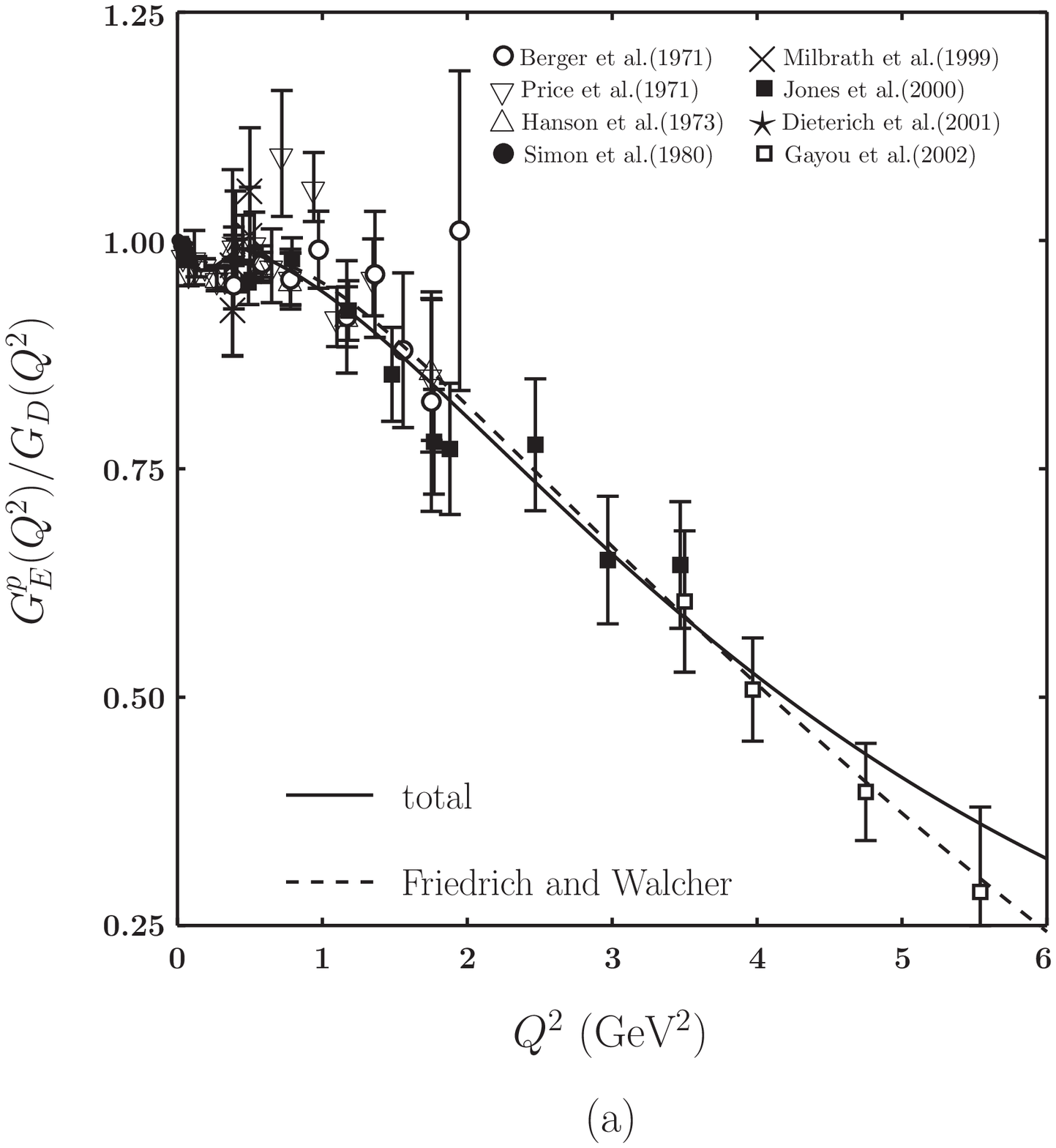,height=8.8cm}
\epsfig{figure=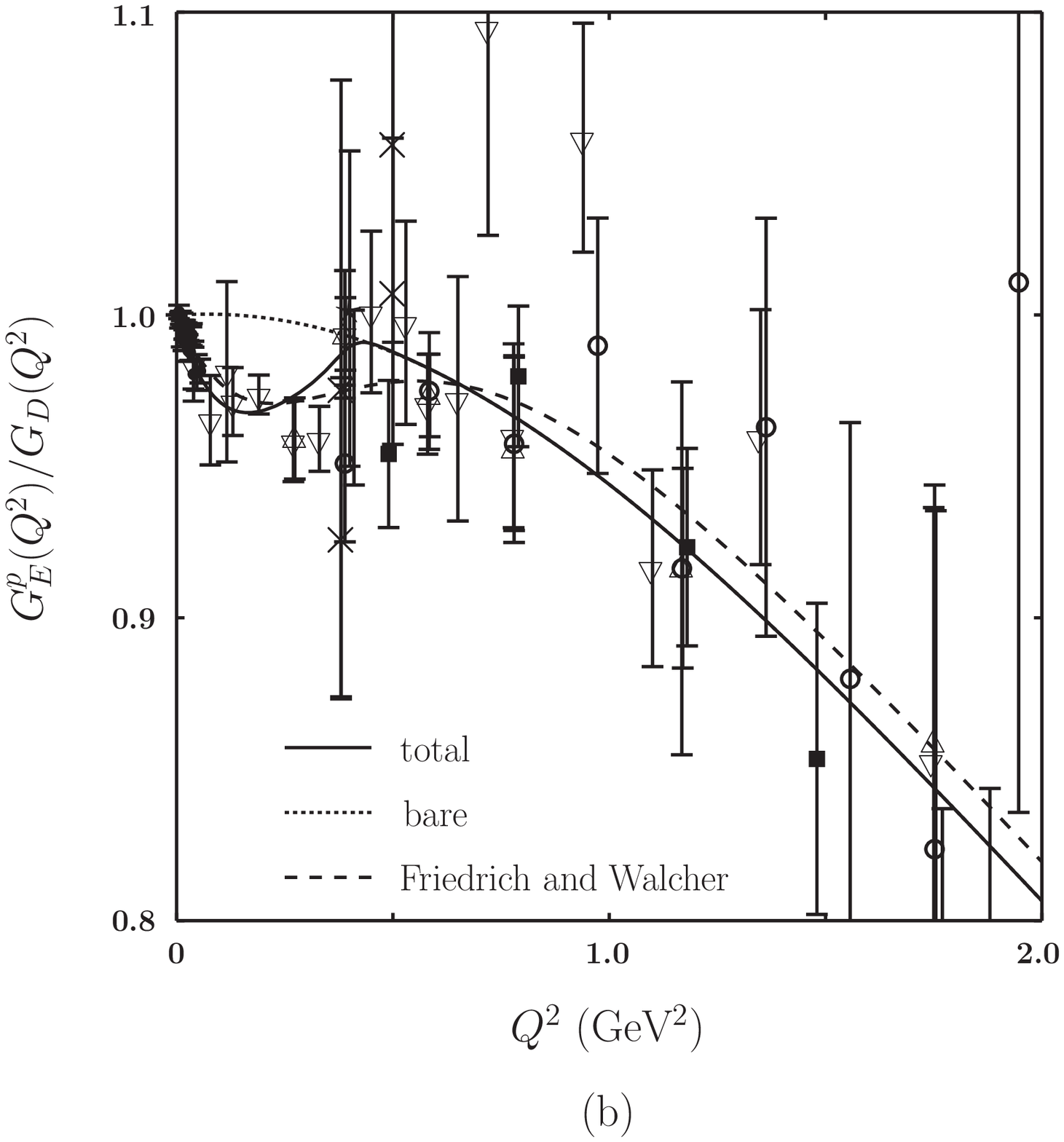,height=8.8cm}
\end{center}
\vspace*{0.5cm}
{\bf Fig. 5.} {\em Ratio $G_E^p(Q^2)/G_D(Q^2)$ : (a) Overall range, 
(b) Up to $Q^2=2\,\,\,{\rm GeV}^2$, the solid line is the total 
contribution, the dotted line is the bare contribution and the 
dashed line is the result reported in Ref.~\cite{Friedrich:2003iz}. 
Experimental data are taken from 
Refs.~\cite{Simon:1980hu,Price:1971zk,Berger:1971kr,Hanson:1973vf, 
Milbrath:1997de,Dieterich:2000mu,Jones:1999rz,Gayou:2001qd}.}
\vspace*{1.0cm}
\begin{center}
\epsfig{figure=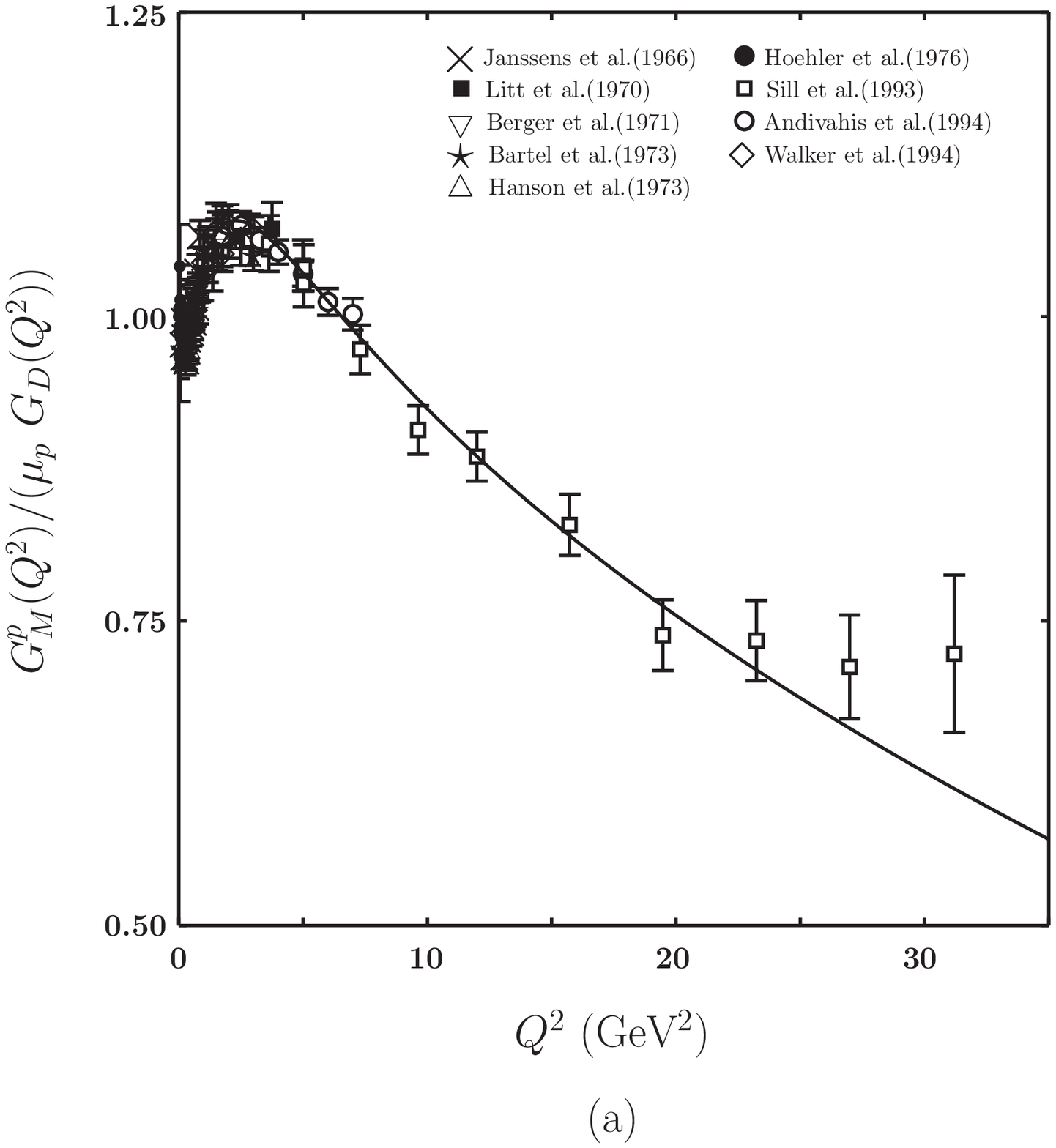,height=8.8cm}
\epsfig{figure=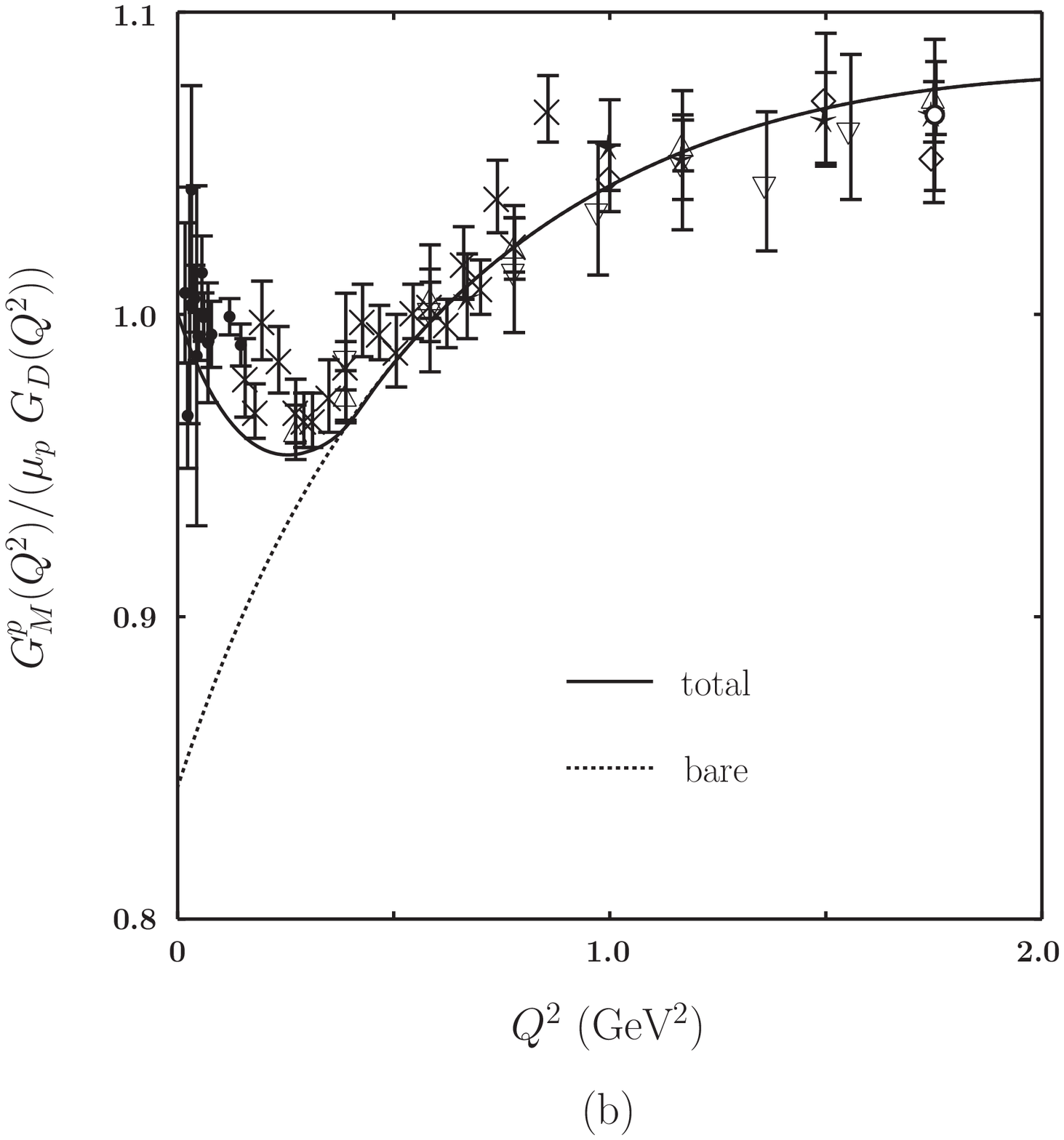,height=8.8cm} 
\end{center} 
\vspace*{0.5cm}
{\bf Fig. 6.} {\em Ratio $G_M^p(Q^2)/(\mu_p\,G_D(Q^2))$ : (a) Overall range, 
(b) Up to $Q^2=2\,\,\,{\rm GeV}^2$, the solid line is the total 
contribution and the dotted line is the bare contribution. Experimental 
data are taken from Refs.~\cite{Berger:1971kr,Hanson:1973vf,Hohler:1976ax,Janssens:1966,Bartel:1973rf,Walker:1993vj,Andivahis:1994rq,Litt:1969my,Sill:1992qw}.} 

\newpage
\vspace*{-1cm}
\begin{center}
\epsfig{figure=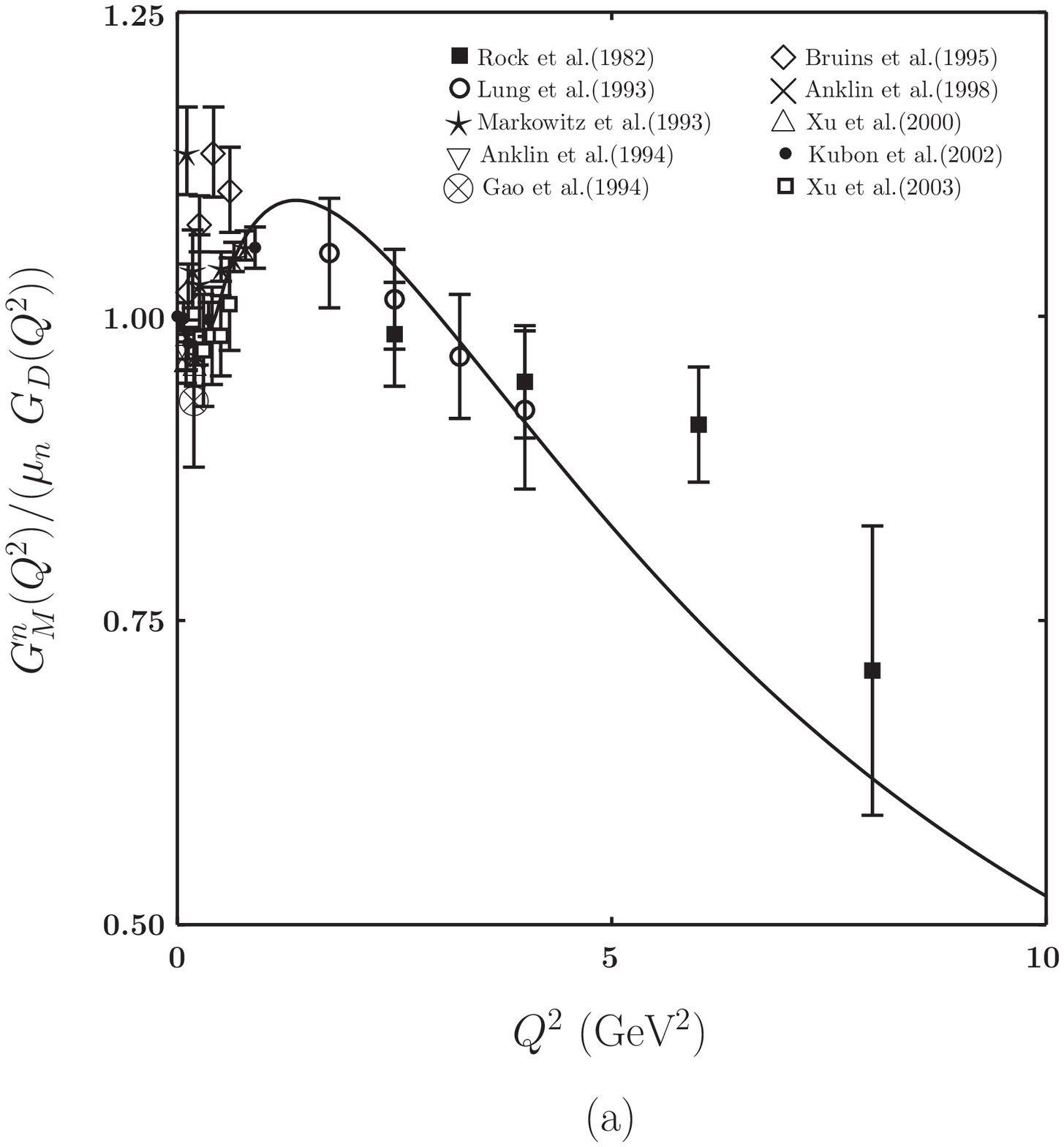,height=8.8cm} 
\epsfig{figure=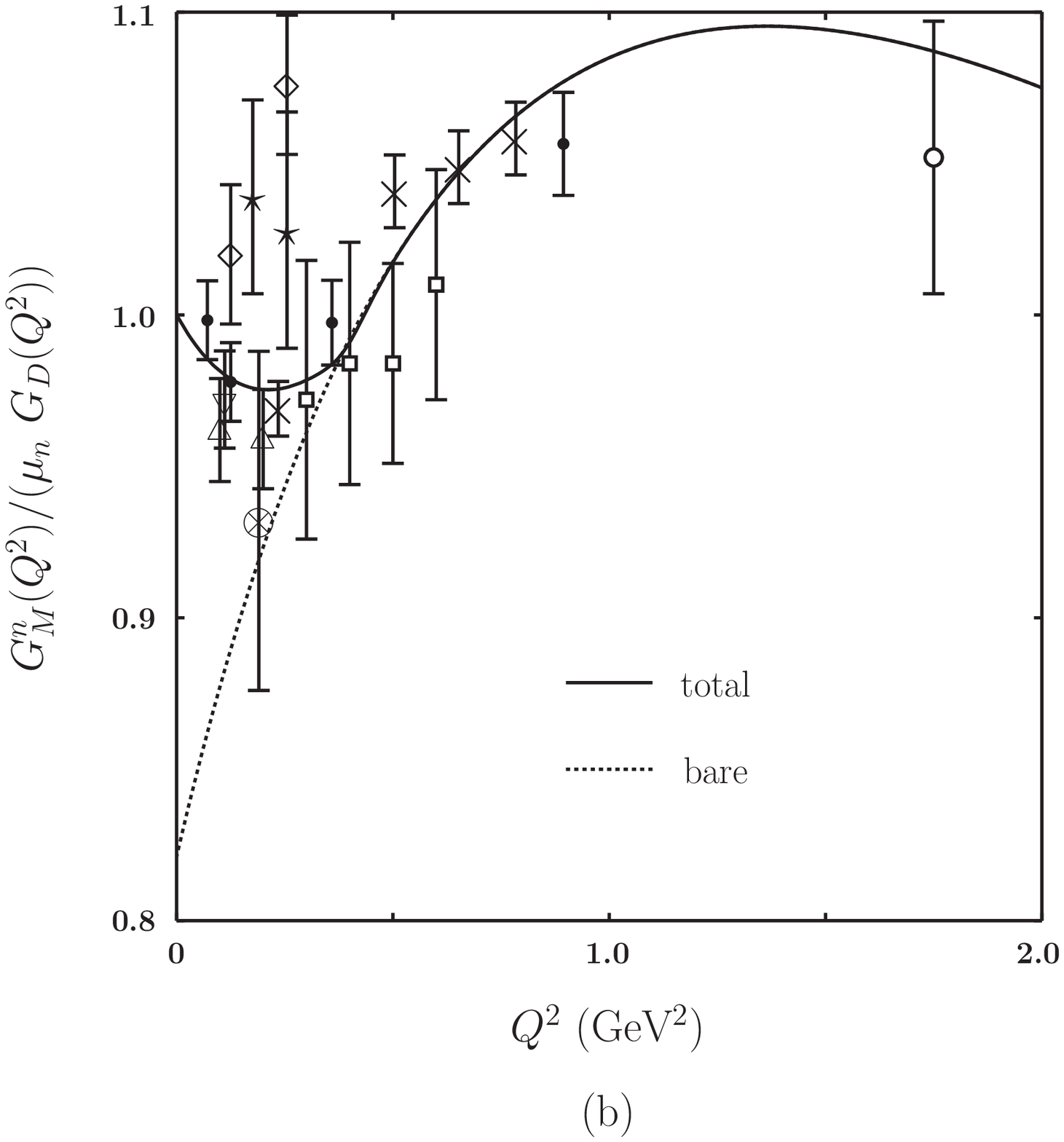,height=8.8cm}
\end{center} 
\vspace*{0.5cm}
{\bf Fig. 7.} {\em Ratio $G_M^n(Q^2)/(\mu_n\,G_D(Q^2))$ : (a) Overall range, 
(b) Up to $Q^2=2\,\,\,{\rm GeV}^2$, the solid line is the total 
contribution and the dotted line is the bare contribution. 
Experimental data are taken from 
Refs.~\cite{Lung:1992bu,Kubon:2001rj,Xu:2000xw,Anklin:1994ae,
Anklin:1998ae,Rock:1982gf,Markowitz:1993hx,Bruins:1995ns,Xu:2002xc,Gao:1994ud}.}
\vspace*{1cm}
\begin{center} 
\epsfig{figure=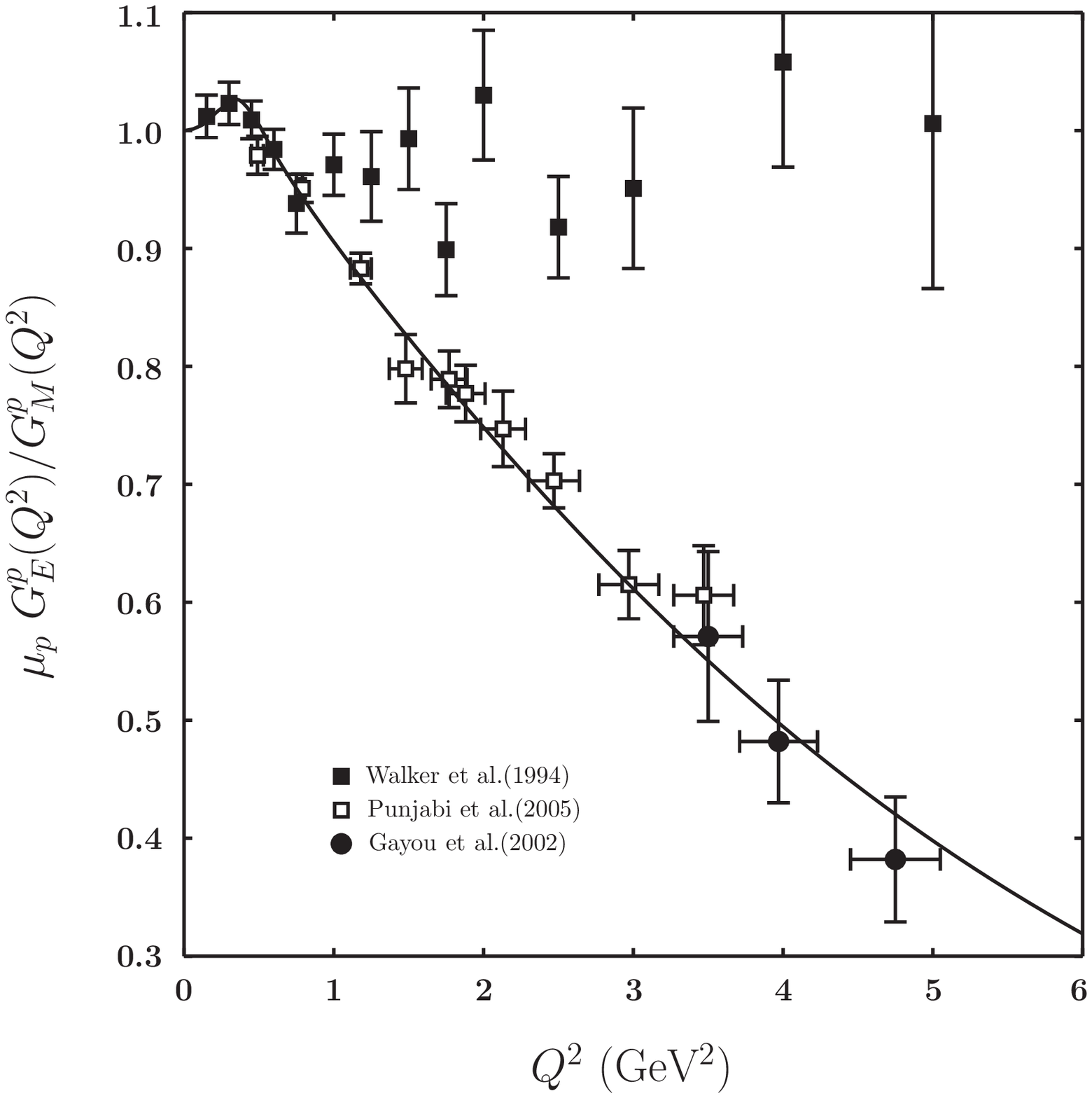,height=8.8cm} 
\end{center}
\vspace*{0.5cm}
{\bf Fig. 8.} {\em Ratio $\mu_p G_E^p(Q^2)/G_M^p(Q^2)$ in comparison to the 
experimental data. Experimental data are taken from 
Refs.~\cite{Walker:1993vj,Gayou:2001qd,Punjabi:2005wq}.}

\newpage 
\vspace*{-1cm}
\begin{center}
\epsfig{figure=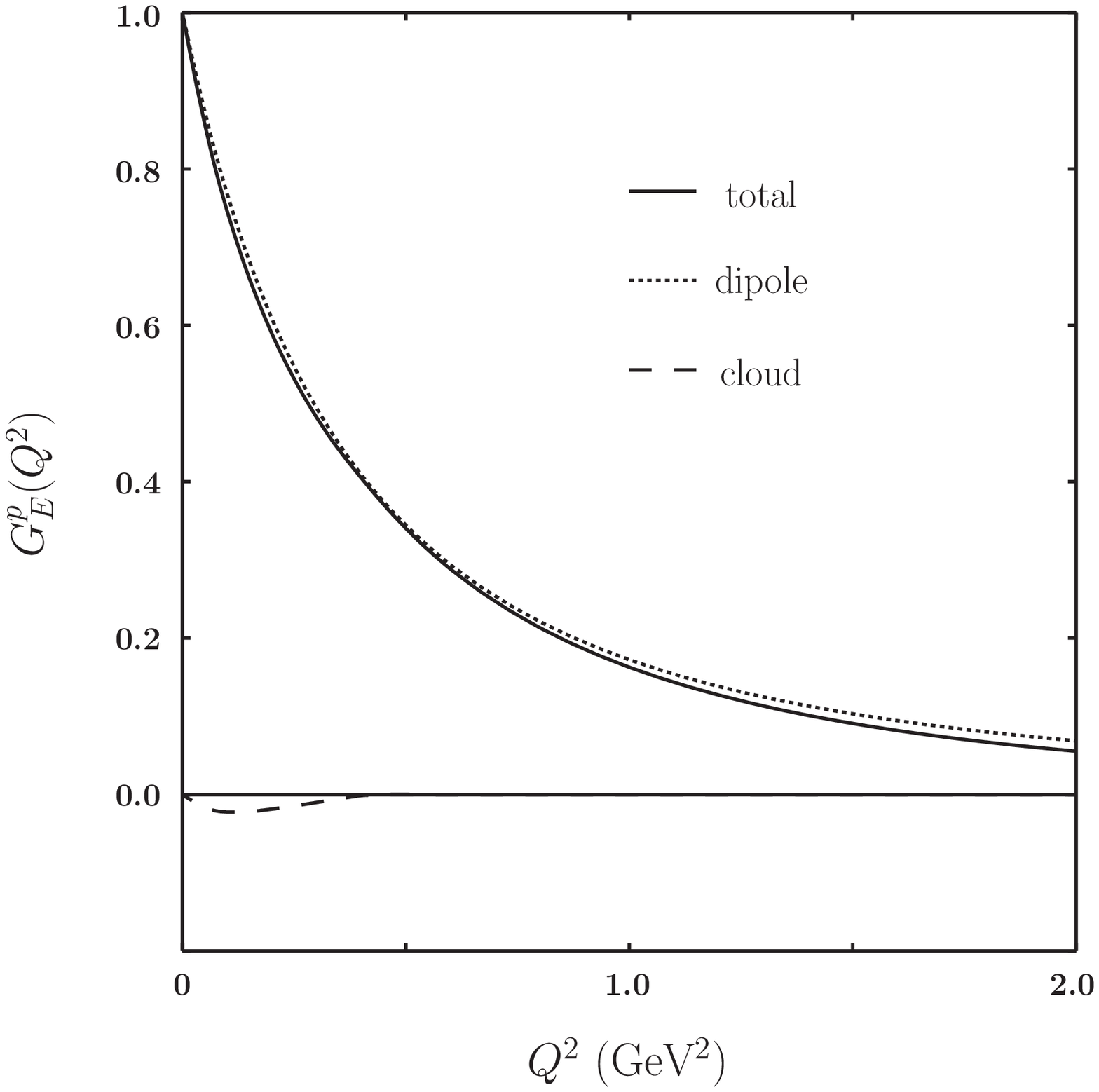,height=8.8cm} 
\epsfig{figure=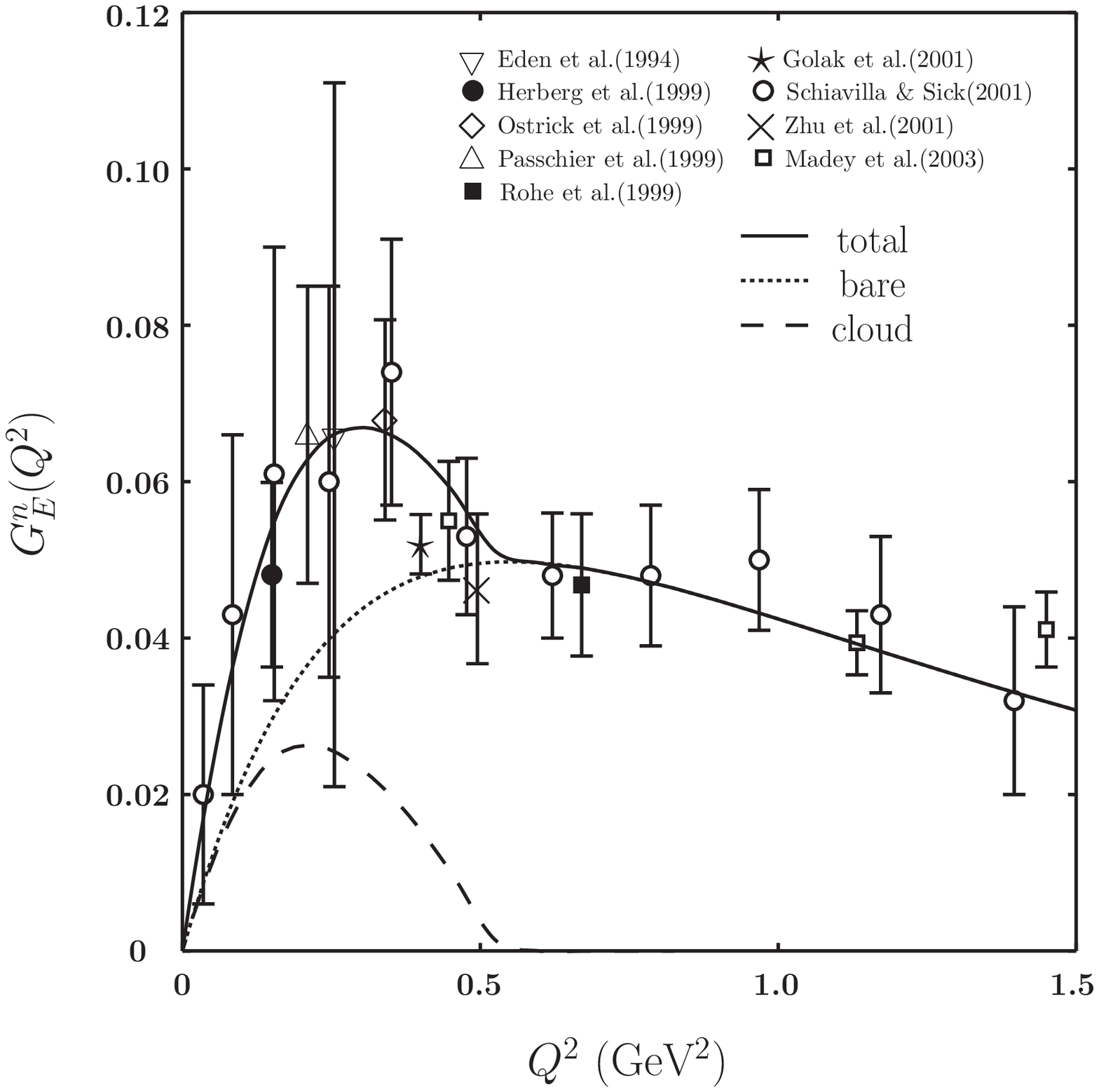,height=8.8cm}

\vspace*{1cm} 
\epsfig{figure=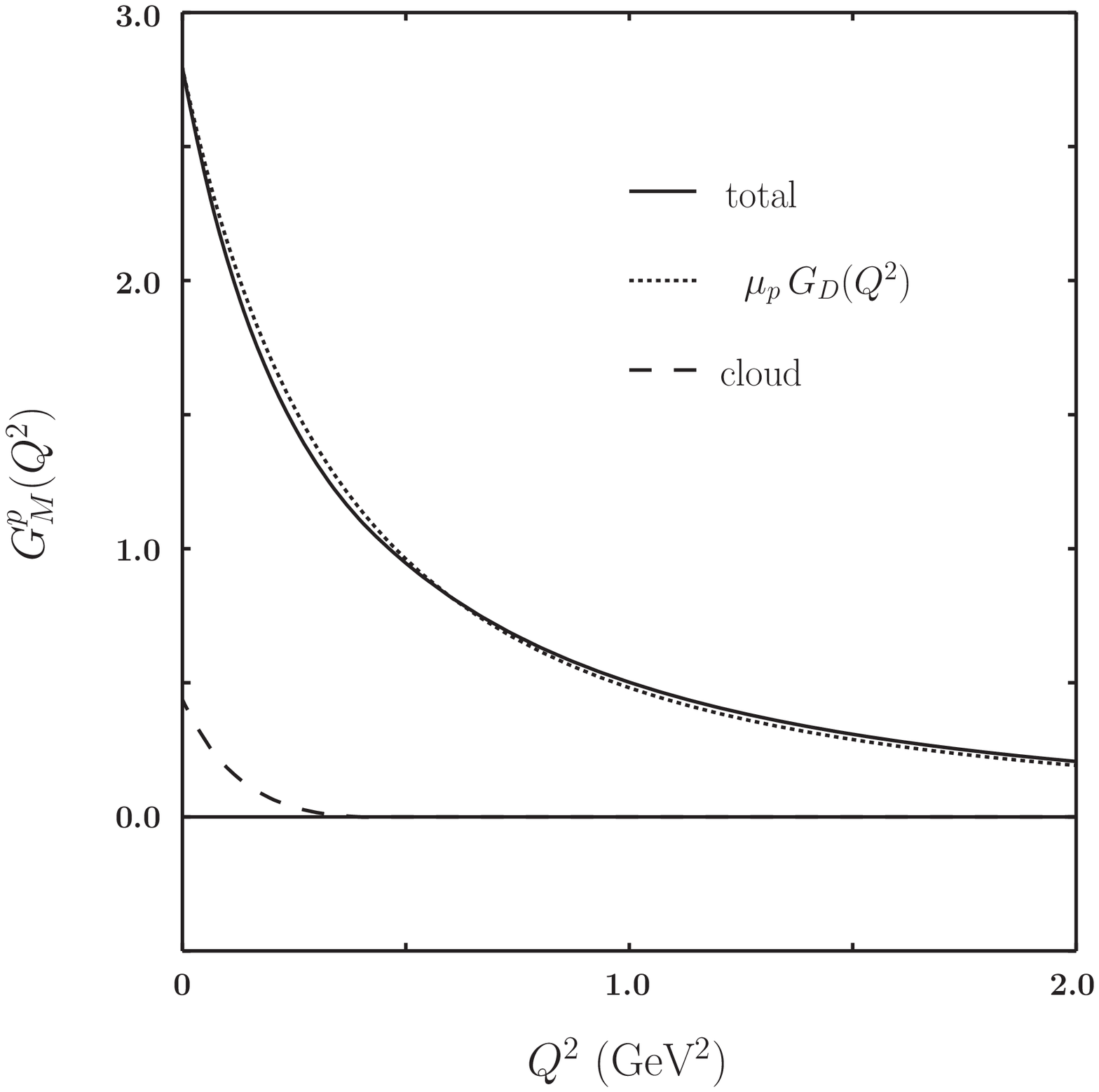,height=8.8cm}
\epsfig{figure=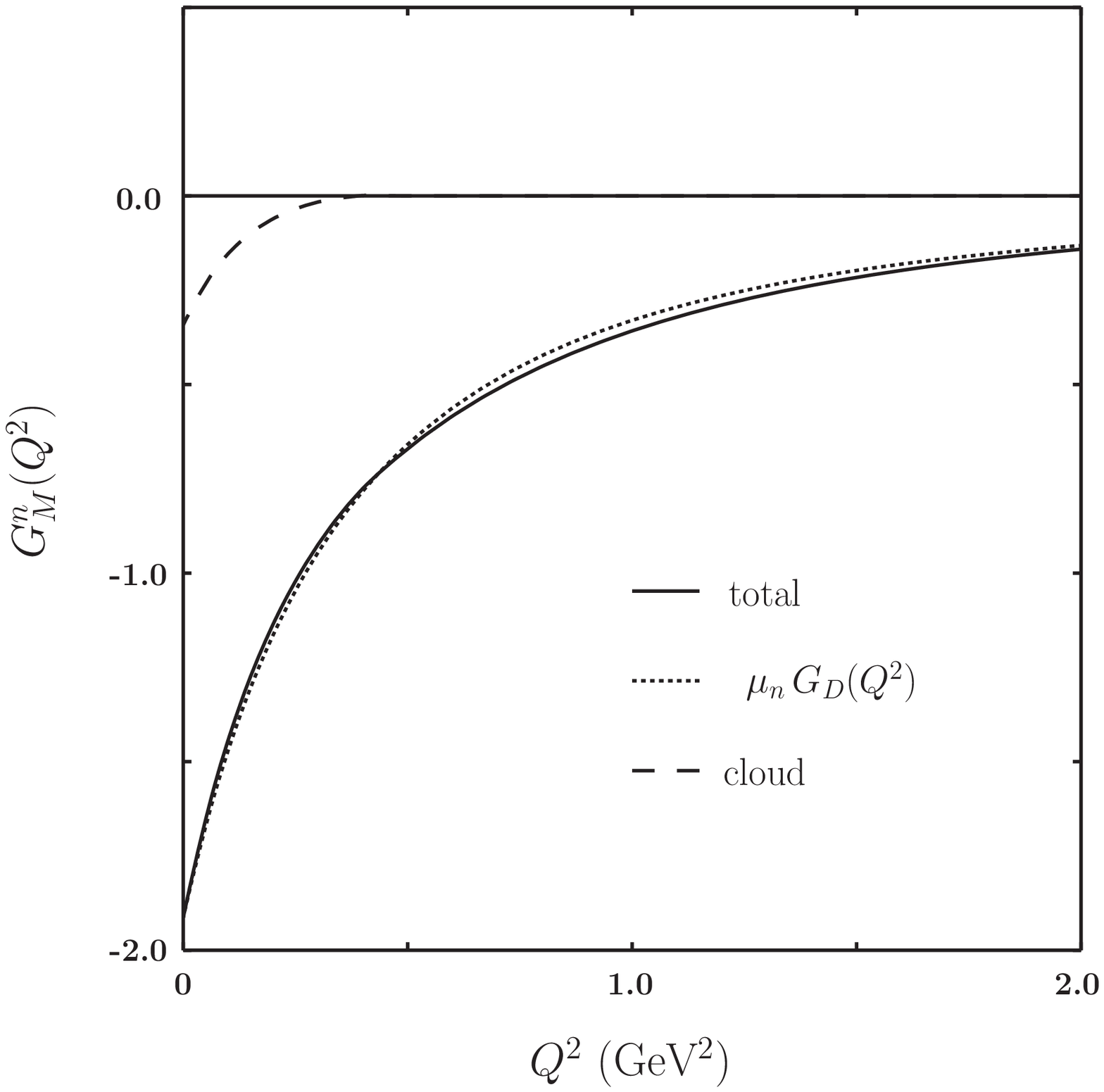,height=8.8cm} 
\end{center} 
\vspace*{0.5cm}
{\bf Fig. 9.} {\em The charge and magnetic form factors of the nucleon and the contribution due to the meson cloud. The corresponding dipole fit for $G_E^p(Q^2)$, $G_M^p(Q^2)$ and $G_M^n(Q^2)$ are also presented. Experimental data for 
$G_E^n(Q^2)$ are taken from 
Refs.~\cite{Herberg:1999ud,Passchier:1999cj,Eden:1994ji,Ostrick:1999xa,Golak:2000nt,Madey:2003av,Zhu:2001md,Rohe:1999sh,Schiavilla:2001qe}.}

\newpage 
\vspace*{-1cm}
\begin{center}
\epsfig{figure=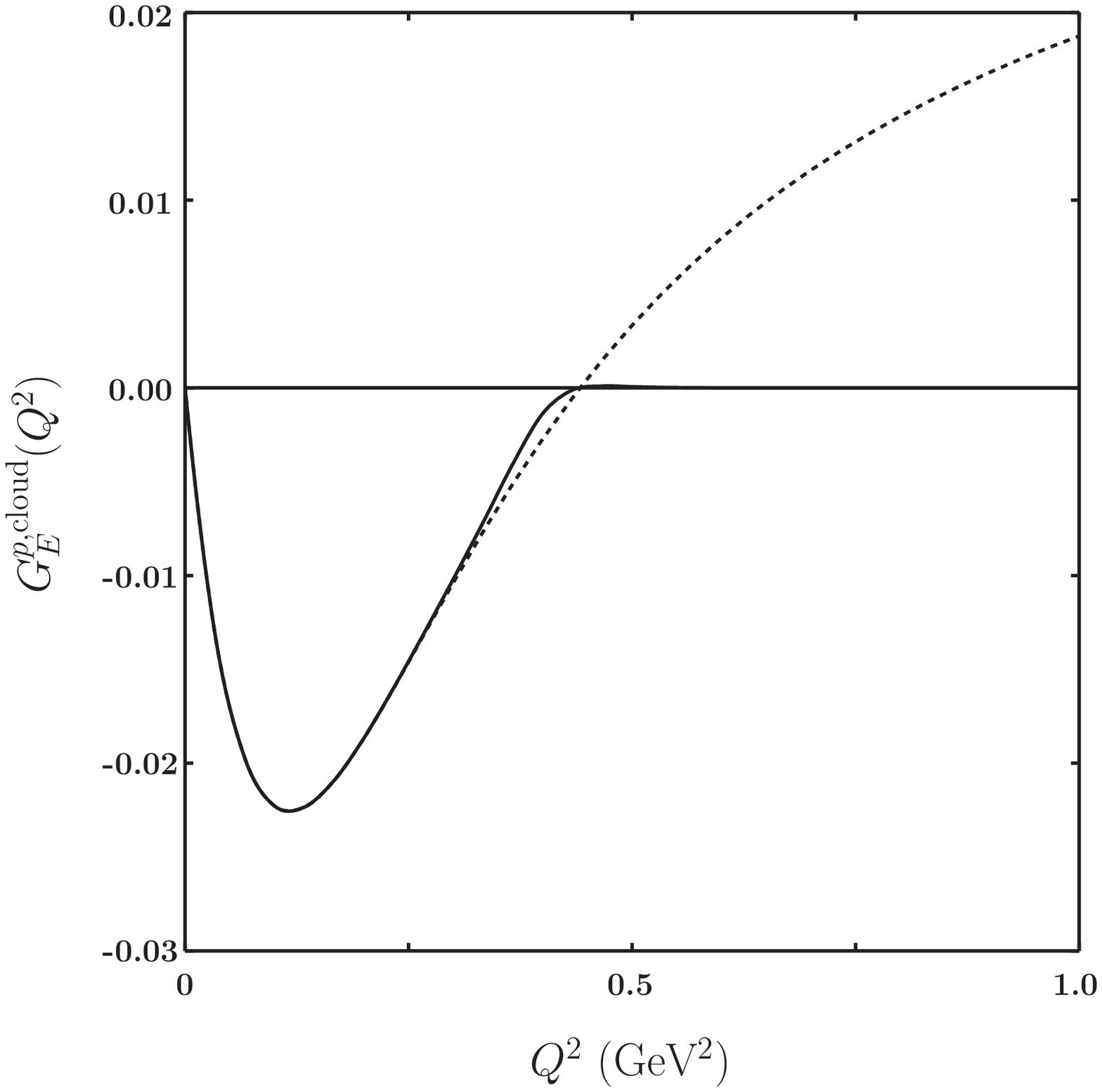,height=8.8cm} 
\epsfig{figure=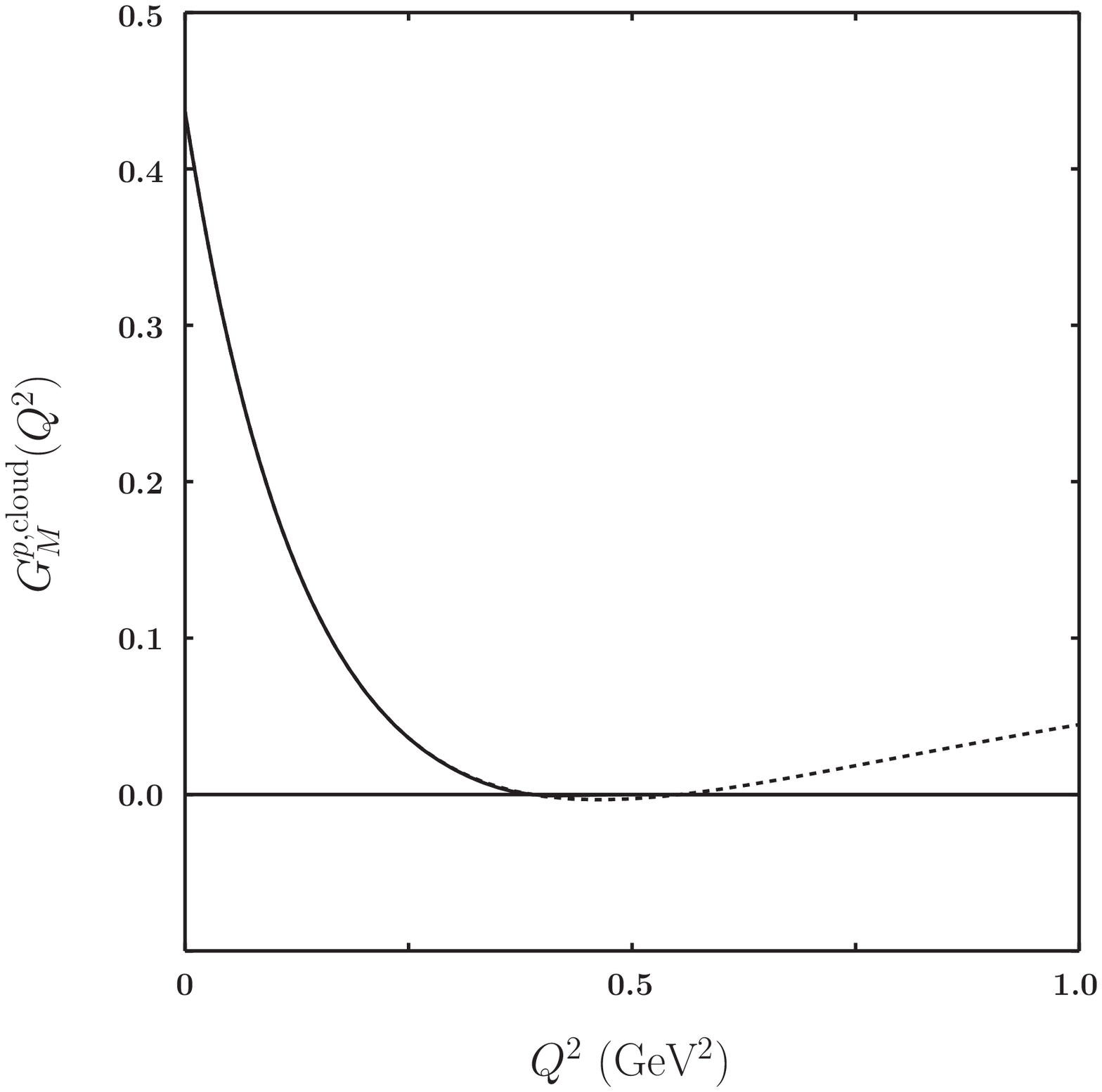,height=8.8cm}

\vspace*{1cm}  
\epsfig{figure=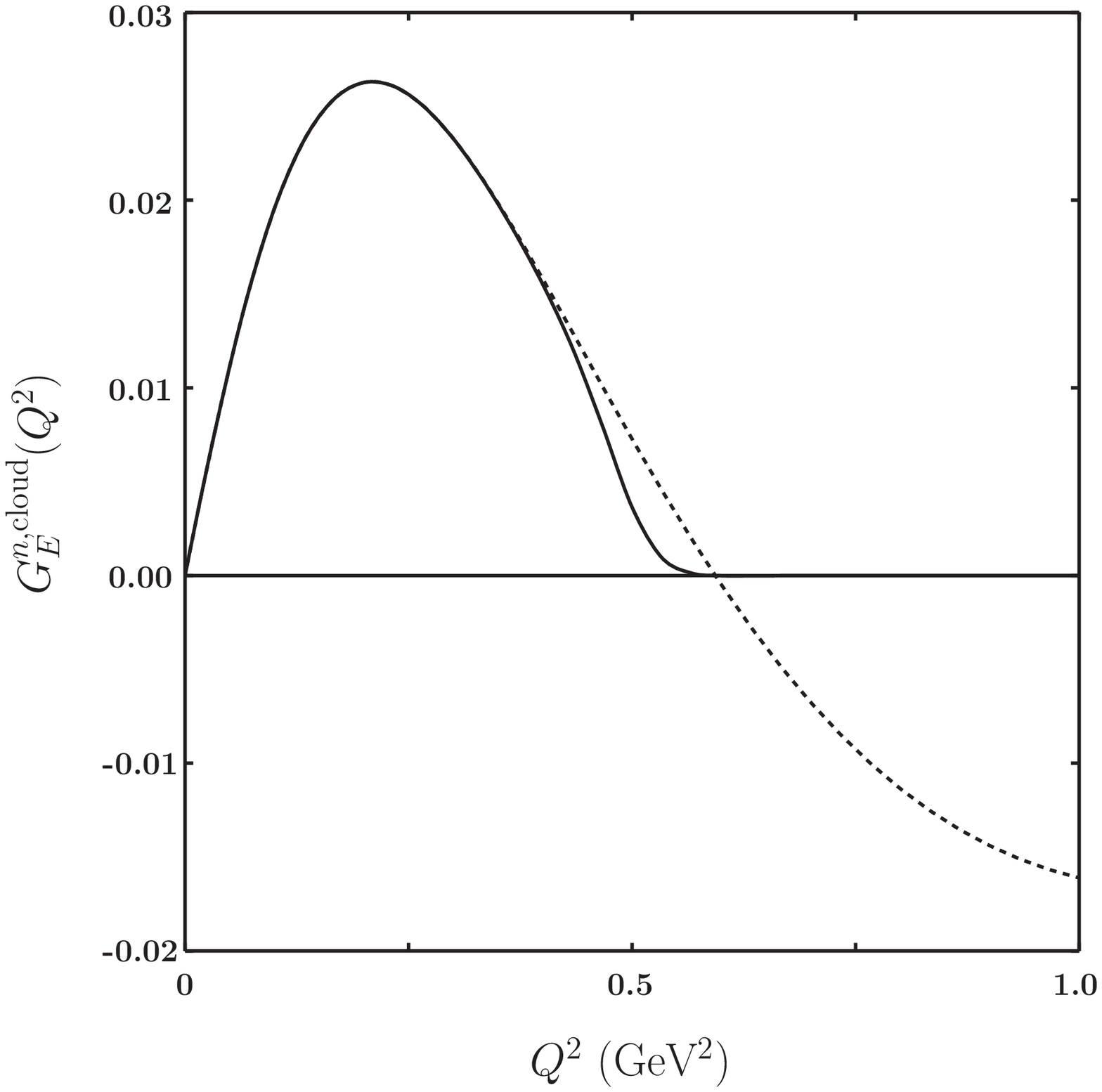,height=8.8cm} 
\epsfig{figure=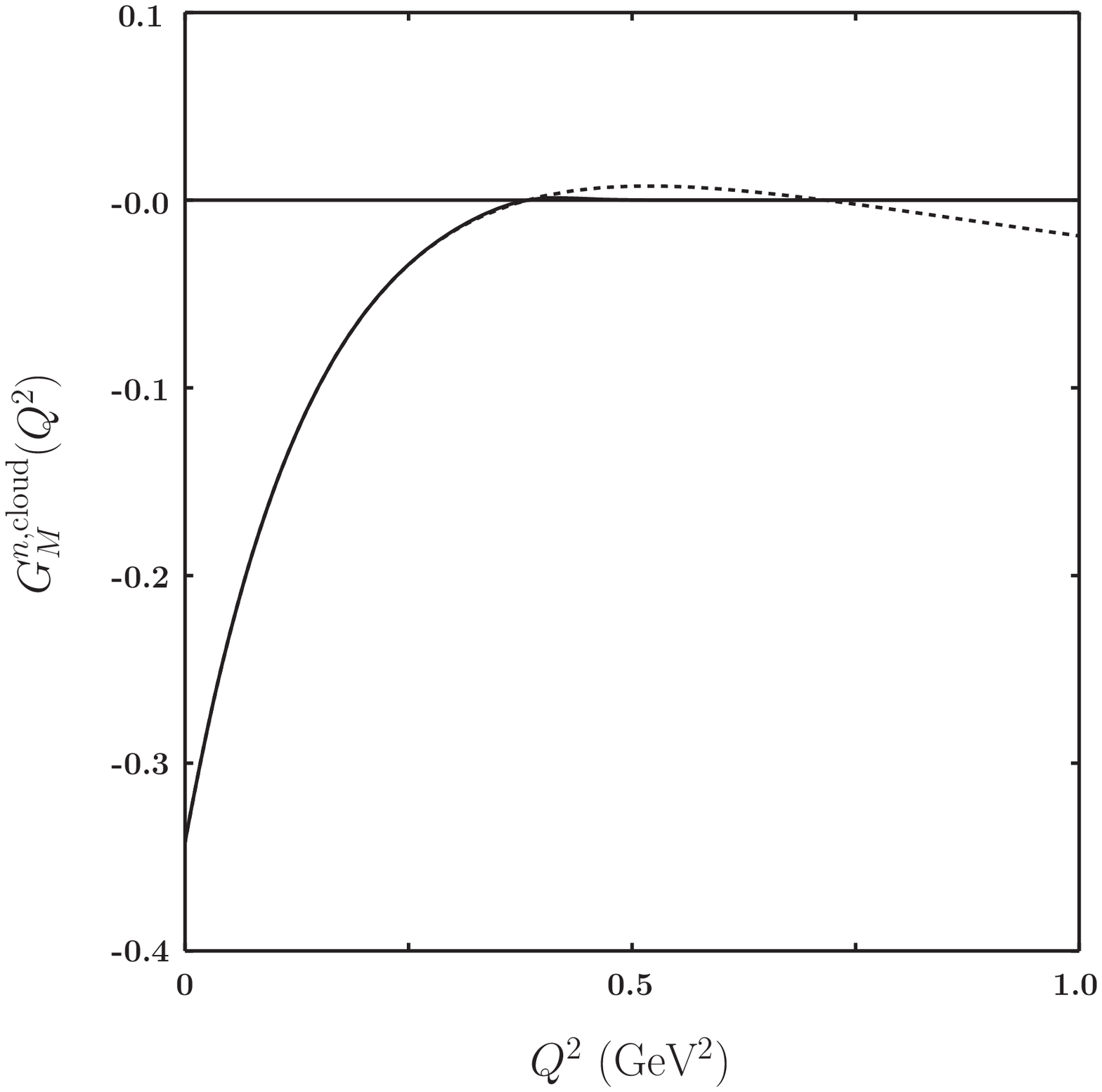,height=8.8cm}  
\end{center} 
\vspace*{0.5cm}
{\bf Fig. 10.} {\em The meson cloud contribution to the nucleon charge and magnetic 
form factors with the cutoff function $f_{cut}(Q^2)$ 
(solid line) and without $f_{cut}(Q^2)$ (dotted line).}

\newpage 
\vspace*{-1cm}
\begin{center} 
\epsfig{figure=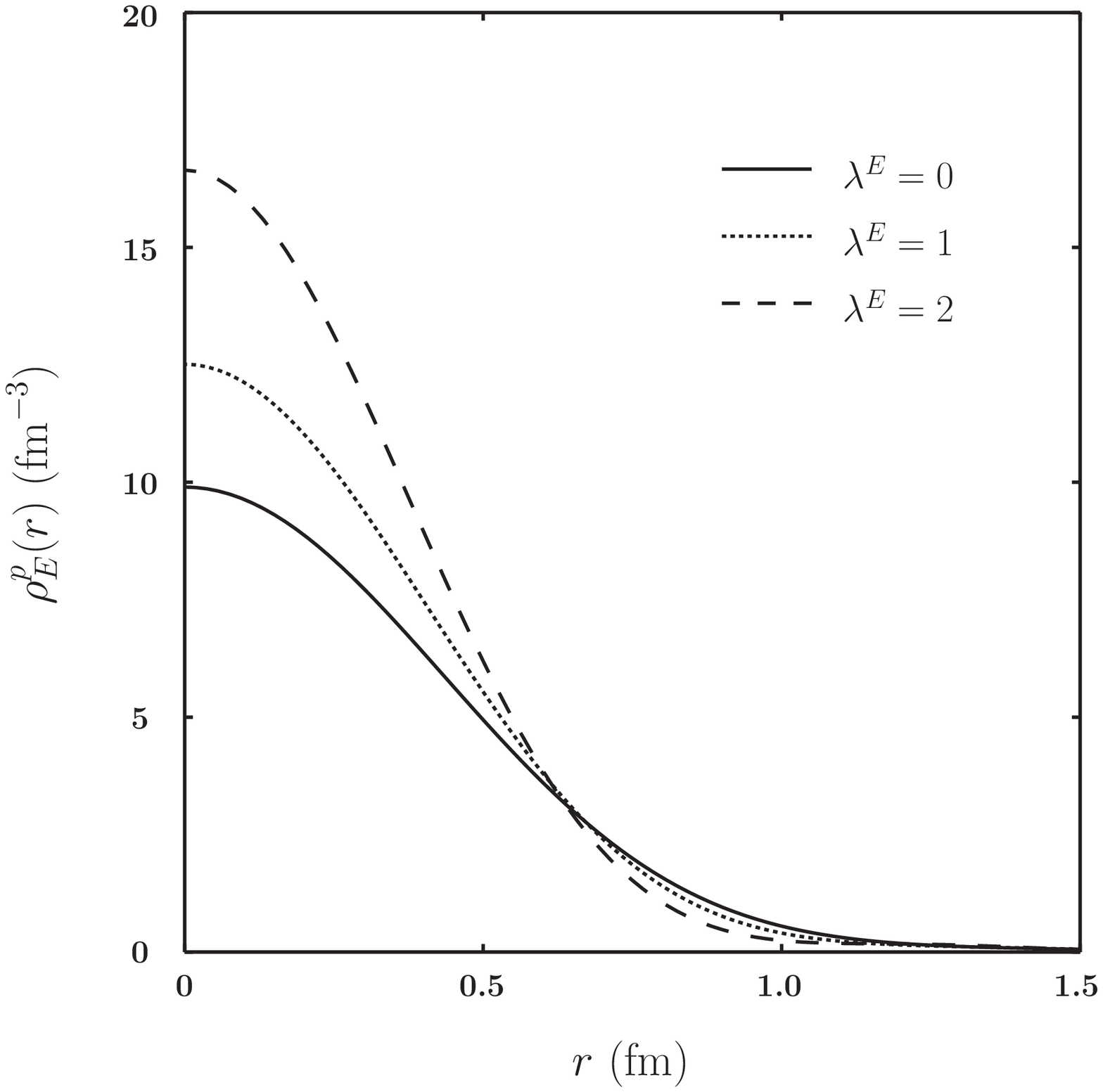,height=8.8cm}  
\epsfig{figure=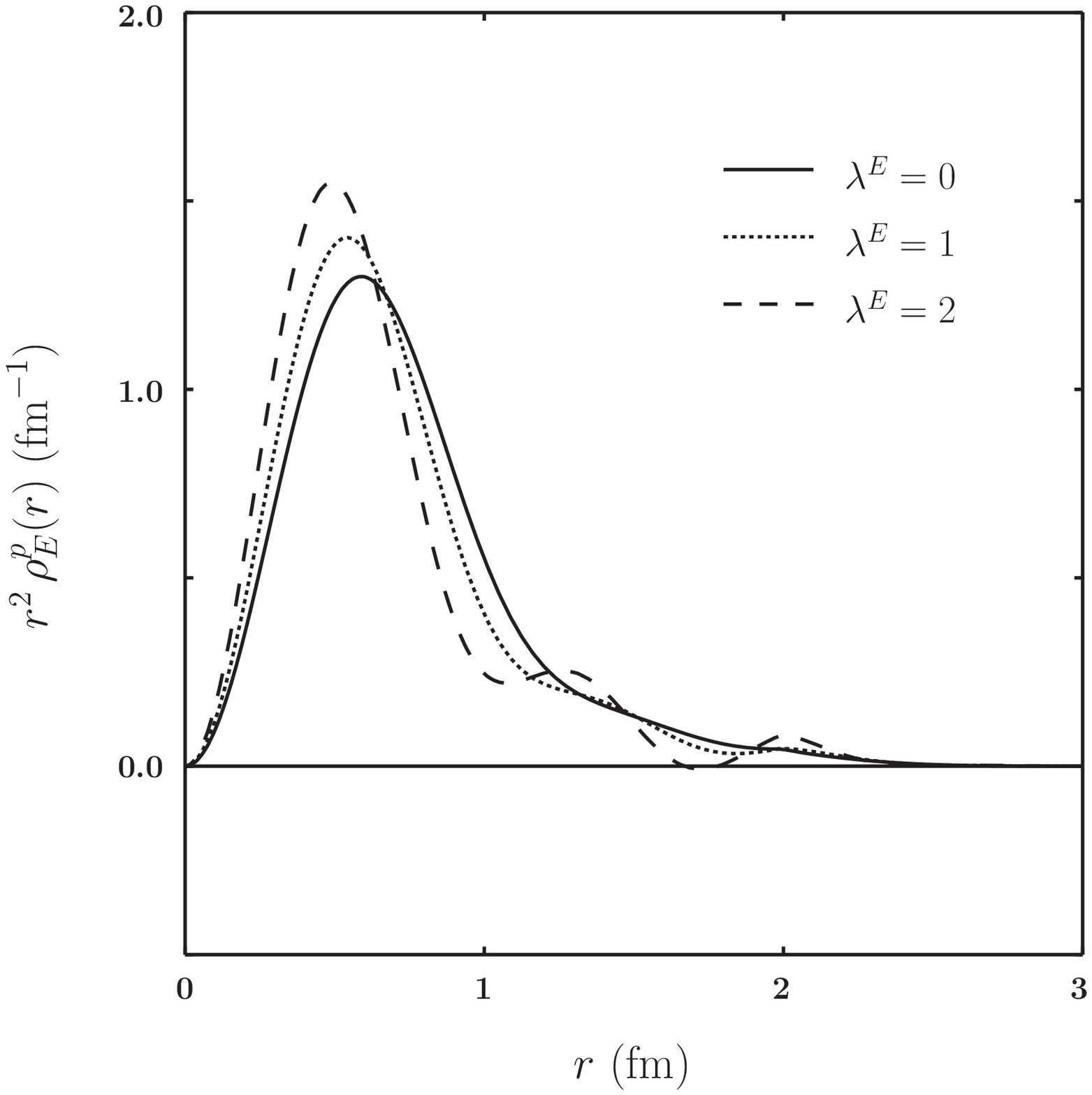,height=8.8cm} 
 
\vspace*{1cm}  
\epsfig{figure=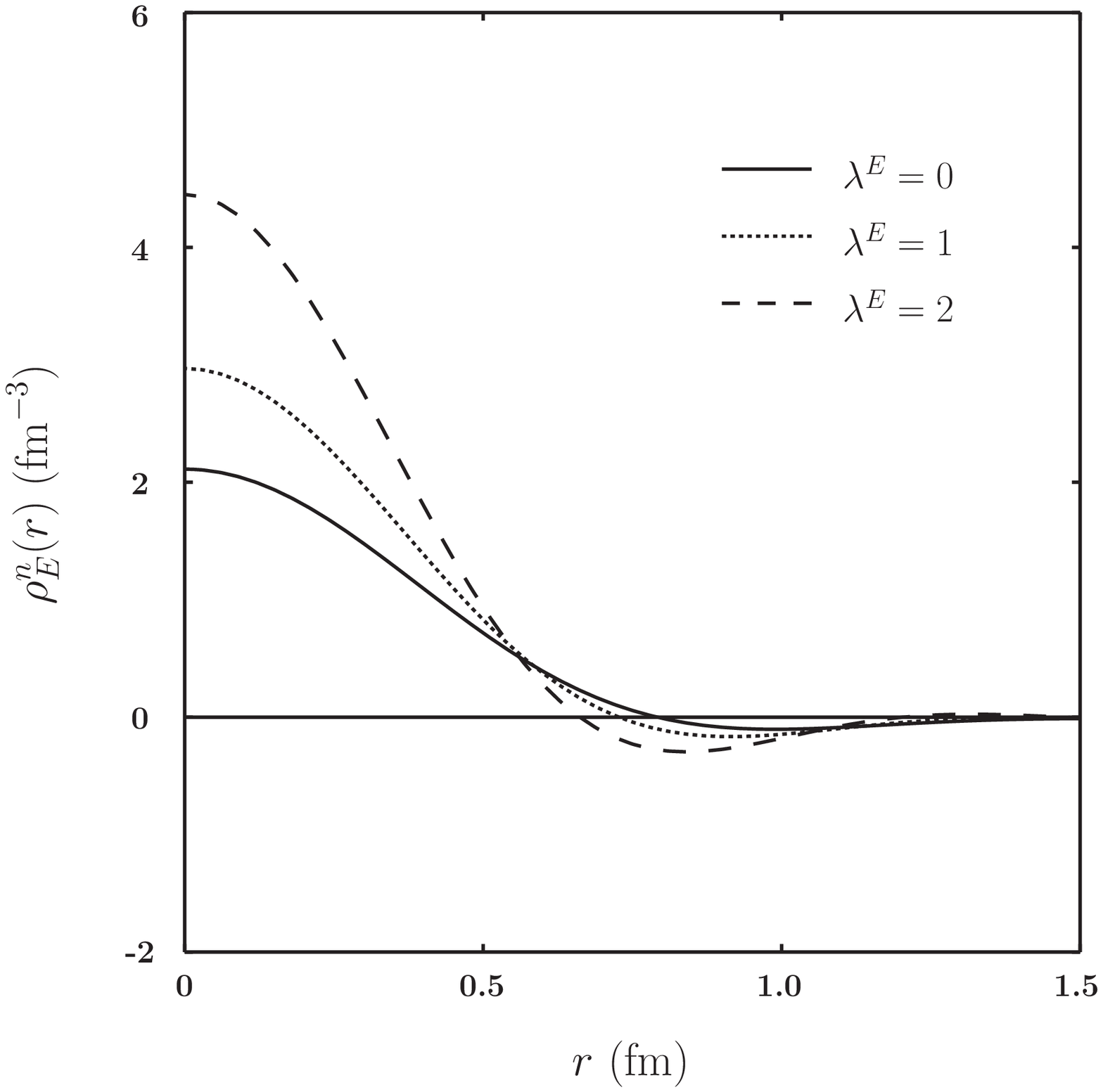,height=8.8cm} 
\epsfig{figure=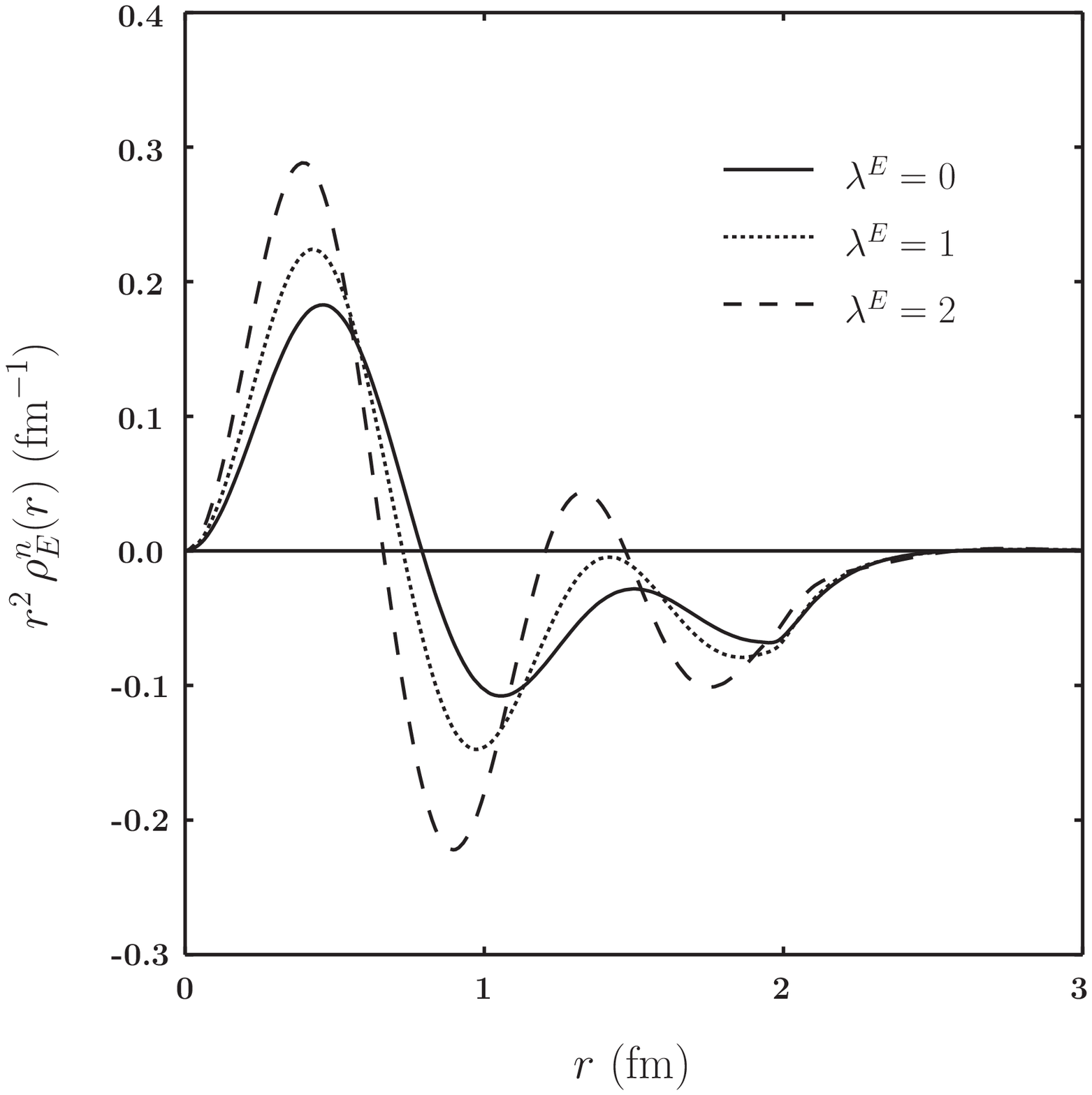,height=8.8cm} 
\end{center}
\vspace*{0.5cm}
{\bf Fig. 11.} {\em Variation of the charge density of the nucleon $\rho_E^N(r)$ 
and $r^2\,\rho_E^N(r)$ with $\lambda^E$: $\lambda^E=0$ (solid line), 
$\lambda^E=1$ (dotted line)}, and $\lambda^E=2$ (dashed line).

\newpage 
\vspace*{-1cm}
\begin{center} 
\epsfig{figure=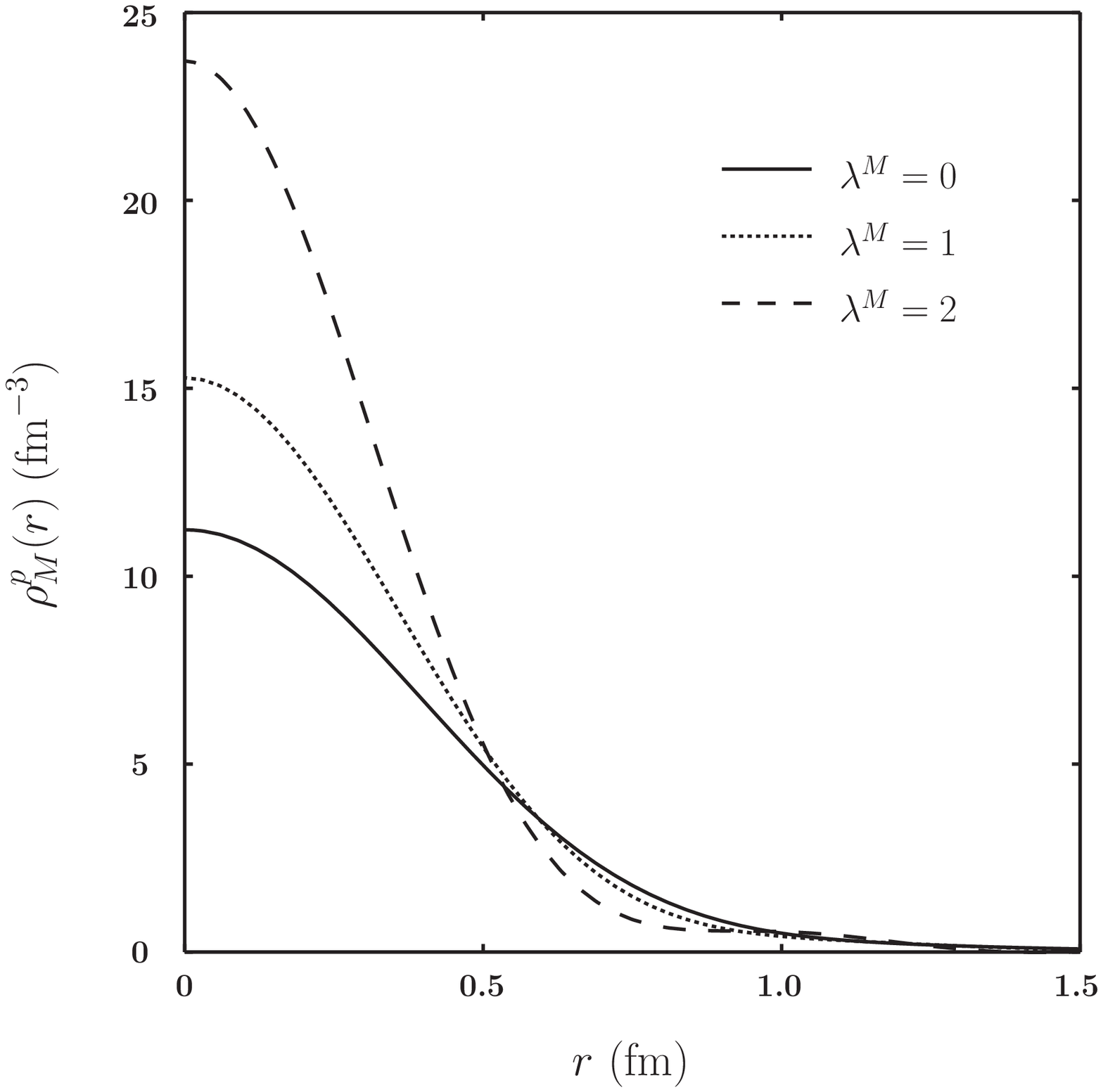,height=8.8cm} 
\epsfig{figure=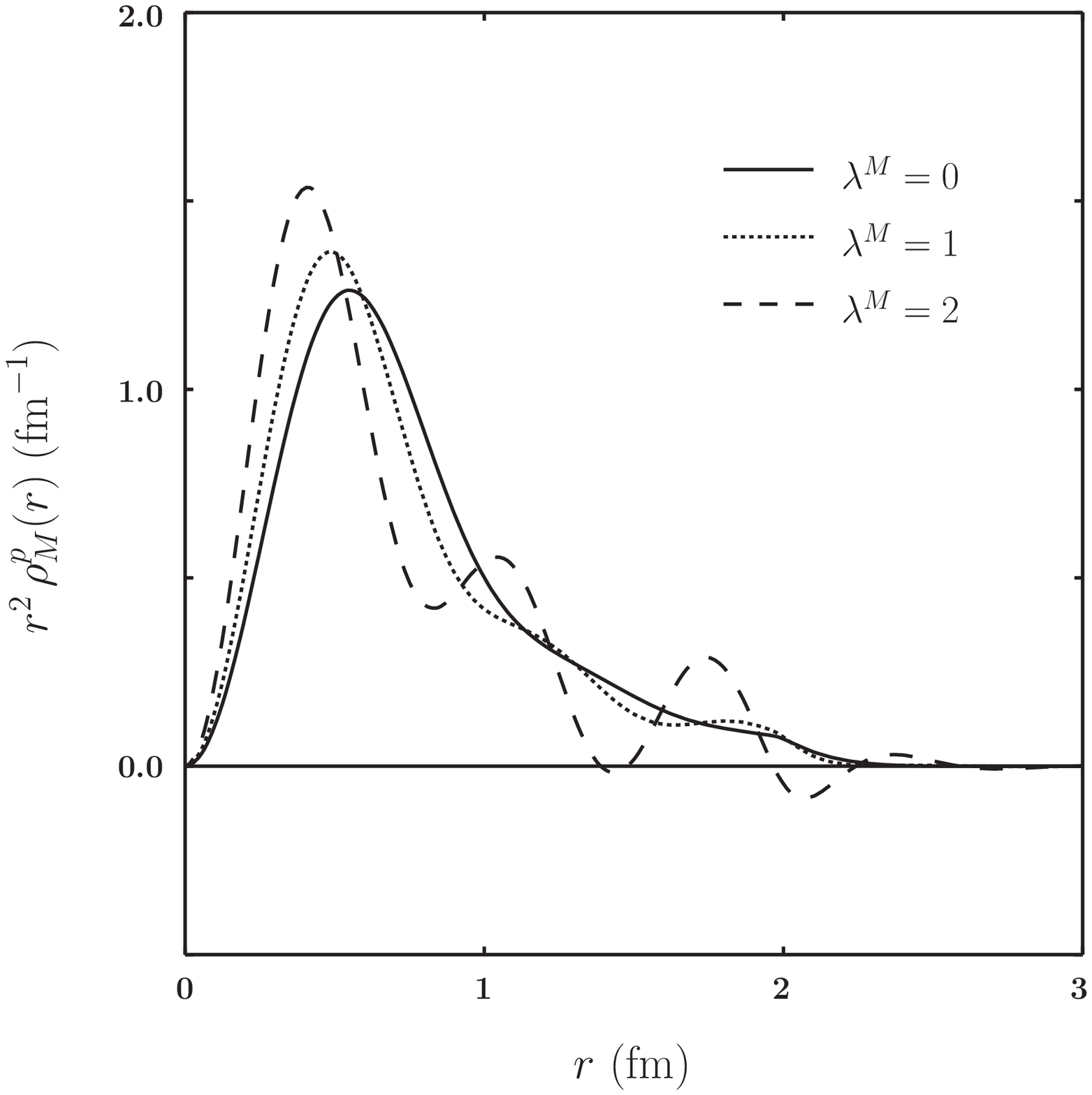,height=8.8cm} 
 
\vspace*{1cm} 
\epsfig{figure=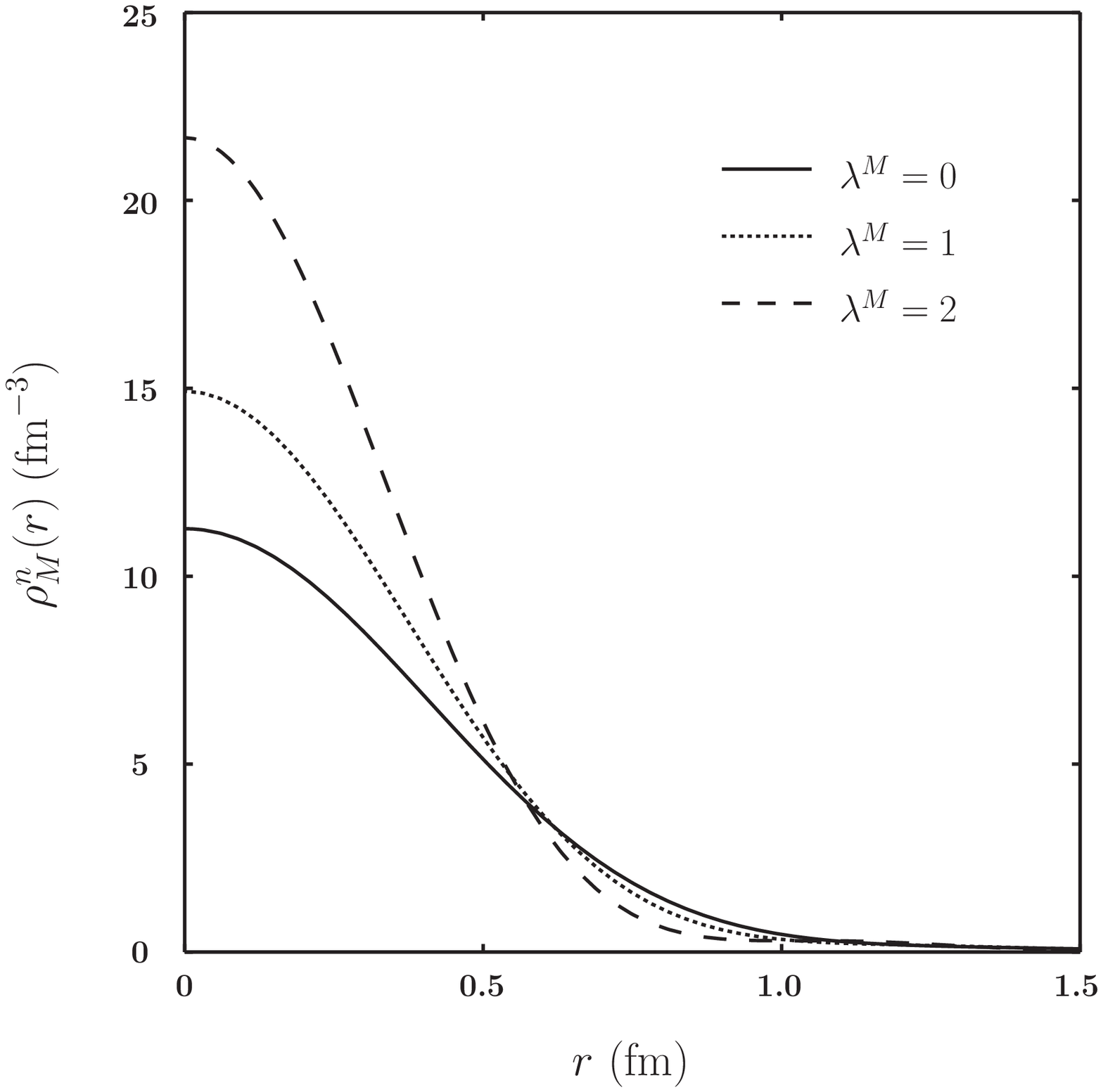,height=8.8cm} 
\epsfig{figure=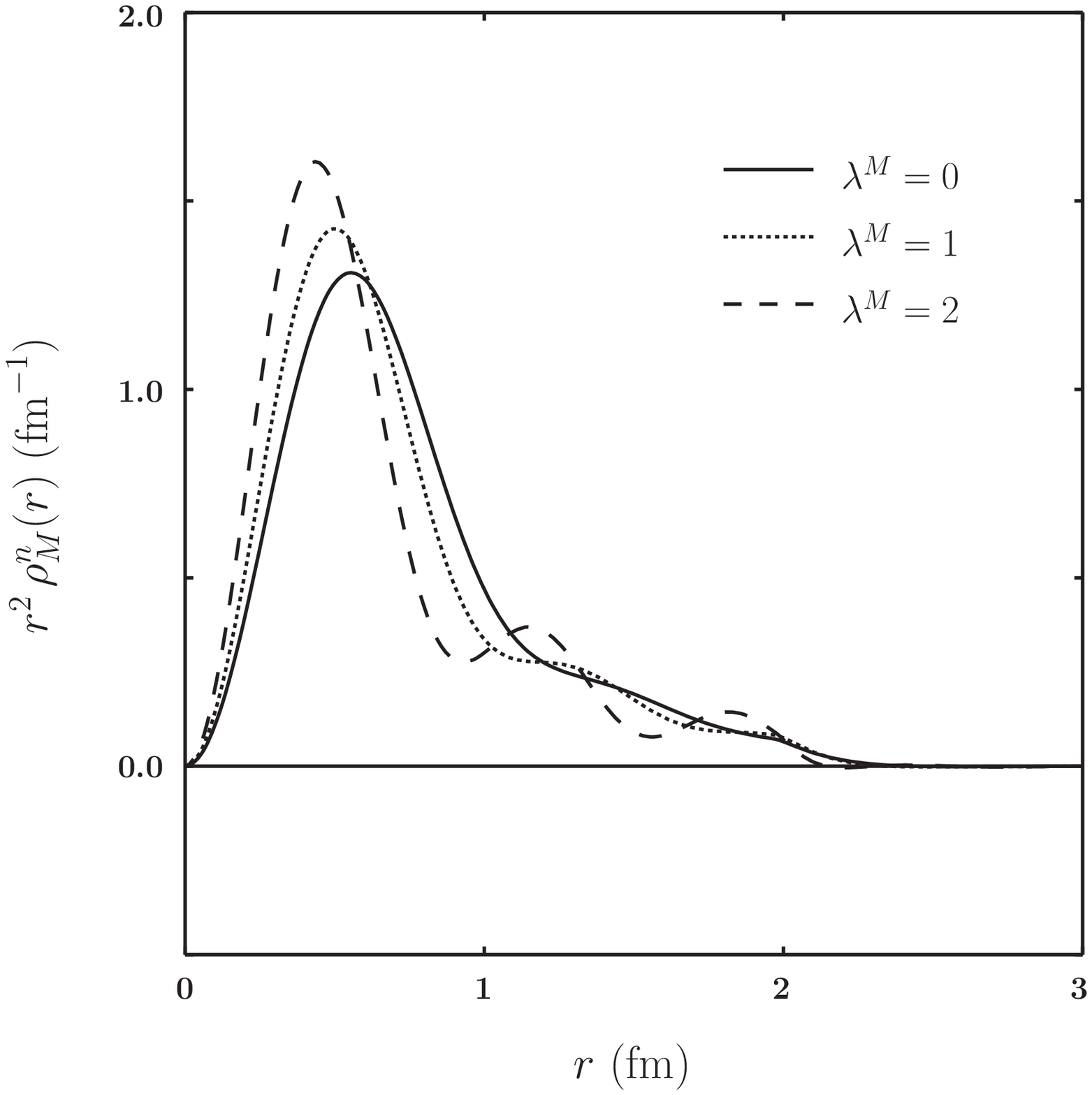,height=8.8cm} 
\end{center}
\vspace*{0.5cm}
{\bf Fig. 12.} {\em Variation of the magnetization density of the nucleon 
$\rho_M^N(r)$ and $r^2\,\rho_M^N(r)$ with $\lambda^M$: $\lambda^M=0$ 
(solid line), $\lambda^M=1$ (dotted line)}, and $\lambda^M=2$ (dashed line).

\end{document}